%% file: STU10.tex
\documentclass[11pt]{myArticle}       

\usepackage{amsmath,amssymb,amsfonts} 
\usepackage{graphicx}
\usepackage{epsfig}
\usepackage{color}                    
\usepackage{hyperref}                 
\usepackage{myFancyhdr}               
\usepackage{makeidx}                  
\usepackage{helvet}                   
\usepackage{setspace}                 
\usepackage{fancybox}                 
\usepackage{float}                    
\usepackage{xspace}                   
\usepackage{wasysym}                  
\usepackage[numbers,sort&compress]{natbib}
\usepackage{hypernat}
\definecolor{mygray}{rgb}{0.3,0.32,0.35}
\definecolor{darkblue1}{rgb}{0,0,.2}
\definecolor{darkblue}{rgb}{0,0,.3}
\definecolor{darkred}{rgb}{0.5,0,0}
\pagecolor{white} 
\color{black}     
\hypersetup{breaklinks=true, 
            colorlinks=true, 
            linkcolor=darkblue1, 
            menucolor=darkblue1, 
            urlcolor=darkblue1,
            citecolor=darkblue1,
            pdftitle={The SM EW fit Revisited},
            pdfauthor={Gfitter Group},
            pdfsubject={The SM EW fit Revisited},
            pdfkeywords={},
            pdfproducer={Gfitter Group}
}
\newcommand\defaultFigureScale{0.6}
\newcommand\HalfPageWidthScale{0.411}
\bibstyle{plain}
\input definitions

\makeindex
\begin{document}
%
%
\input Title

%
%
\pagenumbering{roman}

\setcounter{tocdepth}{3}
\addtolength{\parskip}{-0.40\baselineskip}
{\footnotesize\sl
\tableofcontents
}
\addtolength{\parskip}{+0.34\baselineskip}

\newpage

\pagenumbering{arabic}
%
%
\input Introduction
%
%
\input SMfit
\input Oblparam

%
%

\input Constraints

%
%
\input Conclusions

\input Appendix
%
%
\newpage

\addcontentsline{toc}{section}{References}
\bibliography{References}{}

\end{document}

%% file: definitions.tex
\renewcommand{\headrulewidth}{0.5pt}    
\renewcommand{\footrulewidth}{0.5pt}    

\newcommand\allFontSize{\small}

\newenvironment{myquote}
               {\list{}{\leftmargin0cm}%
                \item\relax}
               {\endlist}

\usepackage[bf]{caption}

\newcommand\detailsSize{\allFontSize}
\newenvironment{details}%
{\begin{myquote}\vspace{-0.2cm}\detailsSize}{\end{myquote}\vspace{-0.2cm}}

\newfloat{codeexample}{H}{loc}
\floatname{codeexample}{\sf\allFontSize Code Example}

\newlength{\gfitterboxwidth}
\definecolor{DarkGray}{rgb}{0.4,0.42,0.45}
\definecolor{LightGray}{rgb}{0.97,0.98,0.98}

{\VerbatimEnvironment\normalsize
\setlength{\gfitterboxwidth}{\textwidth}\addtolength{\gfitterboxwidth}{-2.3\fboxsep}%
\begin{Sbox}\begin{minipage}[t]{\gfitterboxwidth}\begin{Verbatim}}%
{\end{Verbatim}\end{minipage}\end{Sbox}\vspace{1ex} \fcolorbox{DarkGray}{LightGray}{\TheSbox}}

\newfloat{option}{H}{loc}
\floatname{option}{\sf\allFontSize Option Table}

\usepackage{tabularx}
{\setlength{\gfitterboxwidth}{\textwidth}\addtolength{\gfitterboxwidth}{-2.3\fboxsep}%
\begin{Sbox}\begin{tabular*}{\gfitterboxwidth}{>{\rule[-0.5ex]{.0ex}{3.0ex}\tt\small}p{3cm}>{\tt\small}p{4.5cm}>{\small}p{5.7cm}}
&&\\[-0.7cm]
\rm\small Option  & \rm\small Values    & \rm\small Description\\[0.1cm]\hline
&&\\[-0.45cm]}%
{\\[-0.1cm]\end{tabular*}\end{Sbox}\vspace{1ex} \fbox{\TheSbox}}

{\setlength{\gfitterboxwidth}{\textwidth}\addtolength{\gfitterboxwidth}{-2.3\fboxsep}%
\begin{Sbox}\begin{tabular*}{\gfitterboxwidth}{>{\rule[-0.5ex]{.0ex}{3.0ex}\tt\small}p{3cm}>{\tt\small}p{3.5cm}>{\small}p{6.7cm}}
&&\\[-0.7cm]
\rm\small Option  & \rm\small Values    & \rm\small Description\\[0.1cm]\hline
&&\\[-0.45cm]}%
{\\[-0.1cm]\end{tabular*}\end{Sbox}\vspace{1ex} \fbox{\TheSbox}}

{\setlength{\gfitterboxwidth}{\textwidth}\addtolength{\gfitterboxwidth}{-2.3\fboxsep}%
\begin{Sbox}\begin{tabular*}{\gfitterboxwidth}{>{\rule[-0.5ex]{.0ex}{3.5ex}\tt\small}p{3.0cm}>{\tt\small}p{3.0cm}>{\small}p{7.2cm}}
&&\\[-0.7cm]
\rm\small Option  & \rm\small Values    & \rm\small Description\\[0.1cm]\hline
&&\\[-0.45cm]}%
{\\[-0.1cm]\end{tabular*}\end{Sbox}\vspace{1ex} \fbox{\TheSbox}}

{\setlength{\gfitterboxwidth}{\textwidth}\addtolength{\gfitterboxwidth}{-2.3\fboxsep}%
\begin{Sbox}\begin{tabular*}{\gfitterboxwidth}{>{\rule[-0.5ex]{.0ex}{3.0ex}\tt\small}p{2.5cm}>{\tt\small}p{3.5cm}>{\small}p{7.2cm}}
&&\\[-0.7cm]
\rm\small Option  & \rm\small Values    & \rm\small Description\\[0.1cm]\hline
&&\\[-0.45cm]}%
{\\[-0.1cm]\end{tabular*}\end{Sbox}\vspace{1ex} \fbox{\TheSbox}}

{\setlength{\gfitterboxwidth}{\textwidth}\addtolength{\gfitterboxwidth}{-2\fboxsep}%
\begin{Sbox}\begin{tabular*}{\gfitterboxwidth}{>{\rule[-0.5ex]{.0ex}{3.0ex}\tt\small}p{4.5cm}>{\tt\small}p{3.2cm}>{\small}p{5.5cm}}
&&\\[-0.7cm]
\rm\small Option  & \rm\small Values    & \rm\small Description\\[0.1cm]\hline
&&\\[-0.45cm]}%
{\\[-0.1cm]\end{tabular*}\end{Sbox}\vspace{1ex} \fbox{\TheSbox}}

{\setlength{\gfitterboxwidth}{\textwidth}\addtolength{\gfitterboxwidth}{-2\fboxsep}%
\begin{Sbox}\begin{tabular*}{\gfitterboxwidth}{>{\rule[-0.5ex]{.0ex}{3.0ex}\tt\small}p{4.0cm}>{\tt\small}p{3.0cm}>{\small}p{6.2cm}}
&&\\[-0.7cm]
\rm\small Option  & \rm\small Values    & \rm\small Description\\[0.1cm]\hline
&&\\[-0.45cm]}%
{\\[-0.1cm]\end{tabular*}\end{Sbox}\vspace{1ex} \fbox{\TheSbox}}

\newfloat{programs}{H}{loc}
\floatname{programs}{\sf Table}

\usepackage{tabularx}
{\setlength{\gfitterboxwidth}{\textwidth}\addtolength{\gfitterboxwidth}{-2\fboxsep}%
\begin{Sbox}\begin{tabular*}{\gfitterboxwidth}{>{\rule[-0.5ex]{.0ex}{3.0ex}\tt\small}p{4.1cm}>{\small}p{9.5cm}}
&\\[-0.7cm]
\rm\small Macro & \rm\small Description\\[0.1cm]\hline
&\\[-0.45cm]}%
{\\[-0.1cm]\end{tabular*}\end{Sbox}\vspace{1ex} \fbox{\TheSbox}}

\newcommand{\bbar}  {\ensuremath{\overline b}\xspace}
\newcommand{\bbbar} {\ensuremath{b\overline b}\xspace}

\newcommand{\ttbar} {\ensuremath{t\overline t}\xspace}
\newcommand{\ppbar} {\ensuremath{p\overline p}\xspace}
\newcommand{\fbar} {\ensuremath{\overline f}\xspace}
\newcommand{\MSbar}{\ensuremath{\overline{\rm MS}}\xspace}
\mathchardef\Upsilon="7107
\def\Y#1S{\ensuremath{\Upsilon{(#1S)}}\xspace}

\newcommand{\mt}{\ensuremath{m_{t}}\xspace}
\newcommand{\as}{\ensuremath{\alpha_{\scriptscriptstyle S}}\xspace}

\newcommand{\asZ}{\ensuremath{\as(M_Z^2)}\xspace}

\newcommand{\NNNLO}{\ensuremath{\rm 3NLO}\xspace}

\renewcommand\l{\ell}

\newcommand\TC{\ensuremath{{\rm TC}}\xspace}
\newcommand\ETC{\ensuremath{{\rm ETC}}\xspace}
\newcommand\TF{\ensuremath{{\rm TF}}\xspace}
\newcommand\KK{\ensuremath{{K\!K}}\xspace}

\newcommand{\tzw}{\ensuremath{\theta_{\scriptscriptstyle W}}\xspace}

\newcommand{\gv}[1]{{\ensuremath{g_{{\scriptscriptstyle V},#1}}}\xspace}
\newcommand{\ga}[1]{{\ensuremath{g_{{\scriptscriptstyle A},#1}}}\xspace}

\newcommand{\gvz}[1]{{\ensuremath{g_{{\scriptscriptstyle V},#1}^{(0)}}}\xspace}
\newcommand{\gaz}[1]{{\ensuremath{g_{{\scriptscriptstyle A},#1}^{(0)}}}\xspace}

\newcommand{\GF}{{\ensuremath{G_{\scriptscriptstyle F}}}\xspace}

\newcommand{\NC}{{\ensuremath{N_{\scriptscriptstyle C}}}\xspace}

\newcommand{\Kbar    }{\kern 0.2em\overline{\kern -0.2em K}{}\xspace}

\newcommand{\Kz      }{\ensuremath{K^0}\xspace}
\newcommand{\Kzb     }{\ensuremath{\Kbar^0}\xspace}
\newcommand{\KzKzb   }{\ensuremath{\Kz \kern -0.16em \Kzb}\xspace}
\newcommand{\Kp      }{\ensuremath{K^+}\xspace}
\newcommand{\Km      }{\ensuremath{K^-}\xspace}

\newcommand{\KpKm    }{\ensuremath{\Kp \kern -0.16em \Km}\xspace}

\newcommand\Dbar    {\kern 0.18em\overline{\kern -0.18em D}{}\xspace}

\newcommand\Bbar    {\kern 0.18em\overline{\kern -0.18em B}{}\xspace}

\newcommand\Bz      {\ensuremath{B^0}\xspace}

\newcommand\Bzb     {\ensuremath{\Bbar^0}\xspace}
\newcommand\Bu      {\ensuremath{B^+}\xspace}
\newcommand\Bub     {\ensuremath{B^-}\xspace}

\newcommand\BpBm    {\ensuremath{\Bu {\kern -0.16em \Bub}}\xspace}
\newcommand\Bs      {\ensuremath{B^0_{s}}\xspace}
\newcommand\Bsb     {\ensuremath{\Bbar^0_{s}}\xspace}

\newcommand\BzBzb   {\ensuremath{\Bz {\kern -0.16em \Bzb}}\xspace}
\newcommand\BszBszb {\ensuremath{\Bs {\kern -0.16em \Bsb}}\xspace}

\newcommand\invfb{\ensuremath{\:{\rm fb}^{-1}}\xspace}
\newcommand\invpb{\ensuremath{\:{\rm pb}^{-1}}\xspace}
\newcommand\deltatheo{\ensuremath{\delta_{\rm th}}\xspace}

\newcommand\Rfit{{\em R}fit\xspace}
\newcommand{\CP}{\ensuremath{C\!P}\xspace}
\newcommand{\ft}{\footnotesize}
\newcommand{\multic}{\multicolumn}

\newcommand{\ee}{\ensuremath{e^+e^-}\xspace}

\newcommand{\pp}{\ensuremath{\pi^+\pi^-}\xspace}

\newcommand{\tev}{\ensuremath{\mathrm{Te\kern -0.1em V}}\xspace}
\newcommand{\gev}{\ensuremath{\mathrm{Ge\kern -0.1em V}}\xspace}
\newcommand{\mev}{\ensuremath{\mathrm{Me\kern -0.1em V}}\xspace}
\newcommand{\kev}{\ensuremath{\mathrm{ke\kern -0.1em V}}\xspace}
\newcommand{\ev}{\ensuremath{\mathrm{e\kern -0.1em V}}\xspace}
\newcommand{\gevc}{\ensuremath{{\mathrm{Ge\kern -0.1em V\!/}c}}\xspace}
\newcommand{\mevc}{\ensuremath{{\mathrm{Me\kern -0.1em V\!/}c}}\xspace}
\newcommand{\gevcc}{\ensuremath{{\mathrm{Ge\kern -0.1em V\!/}c^2}}\xspace}
\newcommand{\mevcc}{\ensuremath{{\mathrm{Me\kern -0.1em V\!/}c^2}}\xspace}

\newcommand{\bei}{\begin{itemize}}
\newcommand{\eei}{\end{itemize}}
\newcommand{\beq}{\begin{equation}}
\newcommand{\eeq}{\end{equation}}
\newcommand{\beqn}{\begin{eqnarray}}
\newcommand{\eeqn}{\end{eqnarray}}
\newcommand{\beqns}{\begin{eqnarray*}}
\newcommand{\eeqns}{\end{eqnarray*}}
\newcommand{\bitm}{\begin{itemize}}
\newcommand{\eitm}{\end{itemize}}

\newcommand{\amu}{\ensuremath{a_\mu}\xspace}

\newcommand{\dahadZ}{\ensuremath{\Delta\alpha_{\rm had}(M_Z^2)}\xspace}
\newcommand{\dahadZf}{\ensuremath{\Delta\alpha_{\rm had}^{(5)}(M_Z^2)}\xspace}
\newcommand{\dalphaHadMZ}{\ensuremath{\Delta\alpha_{\rm had}^{(5)}(M_Z^2)}\xspace}

\newcommand\ie{{i.e.}\xspace} 
\newcommand\eg{{e.g.}\xspace}

\newcommand\cf{{cf.}\xspace}

\newcommand\rs{\raisebox{1.5ex}[-1.5ex]}

\newcommand{\STU}{$S,\,T,\,U$\xspace}
\def\@citex[#1]#2{\if@filesw\immediate\write\@auxout{\string\citation{#2}}\fi
  \@tempcnta\z@\@tempcntb\m@ne\def\@citea{}\@cite{\@for\@citeb:=#2\do
    {\@ifundefined
       {b@\@citeb}{\@citeo\@tempcntb\m@ne\@citea
        \def\@citea{,\penalty\@m\ }{\bf ?}\@warning
       {Citation `\@citeb' on page \thepage \space undefined}}%
    {\setbox\z@\hbox{\global\@tempcntc0\csname b@\@citeb\endcsname\relax}%
     \ifnum\@tempcntc=\z@ \@citeo\@tempcntb\m@ne
       \@citea\def\@citea{,\penalty\@m}
       \hbox{\csname b@\@citeb\endcsname}%
     \else
      \advance\@tempcntb\@ne
      \ifnum\@tempcntb=\@tempcntc
      \else\advance\@tempcntb\m@ne\@citeo
      \@tempcnta\@tempcntc\@tempcntb\@tempcntc\fi\fi}}\@citeo}{#1}}

\def\@citeo{\ifnum\@tempcnta>\@tempcntb\else\@citea
  \def\@citea{,\penalty\@m}%
  \ifnum\@tempcnta=\@tempcntb\the\@tempcnta\else
   {\advance\@tempcnta\@ne\ifnum\@tempcnta=\@tempcntb \else
\def\@citea{--}\fi
    \advance\@tempcnta\m@ne\the\@tempcnta\@citea\the\@tempcntb}\fi\fi}

\xspace

\newcommand\eps{\varepsilon\xspace}

\newcommand\mini{{\rm min}}

\newcommand\ChiMin{\ensuremath{\chi^2_{\mini}}\xspace}
\newcommand\DeltaChi{\ensuremath{\Delta\chi^2}\xspace}

\newcommand\minchitwo{\ensuremath{\chi^2_{\rm min}}\xspace}

\newcommand{\seffsf}[1]{\sin\!^2\theta^{#1}_{{\rm eff}}}
\newcommand{\rZ}[1]{\rho^{#1}_{Z}}
\newcommand{\kZ}[1]{\kappa^{#1}_{Z}}
\newcommand{\sinfeff}{\sin\!^2\theta^f_{{\rm eff}}}
\newcommand{\sinleff}{\seffsf{\ell}}
\newcommand{\sinthw}{\sin\!^2\theta_W}
\newcommand{\costhw}{\cos\!^2\theta_W}
\newcommand{\sqrtcotthw}{\cot\!\theta_W}
\newcommand{\mc}{\ensuremath{\overline{m}_c}\xspace}
\newcommand{\mb}{\ensuremath{\overline{m}_b}\xspace}

\newcommand{\MHp}{\ensuremath{M_{H^{\pm}}}\xspace}
\newcommand{\tanb}{\ensuremath{\tan\!\beta}\xspace}

%% file: Title.tex
{\small
\color{mygray}
\begin{flushright}
{\sf\em arXiv:1107.0975} \\
{\sf\em CERN-OPEN-2011-033} \\
{\sf\em DESY-11-107} \\
\def\UrlFont{\sf\em}
\url{http://cern.ch/gfitter} 
\end{flushright}
}
\def\UrlFont{\rm}

\vspace{1.3cm}

{\sf\LARGE\bfseries
  Updated Status of the Global Electroweak Fit \\[0.1cm] and Constraints on New Physics  
}

\vspace{1.0cm}

{\Large \em 
  The Gfitter Group \\[0.2cm]
}
{\Large
  M.~Baak$^{a}$, M.~Goebel$^{b}$, J.~Haller$^{c}$, A.~Hoecker$^{a}$, \\
  D.~Kennedy$^{b,c}$, K.~M\"onig$^{b}$, M.~Schott$^{a}$, J.~Stelzer$^{d}$
}

\vspace{0.5cm}

{\normalsize
  $^{a}$CERN, Geneva, Switzerland \\
  $^{b}$DESY, Hamburg and Zeuthen, Germany \\ 
  $^{c}$Institut f\"ur Experimentalphysik, Universit\"at Hamburg, Germany\\
  $^{d}$Department of Physics and Astronomy, Michigan State University, East Lansing, USA

\vspace{1.0cm}

\begin{details}
{\sf\bfseries Abstract} ---
We present an update of the Standard Model fit to electroweak precision data. 
We include newest experimental results on the top quark mass, the $W$ mass 
and width, and the Higgs boson mass bounds from LEP, Tevatron and the LHC. We 
also include a new determination of the electromagnetic coupling strength 
at the $Z$ pole. We find for the Higgs boson mass $91^{\:+30}_{\:-23}\:\gev$ 
and $120^{\:+12}_{\:-5}\:\gev$ when not including and including the 
direct Higgs searches, respectively. From the latter fit we indirectly 
determine the $W$ mass to be $(80.360^{+0.014}_{-0.013})\:\gev$.
We exploit the data to determine experimental constraints on the oblique
vacuum polarisation parameters, and confront these with
predictions from the Standard Model (SM) and selected SM extensions. 
By fitting the oblique parameters to the electroweak data we derive 
allowed regions in the BSM parameter spaces.  We revisit and
consistently update these constraints for a fourth fermion generation, 
two Higgs
doublet, inert Higgs and littlest Higgs models, models with
large, universal or warped extra dimensions and technicolour.  
In most of the models studied a heavy Higgs boson
can be made compatible with the electroweak precision data.
\end{details}

\vfill

\thispagestyle{empty}
\newpage

%% file: Introduction.tex
\section{Introduction}
\label{sec:introduction}

By exploiting contributions from radiative corrections, precision measurements, in line
with accurate theoretical predictions, can be used to probe physics at higher energy 
scales than the masses of the particles directly involved in the experimental reactions.
Theory and experimental data are confronted and unknown model parameters are 
constrained by means of multi-parameter fits. For cases where the parameter space is 
overconstrained it is possible to derive p-values for the compatibility between data 
and theoretical model~\cite{NakamuraCowan:2010zzi}, and hence to directly assess the 
validity of the model. Such an 
approach has been used in the Gfitter analysis of the Standard Model (SM) in light of 
the electroweak precision data~\cite{Flacher:2008zq}, which we revisit in this paper with 
updated experimental constraints. Global electroweak SM fits are also routinely performed 
by the LEP Electroweak Working Group~\cite{LEPEWWG} and for the electroweak review of 
the Particle Data Group~\cite{ErlerNakamura:2010zzi}.

Assuming that the dominant virtual contributions to the electroweak observables arise 
through vacuum polarisation loops, and that other corrections, such as vertex diagrams 
involving light quarks, or box and bremsstrahlung diagrams,
are scale suppressed, physics beyond the SM (BSM) can be parametrised through so-called 
quantum {\em oblique corrections}, for which several parametrisations exist in the 
literature~\cite{Peskin:1990zt,Peskin:1991sw,Marciano:1990dp,Kennedy:1990ib,Kennedy:1991sn,Holdom:1990tc,Golden:1990ig,Altarelli:1990zd,Altarelli:1991fk}. A popular choice are the $S,T$ and $U$ 
parameters~\cite{Peskin:1990zt,Peskin:1991sw}, which have been computed for most of
the prevailing BSM models. 
The \STU parameters are defined with respect to a canonical SM reference so that, 
for SM parameters identical to the reference point values, the parameters vanish in 
the SM. In that case, any significant non-zero value in at least one parameter 
would hint at BSM physics. 

In this paper we derive, for a chosen SM reference point, experimental
constraints on the \STU parameters, and compare them with predictions
from the SM and various BSM models. We study a fourth fermion
generation, two Higgs and inert Higgs doublet models, the littlest
Higgs model and models with large, universal and warped extra
dimensions as well as technicolour.  We also use the experimental
constraints to derive allowed regions in the relevant parameter spaces
of these models. Several similar analyses have been performed and
published in the past. We refer to these in the corresponding BSM
sections. The current analysis revisits these works and provides a
consistent set of BSM constraints derived from the most recent
electroweak data and using the statistics tools of the Gfitter
framework~\cite{Flacher:2008zq}. Its modular design allows us to
determine these constraints directly in the fit, thus invoking 
known two-loop and beyond two-loop SM corrections.

The paper is organised as follows. The updated SM fit to the electroweak
precision data is discussed in Section~\ref{sec:smfit}. An introduction 
of the oblique parameter formalism is given in Section~\ref{sec:oblique}, 
where we also present the experimental results, and discuss the predictions
from the SM. Additional formulas are provided in the Appendix.
In Section~\ref{sec:constraints} and subsections we discuss the oblique
corrections for the aforementioned BSM models and the
corresponding constraints in the relevant parameter spaces.

\newpage

%% file: SMfit.tex
\section{The Global Fit of the Electroweak Standard Model}
\label{sec:smfit}

We present  an update of the SM fit to electroweak precision data, the results
of which will be used as a reference throughout this paper. A detailed
description of the experimental data, the theoretical calculations,
and the statistical methods used in the Gfitter analysis is given in
our reference paper~\cite{Flacher:2008zq}. Since its publication, the
fit software has been continuously maintained and kept in line with the
experimental and theoretical progress.  Here, we shall recall only the
most important aspects of the fit, outline recent changes, which
mainly concern updates of the experimental or phenomenological input
data, and present a full result table together with representative plots
and a discussion of selected results. 

\subsection{Fit inputs}
\label{sec:smfit-inputs}

\subsubsection*{Standard Model predictions}

The SM predictions for the electroweak precision observables measured by the LEP, SLC, 
and Tevatron experiments are fully implemented. State-of-the-art 
calculations are used, in particular the full two-loop and leading beyond-two-loop 
corrections for the prediction of the $W$ mass and the effective weak mixing 
angle~\cite{Awramik:2003rn,Awramik:2004ge,Awramik:2006uz}, which exhibit the strongest 
constraints on the Higgs mass. 
A modification to Ref.~\cite{Flacher:2008zq} is the usage of accurate
parametrisations~\cite{Hagiwara:1994pw,Hagiwara:1998yc,Cho:1999km,Cho:2011rk} for
the calculation of the vector and axial-vector couplings, $g_A^f$ and
$g_V^f$, which are computed at one-loop level and partly at two-loop
level for $ {\cal O}(\alpha\alpha_s)$.\footnote
{
   Up to two-loop electroweak corrections are available in
   Refs.~\cite{Akhundov:1985fc,Arbuzov:2005ma,Bardin:1986fi,Barbieri:1992dq,Fleischer:1993ub,Bardin:1999yd,
    Degrassi:1994tf,Degrassi:1995mc,Degrassi:1996mg,Degrassi:1999jd,Bardin:1997xq,Bardin:1999ak}.
   All known QCD corrections are given in Refs.~\cite{Arbuzov:2005ma,Bardin:1999yd,Kniehl:1989yc}.  
}
Small additional correction factors, determined from a comparison 
with the Fortran ZFITTER package~\cite{Arbuzov:2005ma,Bardin:1999yd}, 
are used to accommodate heavy Higgs masses~\cite{goebel_phd}. These
couplings enter the calculations of the partial and total
widths of the $Z$ and the total width of the $W$ boson, which, due to 
the insufficient experimental precision, display a weak constraint on the 
Higgs mass only.
In the radiator functions~\cite{Bardin:1999ak, Bardin:1999yd} 
the fourth-order (3NLO) perturbative calculation of
the massless QCD Adler function~\cite{Baikov:2008jh}
is also included,
allowing the fit to determine the strong coupling constant with very small theoretical 
uncertainty. 

The SM parameters relevant for the prediction of the electroweak observables are the 
coupling constants of the electromagnetic ($\alpha$), weak (\GF) and strong interactions 
($\as$), and the masses of the elementary bosons ($M_Z$, $M_W$, $M_H$) and fermions ($m_f$), 
where neutrino masses are set to zero. Electroweak unification results in a massless 
photon and a relation between the electroweak gauge boson masses and couplings, thus 
reducing the number of unknown SM parameters by two. 
The SM gauge sector is left with four free parameters taken to be $\alpha$,
$M_Z$, \GF and $\as$. Simplification of the fit is achieved by fixing parameters with 
insignificant uncertainties compared to the sensitivity of the fit. The final list of 
floating fit parameters is: $M_Z$, $M_H$, $m_t$, $\mb$, $\mc$,\footnote
{
   In the analysis and throughout this paper we use the 
   \MSbar renormalised masses of the $c$ and $b$ quarks, $\mc(\mc)$ and $\mb(\mb)$, at 
   their proper scales. In the following they are denoted with $\mc$ and $\mb$ 
   respectively, and their values are taken from~\cite{Nakamura:2010zzi}.
}
$\dahadZf$ and $\asZ$, where only the latter parameter is kept fully unconstrained 
allowing an independent measurement.\footnote
{
   Using an external precision measurement of \asZ in the fit has been studied in 
   Ref.~\cite{Flacher:2008zq} and found to have a negligible impact on the $M_H$ result. 
} 

Theoretical uncertainties due to unknown higher order terms affecting the predictions 
of $M_W$ and $\sinleff$~\cite{Awramik:2003rn,Awramik:2006uz} are parametrised 
by $\deltatheo M_W\simeq 4\:\mev$ and $\deltatheo \sinleff \simeq 4.7\cdot 10^{-5}$. 
They are treated according to the \Rfit scheme~\cite{Hocker:2001xe,Charles:2004jd} 
as freely varying but bound parameters in the fit. 

\subsubsection*{Experimental input}

The experimental results used in the fit include the electroweak precision data 
measured at the $Z$ pole~\cite{:2005ema} ($Z$ resonance parameters, partial 
$Z$ cross sections, neutral current couplings\footnote
{
   We do not include the CDF and D0 measurements of the forward-backward charge 
   asymmetry in $\ppbar\to Z/\gamma^\star+X\to\ee+X$ events, used to extract the 
   $\seffsf e$ values $0.2238 \pm 0.0040 \pm 0.0030$ by CDF~\cite{Acosta:2004wq},
   and $0.2326 \pm 0.0018 \pm 0.0006$ by D0~\cite{Abazov:2008xq}, as their 
   impact so far is negligible compared to the precision of the combined $Z$ pole 
   data in the fit, $\seffsf \ell=0.23143 \pm 0.00013$.
   Also due to lack of precision, we do not include results from atomic parity 
   violation measurements, and from parity violation left-right asymmetry measurements 
   using fixed target polarised M\o ller scattering at low $Q^2$ (see~\cite{Flacher:2008zq}
   for references). 

   The NuTeV Collaboration measured ratios of neutral and charged current cross sections in
   neutrino-nucleon scattering at an average $Q^2\simeq 20\gev^2$ using both muon neutrino 
   and muon anti-neutrino beams~\cite{Zeller:2001hh}. The results derived for the effective 
   weak couplings are not included in this analysis because of unclear theoretical 
   uncertainties from QCD effects such as next-to-leading order corrections 
   and nuclear effects of the bound nucleon parton distribution functions~\cite{Eskola:2006ux}
   (for reviews see, \eg, Refs.~\cite{Davidson:2001ji, McFarland:2003jw}).
}),
their experimental correlations~\cite{:2005ema}, and
the latest $W$ mass world average $M_W=80.399\pm0.023\:\gev$~\cite{:2009nu}  
and width $\Gamma_W=2.098\pm0.048\:\gev$~\cite{:2010in}.

Furthermore we use the newest average of the direct Tevatron top mass measurements 
$m_t=173.3\pm0.9\pm0.6\:\gev$~\cite{:1900yx}, which is interpreted in terms of a pole mass. 
It should be noted that the theoretical 
uncertainties arising from nonperturbative colour-reconnection effects in the fragmentation 
process~\cite{Skands:2007zg,Wicke:2008iz}, and from ambiguities in the top-mass 
definition~\cite{Hoang:2008yj,Hoang:2008xm}, affect the (kinematic) top mass measurement.  
Their quantitative estimate is difficult and may reach roughly 0.5\:\gev each, 
where the systematic error due to shower effects could be larger~\cite{Skands:2007zg}. 
Especially the colour-reconnection and shower uncertainties, estimated by means 
of a toy model, need to be verified with experimental data and should be included 
in the top mass result and uncertainty published by the experiments. None of these 
additional theoretical uncertainties on $m_t$ is included in the fit.

The top mass definition entering the SM $\ppbar\to\ttbar+X$ inclusive cross
section prediction is unambiguous once a renormalisation procedure is defined
and assuming no contributions from new physics to the measured cross section.\footnote
{
  In Ref.~\cite{Langenfeld:2009wd} the \MSbar scheme is used to predict the QCD
  scaling function versus scale ratios (including the dependence on the top mass) that,
  convolved with the parton luminosity and multiplied by $(\as/\overline m_t)^2$, determines
  the inclusive $\ttbar$ production cross section. The experimental cross section
  measurement thus allows one to infer $\overline m_t$ and hence the pole mass (being the 
  renormalised quark mass in the on-shell renormalisation scheme) from the ratio of 
  the corresponding renormalisation factors known to three loops~\cite{Melnikov:2000qh}. 
  The numerical analysis must account for the dependence of the experimental cross section
  value on the top mass used to determine the detector acceptance and reconstruction
  efficiencies.
}
The latest extraction of the top mass from the $\ttbar$ cross 
section was performed by the D0 Collaboration.  Using a theoretical 
$\sigma_{\ttbar}(m_t)$ prediction based on approximate NNLO QCD that includes all 
next-to-next-to-leading logarithms relevant in NNLO QCD~\cite{Langenfeld:2009wd}, 
the top pole mass, derived from the measured cross section 
$\sigma_{\ttbar}(m_t=172.5\:\gev) = 8.13^{\:+1.02}_{\:-0.90}\:{\rm pb}$~\cite{Abazov:2011mi},
was found to be $m_t=167.5^{\:+5.0}_{\:-4.5}\:\gev$~\cite{Abazov:2011pt}.\footnote
{
  The quoted error on the extracted top mass does not include the ambiguity in
  the Monte Carlo top mass interpretation. 
}
\label{page:mtfromtopcrosssection}
A similar value for $m_t$ is obtained when using the cross section prediction
of Ref.~\cite{Kidonakis:2010dk}. While the nominal electroweak fits in this work 
use the direct Tevatron top mass average, we will employ the cross section 
based value for comparison.

For the vacuum polarisation contribution from the five lightest quark flavours to the 
electromagnetic coupling strength at $M_Z$ we use the evaluation 
$\dahadZf=(2757\pm10)\cdot10^{-5}$~\cite{Davier:2010nc}.\footnote
{
   A mistake has been found in the published result of $\dahadZf$~\cite{Davier:2010nc}.
   The corrected result used here is reported in Version 2 of the arXiv submission [1010.4180].
} 
It includes new \pp cross section data from BABAR and KLOE, new multi-hadron 
data from BABAR, a reestimation
of missing low-energy contributions using results on cross sections
and process dynamics from BABAR (cf. references in~\cite{Davier:2010nc}), 
and a reevaluation of the continuum
contribution from perturbative QCD at four loops . Mostly the
reevaluation of the missing low-energy contributions has led to a
smaller $\dahadZf$ estimate compared to that of Ref.~\cite{Hagiwara:2006jt} 
used in our previous fits. Reference~\cite{Davier:2010nc} quotes a
functional dependence of the central value of \dahadZ on 
\asZ of $0.37\cdot10^{-4}\times(\asZ-0.1193)/0.0028$ around the given
central value of \dahadZ. This dependence is included via the rescaling
mechanism~\cite{Flacher:2008zq} in the Gfitter software.  

\begin{figure}[t]
  \vspace{0.9cm}
  \centerline{\epsfig{file=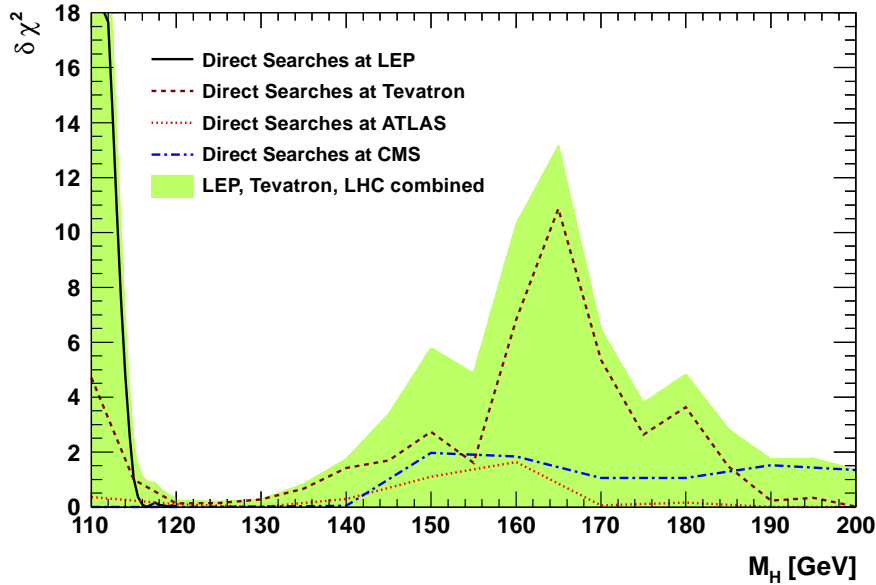,scale=\defaultFigureScale}}
   \vspace{0.1cm}
   \caption{Contribution to the $\chi^2$ test statistic versus $M_H$ derived from 
            the experimental information on direct Higgs boson searches made available 
            by the LEP Higgs Boson and the Tevatron New Phenomena and Higgs Boson 
            Working Groups~\cite{Barate:2003sz,:2010ar,CDF:1336706} and the ATLAS~\cite{Collaboration:2011qi}  
            and CMS Collaborations~\cite{Chatrchyan:2011tz}. 
            The solid (black) and dashed (dark red) lines show the contribution from LEP and Tevatron, 
            while the dotted (light red) and dashed-dotted (blue) lines indicate the constraints obtained from the 
            2010 data by ATLAS and CMS, respectively.
            Following the original figures they have been interpolated by straight lines 
            for the purpose of presentation and in the fit. 
            The light green area gives the combination of these measurements.
            Correlations due to common systematic errors have been neglected in 
            this combination. See text for a description of the method applied. }
   \label{fig:directHiggsSearches}
\end{figure}

This setup defines the {\em standard} electroweak fit. 

The {\em complete} electroweak fit also includes 
the information from the direct Higgs searches at LEP~\cite{Barate:2003sz}, 
Tevatron~\cite{:2010ar,CDF:1336706} and -- for the first time --  the LHC. We
include results from the 2010 LHC run published by ATLAS (combining 
six different final states)~\cite{Collaboration:2011qi} and CMS
($H\to WW\to \ell\nu\ell\nu$)~\cite{Chatrchyan:2011tz}, where correlations due to 
common systematic errors between these results are neglected. Because in the electroweak 
fit we are interested in the compatibility of the SM (assumed to be true) with the data,
we transform the one-sided confidence level, ${\rm CL}_{\rm s+b}$, reported 
by the experiments,\footnote
{
   In lack of published ${\rm CL}_{\rm s+b}$ values by ATLAS~\cite{Collaboration:2011qi} 
   we approximate a chi-squared behaviour of the $\tilde q_\mu$ test statistics 
   used~\cite{Cowan:2010js} and compute ${\rm CL}_{\rm s+b} \simeq {\rm Prob}(\tilde q_1,1)$, 
   where the published $p_0$ values have been converted into $\tilde q_0$, and 
   $\tilde q_1 = \tilde q_0 -2{\rm LLR}$ with the definition
   ${\rm LLR}=-2\ln({\cal L}(1,\hat{\hat{\theta}})/{\cal L}(0,\hat{\theta}))$. 
   A nearly identical result is found with 
   ${\rm CL}_{\rm s+b} \simeq {\rm Prob}(-2{\rm LLR}+{\rm offset},1)$, 
   where the offset of $1.1$ has been added to ensure positive values over 
   the Higgs mass range, and using the published LLR numbers only. 
   These are the numbers used in the fit.
}
into a two-sided confidence level, ${\rm CL}^{\rm 2-sided}_{\rm s+b}$. This transformation 
reduces the statistical constraint from the direct Higgs searches compared to that of 
the one-sided ${\rm CL}_{\rm s+b}$, because positive fluctuations (or signals), 
beyond the signal plus background expectation, are penalised by the test statistics 
as are negative fluctuations (or absence of signals).
The contribution to the $\chi^2$ test statistic minimised in the fit is obtained from
$\delta\chi^2(M_H)=2\cdot [{\rm Erf}^{-1}(1-{\rm CL}^{\rm 2-sided}_{\rm s+b}(M_H))]^2$, where we 
add up the terms from the LEP, Tevatron and LHC experiments ignoring the correlations
among these. As the LHC results are statistics dominated this assumption 
should not be too inaccurate. Nevertheless, an official combination of all the results 
by the experiments should be encouraged. A more detailed discussion of our combination 
method is given in Ref.~\cite{Flacher:2008zq}. 
The resulting $\delta\chi^2$ versus $M_H$ is shown in Fig.~\ref{fig:directHiggsSearches} 
for LEP, Tevatron, ATLAS and CMS individually (lines) as well as their combination 
(shaded/green area). Note that the minimum $\delta\chi^2$ at $M_H\sim 125\:\gev$ is 
not to be interpreted (in a Bayesian sense) as a ``most probable Higgs mass'', but as
an area where the experimental sensitivity is not sufficient to either exclude nor confirm 
a Higgs boson.  

The second column in Table~\ref{tab:SMresults} gives an overview of all the input quantities
used in the fit.

\subsection{Fit results}
\label{sec:smfit-results}

The standard and complete fits converge with global minimum values of the test statistics 
of respectively $\ChiMin=16.6$ and $\ChiMin=17.8$ for 13 and 14 degrees of freedom,
giving the naive p-values ${\rm Prob}(\ChiMin,13)=0.21$ and ${\rm Prob}(\ChiMin,14)=0.23$,
which have been confirmed by pseudo experiments generated with Monte Carlo techniques.\footnote
{
   The ${\rm CL}_{\rm s+b}$ obtained from the direct Higgs searches has been left 
   unaltered during the Monte Carlo based p-value evaluation of 
   the complete fit. This is justified by the strong statistical significance of 
   the LEP constraint, which drives the contribution of the direct Higgs searches to 
   the $\ChiMin$.
} 
The minor improvement in the
p-value of the complete fit with respect to our earlier result~\cite{Flacher:2008zq} 
arises from the increased best-fit value of the Higgs mass in the standard fit (see below), 
owing to the reduced electromagnetic coupling strength at $M_Z^2$~\cite{Davier:2010nc}. The 
new result reduces the tension with the direct Higgs boson searches.

\input{SMtable_para_new}

\begin{figure}[p]
   \centerline{\epsfig{file=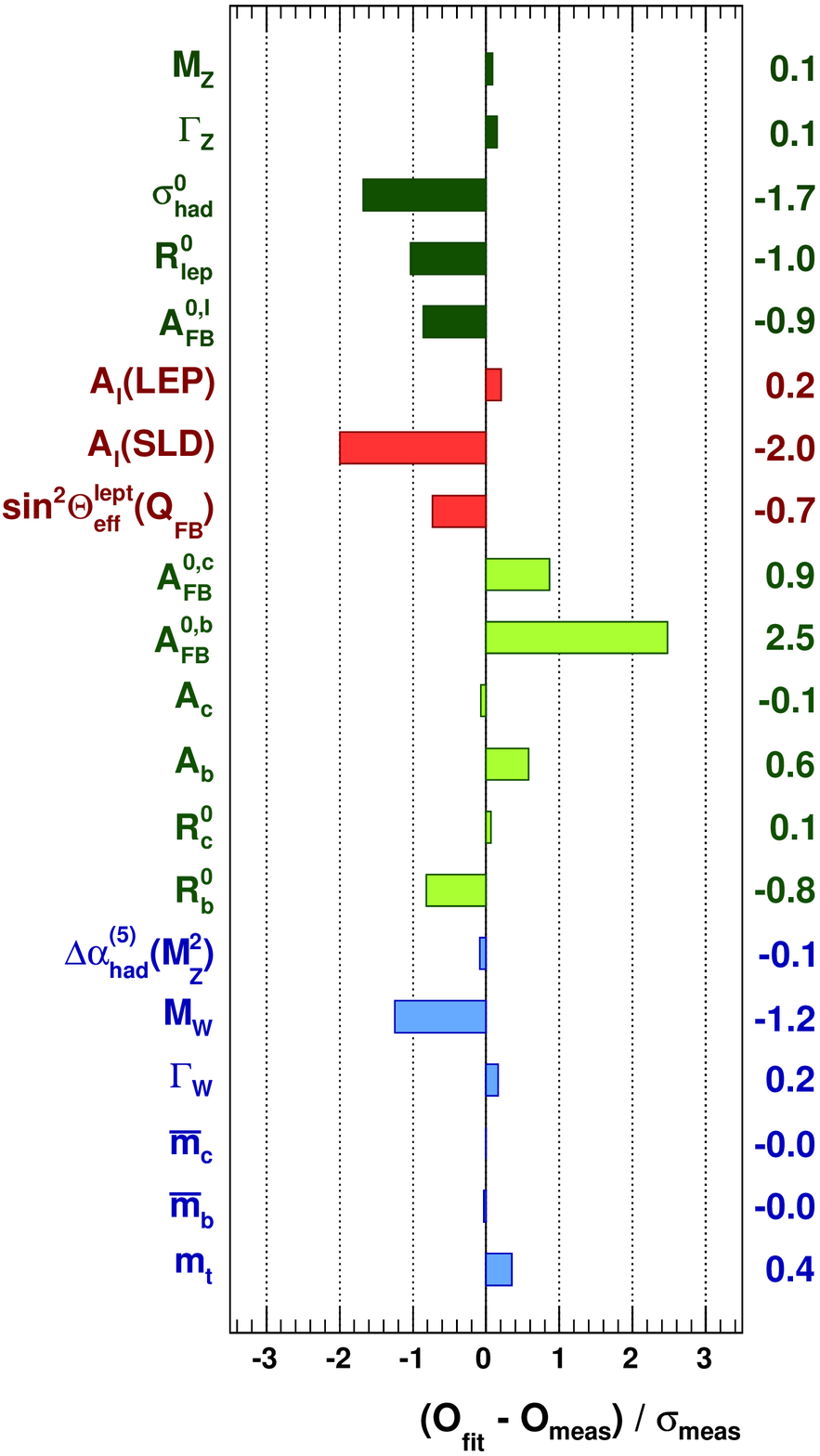, scale=0.53}
               \epsfig{file=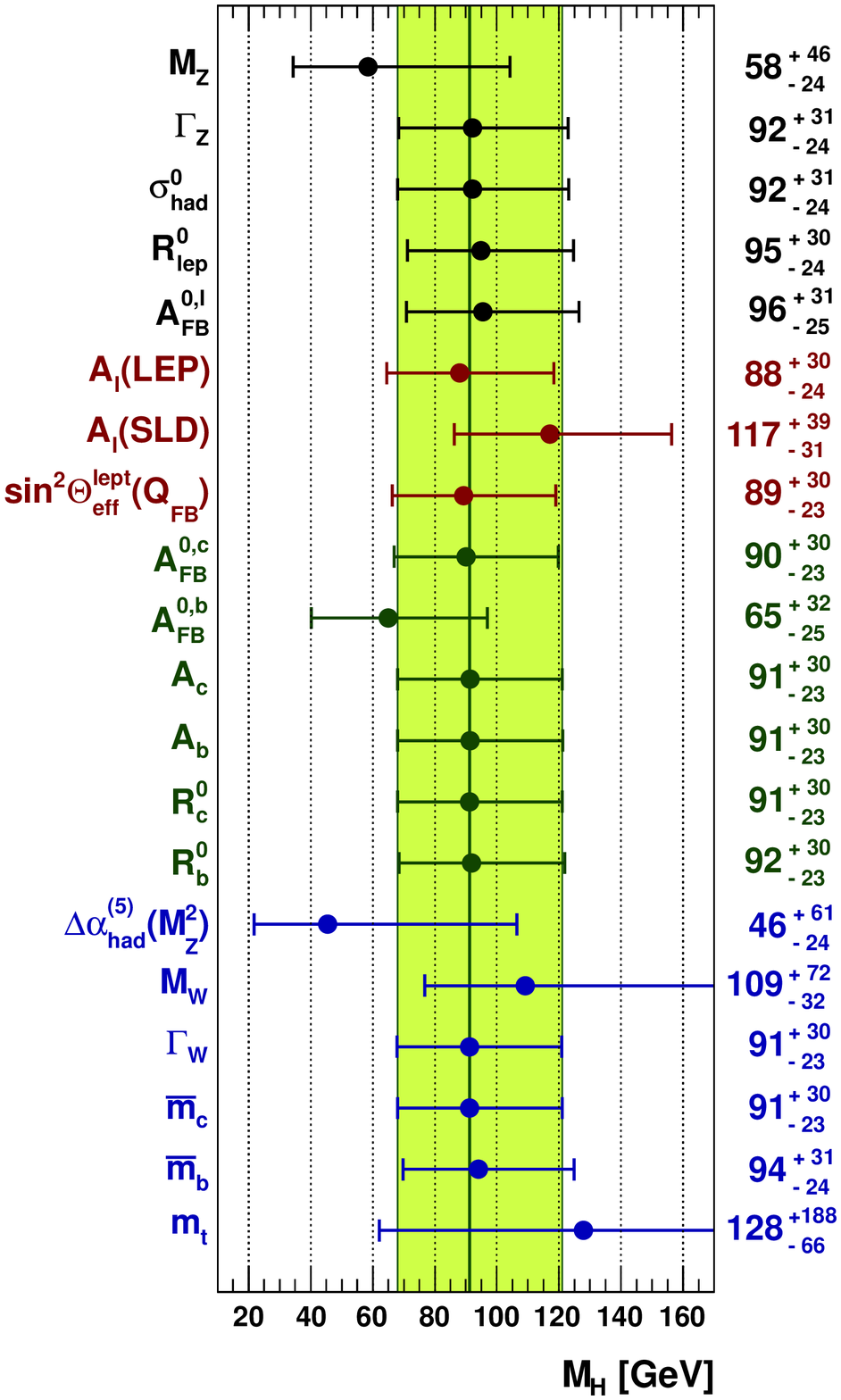, scale=0.53}}
   \vspace{-0.35cm}
   \caption{Comparing fit results with direct measurements: pull values for the 
            complete fit (left), and results for $M_H$ from the standard fit 
            excluding the respective measurements from the fit (right). }
   \label{fig:pullplot}
  \vspace{0.8cm}
  \centering
   \epsfig{file=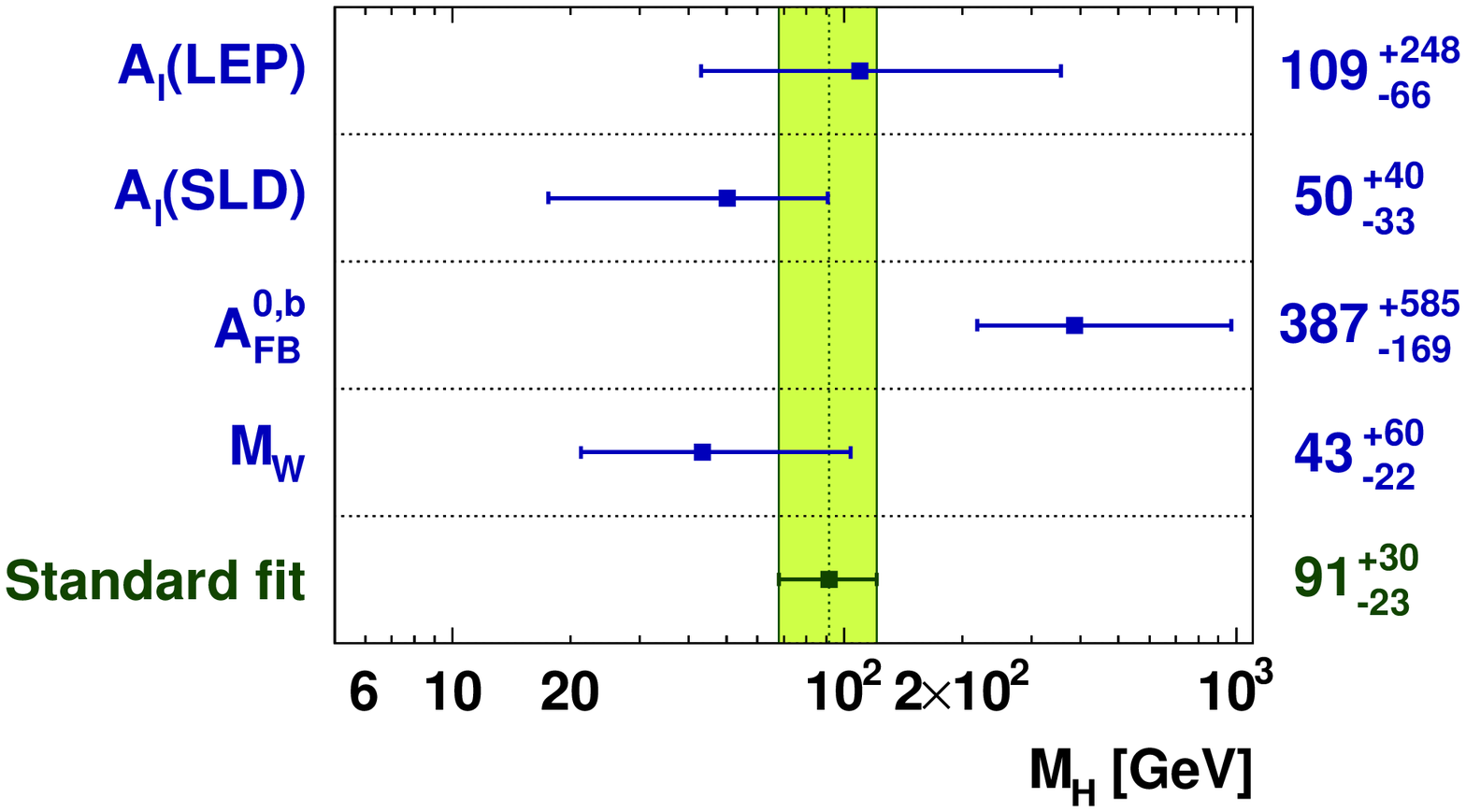, scale=0.45}
   \vspace{0.05cm}
   \caption{Determination of $M_H$ excluding all the sensitive observables from the 
            standard fit except the one given. Note that the results shown are not 
            independent. The information in this figure is complementary to that
            of the right hand plot of Fig.~\ref{fig:pullplot}. 
            } 
   \label{fig:mainObservables}
\end{figure}
\begin{figure}[p]
   \def\MyHiggsFigScale{}
   \centerline{\epsfig{file=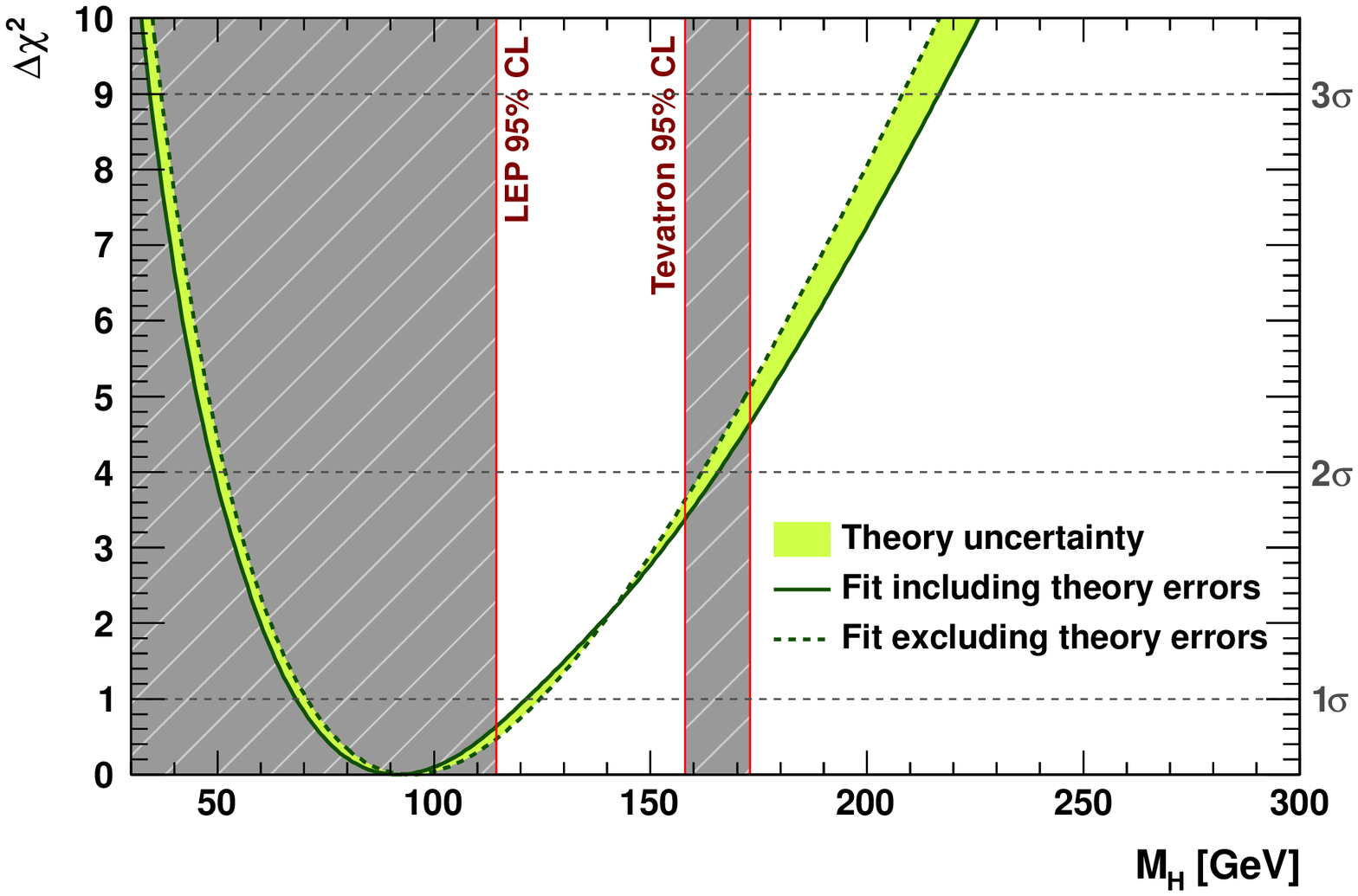, scale=\defaultFigureScale}}
   \vspace{0.4cm}
   \centerline{\epsfig{file=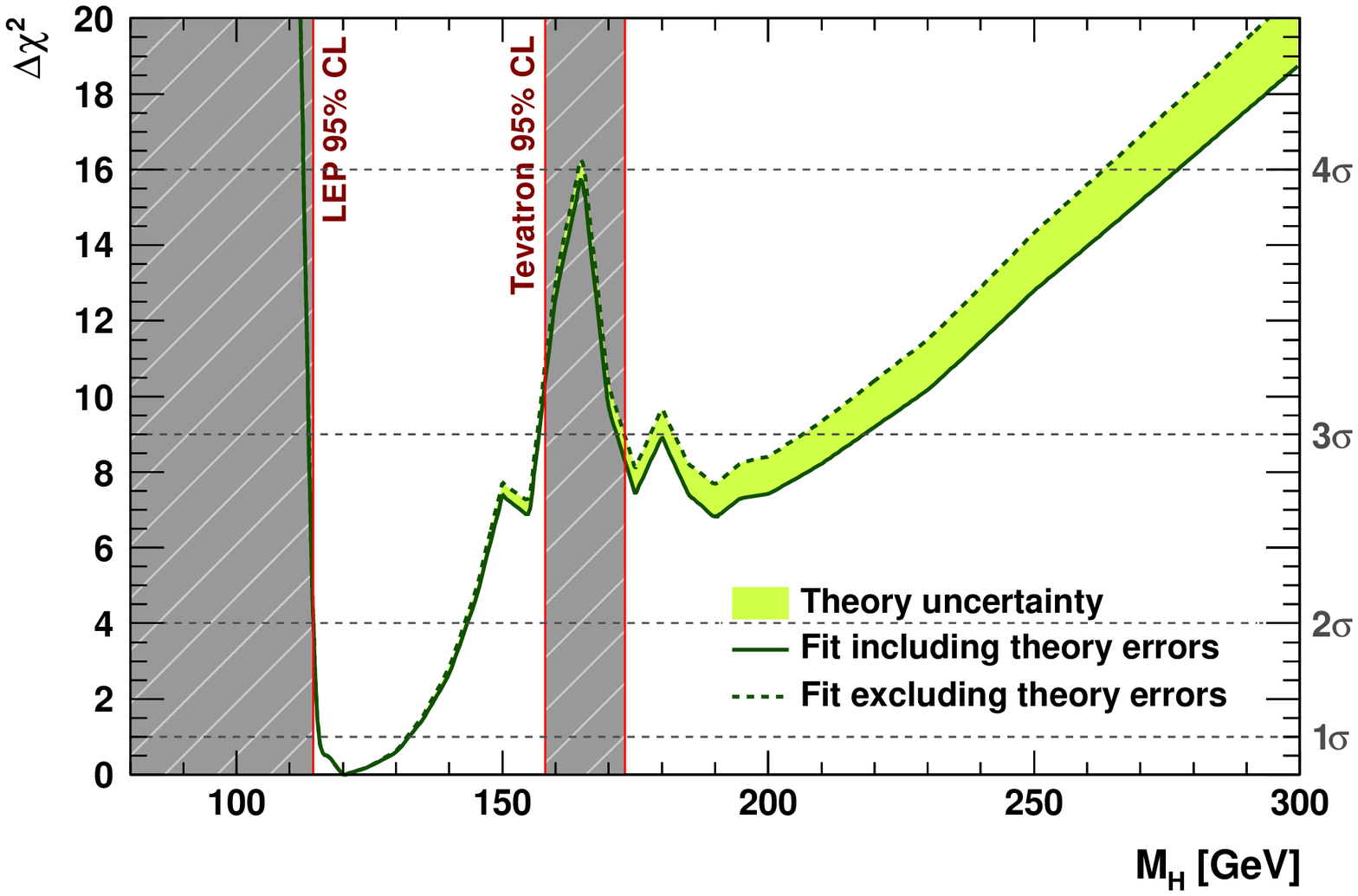, scale=\defaultFigureScale}}
  \vspace{ 0.3cm}
   \caption{Indirect determination of the Higgs boson mass: 
            \DeltaChi as a function of $M_H$ for the standard fit (top) and the 
            complete fit (bottom). The solid (dashed) lines give the results when 
            including (ignoring) theoretical errors. Note that we have modified the 
            presentation of the theoretical uncertainties here with respect to our earlier
            results~\cite{Flacher:2008zq}. Before, the minimum \ChiMin of the 
            fit including theoretical errors was used for both curves 
            to obtain the offset-corrected \DeltaChi. We now individually subtract
            each case so that both \DeltaChi curves touch zero. In spite of the 
            different appearance, the theoretical errors used in the fit are unchanged
            and the numerical results, which always include theoretical uncertainties,
            are unaffected.
         }
   \label{fig:greenband}
\end{figure}

The results for the parameters and observables of the two fits are given in columns 
three and four of Table~\ref{tab:SMresults},\footnote
{
   It is noticable that the values of the four theoretical uncertainty parameters 
   converge at the limits of their allowed intervals. This is explained by their
   uniform contribution to the $\chi^2$ function
   but the necessarily non-uniform values of the global $\chi^2$ function that depends
   on these theory parameters. The fit thus converges at the extrema of the allowed
   ranges. 
}
together with their one standard deviation 
($\sigma$) intervals derived from the $\Delta\chi^2$ test statistics.\footnote
{
   We have verified the chi-squared property of the test statistics by sampling pseudo 
   MC experiments.
}
The correlation coefficients are given in Table~\ref{tab:SMcormat} (for the standard fit).

The left-hand plot of Fig.~\ref{fig:pullplot} gives for the complete fit the pull 
values obtained from the difference between the fit result and the measurement in units of 
the total experimental error (not including the error from the fit performed here). 
They reflect the known tension between the left-right asymmetry and $A_{\rm FB}^{0,b}$, 
though it is noticeable that no single pull value exceeds $3\sigma$. 

\subsubsection*{Higgs mass constraints}

The top and bottom plots of Fig.~\ref{fig:greenband} show the profile curves of the 
\DeltaChi test statistic for the standard and complete fits versus the $M_H$ parameter. 
In this, as in the following \DeltaChi graphs, the test statistic is minimized 
with respect to all other freely varying fit parameters for each (fixed) value of 
the parameter being plotted. We find from this scan
\beqn
\label{eq:mhfit}
M_H = \bigg\{
\begin{array}{rl}
     91^{\:+30}_{\:-23}\: \gev   &~~~ {\rm (standard~fit)}\,, \\
     120^{\:+12}_{\:-5}\: \gev   &~~~ {\rm (complete~fit)}\,,
\end{array}
\eeqn 
with the 95\% (99\%) upper bounds of 163\:\gev (194\:\gev) for
the standard fit, and 143\:\gev (149\:\gev) for the complete fit,
respectively. The errors and limits include the various theory uncertainties 
that taken together amount to approximately $8\:\gev$ on $M_H$.\footnote
{
   Repeating the standard fit with all theory uncertainties fixed 
   to zero gives $\ChiMin=17.2$ and $M_H=94^{\:+30}_{\:-24}\: \gev$.
   A direct comparison of this result with Eq.~(\ref{eq:mhfit}) is 
   not straightforward as the fit uses the additional nuisance parameters, 
   when let free to vary, to improve the test statistics (recall the value 
   of $\ChiMin=16.7$
   for the standard fit result). The impact on the parameter errors 
   would become noticeable once the input observables exhibit better 
   compatibility (cf. discussion in Ref.~\cite{Flacher:2008zq})
}

The standard fit value for $M_H$ has moved by +8\:\gev
as a consequence of the new \dahadZf
evaluation~\cite{Davier:2010nc}. Using instead the preliminary result
$\dahadZf=(2762.6\pm10.3)\cdot10^{-5}$~\cite{Hagiwara:2011af}, 
obtained with the use of similar experimental data but less
reliance on perturbative QCD, we find $M_H=88^{\:+29}_{\:-23}\:\gev$.

The results~(\ref{eq:mhfit}) are obtained using the experimental world average of 
the direct Tevatron top mass measurements~\cite{:1900yx}
whose interpretation as pole mass in theory calculations is affected 
with additional uncertainties (\cf discussion in Sec.~\ref{sec:smfit-inputs}). 
Using, for comparison, the pole mass value $m_t=167.5^{\:+5.0}_{\:-4.5}\:\gev$~\cite{Abazov:2011pt}, 
as determined from the $\ppbar\to\ttbar+X$ inclusive cross section, the standard
electroweak fit returns for the Higgs boson mass $72^{\:+38}_{\:-25}\:\gev$, 
which is appreciably lower than the result~(\ref{eq:mhfit}).

The contributions from the various measurements to the central value and uncertainty of $M_H$ in 
the standard fit are given in the right hand plot of Fig.~\ref{fig:pullplot}, where 
all input measurements except for the ones listed in a given line are used in the fit.
It can be seen that, \eg, precise measurements of $\mt$ and $M_W$ and a precise determination of 
\dahadZf are essential for an accurate constraint of $M_H$.

Figure~\ref{fig:mainObservables} displays complementary information. Among the four observables 
providing the strongest constraint on $M_H$, namely $A_\ell$(LEP), $A_\ell$(SLD), $A_{\rm FB}^{0,b}$
and $M_{W}$, only the one indicated in a given row of the plot is included in the fit.\footnote
{
   The uncertainties in the free fit parameters that are correlated to $M_H$ (mainly \dalphaHadMZ
   and \mt) contribute to the errors shown in Fig.~\ref{fig:mainObservables}, and generate
   correlations between the four $M_H$ values found. 
}
The compatibility among these measurements is estimated by performing a reduced standard fit 
in which the least compatible of the measurements (here $A_{\rm FB}^{0,b}$) is removed, and by 
comparing the $\ChiMin$ test statistic obtained in that fit to the one of the standard fit. 
The p-value of the $\ChiMin$ difference of 8.0 between these fits is evaluated by means of pseudo 
experiments, with observables fluctuating according to their experimental errors around 
a consistent set of SM predictions corresponding to the best-fit results for the SM parameters. 
In each of the pseudo experiments we follow the same procedure as for data, that is, we 
remove the least compatible of the measurements, refit, and compute the $\ChiMin$ difference
with respect to the standard fit corresponding to that pseudo experiment. We find that in 
$(1.4\pm0.1)\%$ (``$2.5\sigma$'') of the pseudo experiments the $\ChiMin$ difference 
exceeds that observed in data.

\begin{figure}[t]
  \centering
   \epsfig{file=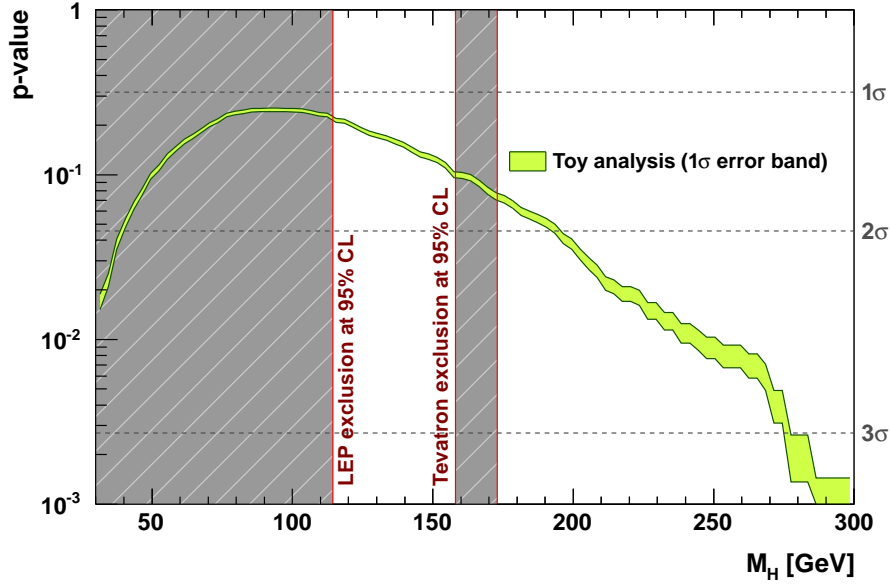, scale=\defaultFigureScale}
   \vspace{0.1cm}
   \caption{P-value versus $M_H$ of the standard electroweak fit as obtained from 
            pseudo-MC simulation. The error band represents the statistical error 
            from the MC sampling size.}
   \label{fig:pval}
\end{figure}
Finally, Fig.~\ref{fig:pval} shows the p-value obtained from Monte Carlo samples 
of the standard fit as a function of the true Higgs mass.\footnote
{
   Note that by fixing $M_H$ the number of degrees of freedom of the fit is increased compared to
   the standard fit resulting in a larger average \minchitwo and thus in a larger
   p-value.
} 
At the best-fit value of $91\:\gev$ the plot reproduces the goodness of the 
standard fit. With increasing $M_H$ the p-value drops reaching the 
$2\sigma$ level at $M_H=190\:\gev$ and the $3\sigma$ level at $M_H=275\:\gev$.

\subsubsection*{Constraints on other parameters}

The two rightmost columns of Table~\ref{tab:SMresults} give the 
results of, respectively, the complete fit and when assuming $M_H=120\:$GeV
(fixed) in the fit for each parameter or observable, obtained
by scanning the profile likelihood without using the corresponding
experimental or phenomenological constraint in the fit (indirect
determination -- similar to the $M_H$ determinations in the right-hand
plot of Fig.~\ref{fig:pullplot}).  Apart from the intrinsic interest
of having an indirect determination of the observables, this procedure
provides interesting insight into the requirements of the fit. If the
direct knowledge (first column in Table~\ref{tab:SMresults}) of an
observable is much more precise than the indirect one (last column),
for example $M_Z$, the variable could have been fixed in the fit
without impacting the results, and thus there is no need, for the purpose of
the global electroweak fit, for an
improved direct determination.  On the other hand, if the indirect
constraint of an observable strongly outperforms the direct one, as is
the case for $\Gamma_W$ or the direct measurement of $\sinleff(Q_{\rm
FB})$, the observable is irrelevant for the fit and can be removed.
To improve the indirect constraint on $M_H$, the experimental efforts
must focus on observables with good sensitivity to $M_H$, and with
competing accuracy between direct and indirect constraints, as is the
case for $M_W$ and the $Z$-pole asymmetries.

\begin{figure}[!t]
  \centerline{\epsfig{file=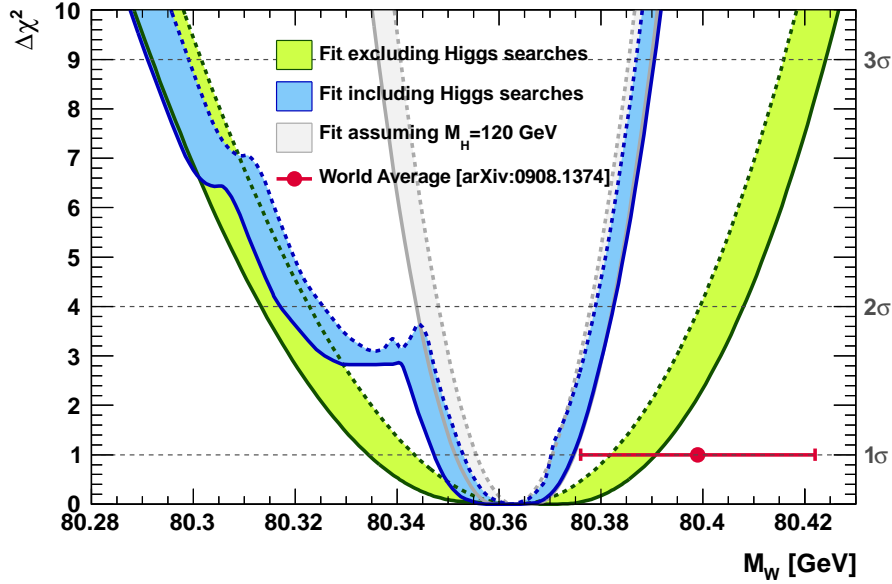, scale=\defaultFigureScale}}
  \vspace{0.1cm}
  \caption[]{Indirect determination of the $W$ boson mass: 
             profile of $\DeltaChi$ versus $M_W$ for the complete fit (blue 
             shaded curve) and the standard fit (green shaded curve). In both fits
             the direct $M_W$ measurement, indicated by the dot with $1\sigma$ error 
             bar, is not included. The widths of the bands indicate the size of 
             the cumulative theoretical uncertainty in the fit. The grey shaded curve
             shows the constraint one would obtain for a hypothetical Higgs
             discovery at 120\:\gev (with negligible error on $M_H$).}
\label{fig:wmassscan}
\end{figure}
From these scans we obtain the following results.
\begin{itemize}
\item We indirectly determine the $W$ mass from the complete fit to be 
\beq 
   M_W = 80.360 ^{\:+0.014}_{\:-0.013}\:\gev\,, 
\eeq 
which is $1.6\sigma$ below and exceeds in precision the experimental 
world average~\cite{:2009nu}. Using the cross section derived \mt value 
(\cf Sec.~\ref{page:mtfromtopcrosssection}, page~\pageref{page:mtfromtopcrosssection}) 
we find $M_W = 80.340 ^{\:+0.029}_{\:-0.021}\:\gev$ from the complete fit.
Figure~\ref{fig:wmassscan} shows the
$\DeltaChi$ profile versus $M_W$ for the standard fit (green band) and 
the complete fit (blue band). Also shown is the world average of the direct 
$M_W$ measurements (dot with error bar).  For both fits the theoretical
uncertainty in the $M_W$ prediction ($\deltatheo M_W\simeq 4\:\mev$)
and its treatment via the \Rfit scheme leads to a broadening of the fit minima.
The inclusion of the direct Higgs searches provides a considerably improved
indirect $M_W$ determination. The grey-shaded
band shows the constraint one would obtain for a hypothetical Higgs
discovery at 120\:\gev with negligible error on $M_H$. The precision
of the indirect $M_W$ determination would reach $11\:\mev$. The uncertainty would
increase to approximately $25\:\mev$ when using the cross section derived
top mass value. 

\begin{figure}[!t]
  \centerline{\epsfig{file=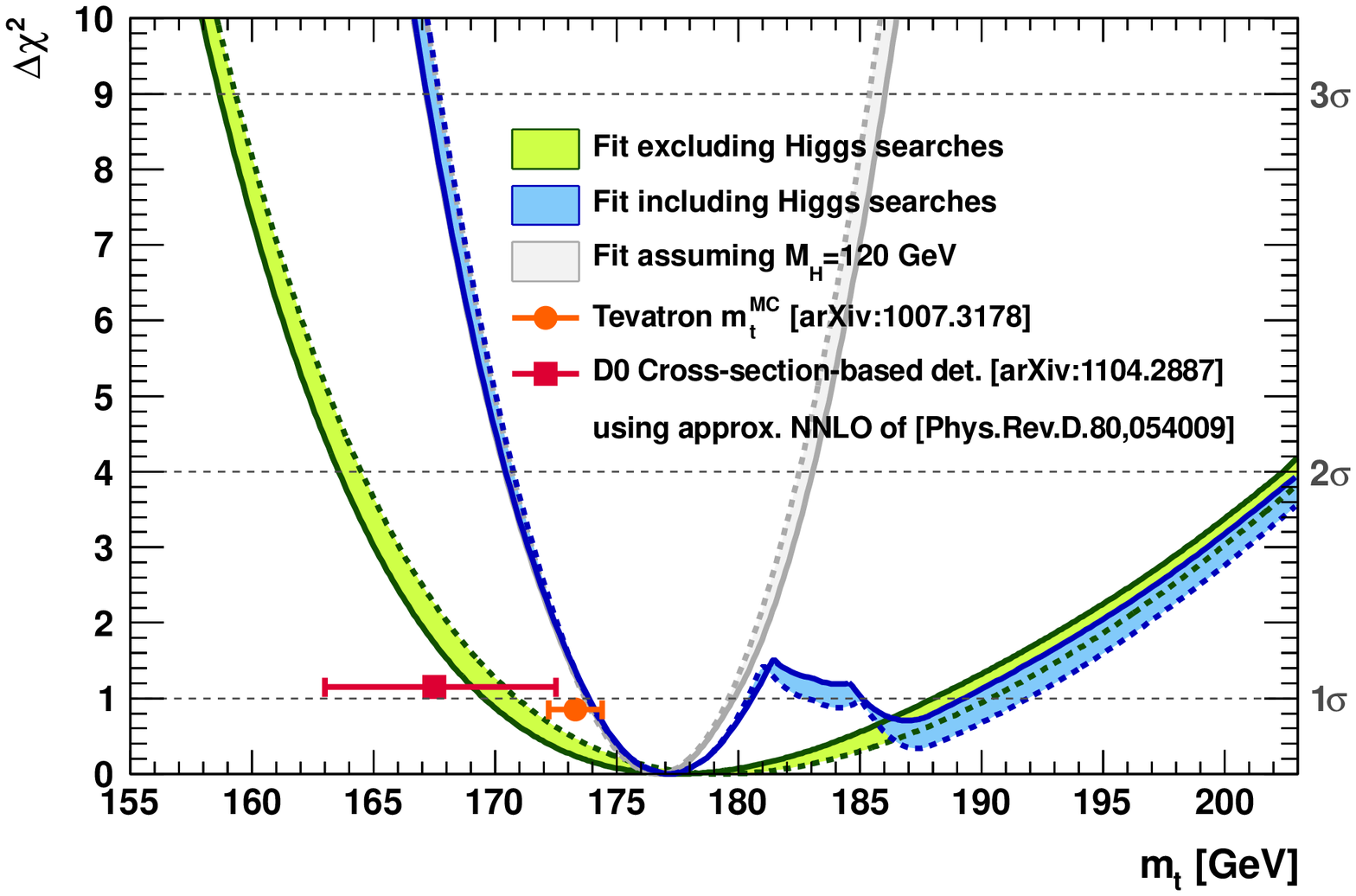, scale=\defaultFigureScale}}
  \vspace{0.1cm}
  \caption[]{Indirect determination of the top quark pole mass: 
             profile of $\DeltaChi$ versus $m_t$ for the complete fit 
             (blue shaded curve) and the standard fit (green shaded curve). In both fits the 
             direct $m_t$ measurement, indicated by the dot with $1\sigma$ error 
             bar, is not included. The widths of the bands indicate the size of 
             the cumulative theoretical uncertainty in the fit. Also shown is the pole
             mass result inferred by D0 from the measurement of the $\ppbar\to\ttbar+X$ 
             cross section~\cite{Abazov:2011pt} (square dot, see text). The grey 
             shaded curve shows the constraint one would obtain for a hypothetical 
             Higgs discovery at 120\:\gev (with negligible error on $M_H$).
}
  \label{fig:topscan}
\end{figure}
\item The indirect determination $\dahadZf=(2729^{\:+57}_{\:-50})\cdot10^{-5}$ comes out 
      slightly smaller but fully compatible with the phenomenological evaluation, while 
      being a factor of almost five less accurate. Knowing the Higgs boson mass would 
      only marginally improve the precision of the indirect determination (\cf last 
      column in Table~\ref{tab:SMresults}).

\item The strong coupling constant at the $Z$ pole to four loop perturbative order 
      for the massless fermion propagators is found to be 
      \beq
          \asZ = 0.1194\pm0.0028\,,
          \label{eq:asZresult}
      \eeq
      with negligible theoretical uncertainty due to the good convergence of the 
      perturbative series at that scale (cf. Ref.~\cite{Flacher:2008zq}).

\item Two local \DeltaChi minima are found from the indirect
      constraint of the top quark pole mass in the complete fit, giving the 
      $\DeltaChi<1$ ranges
\beq
\label{eq:mtop_completefit}
     \mt = [173.8, 180.6]\:\gev~~\rm{and}~~[185.1, 189.3]\:\gev\, .  
\eeq 
The first region agrees within $1.1\sigma$ with the experimental world
average of the direct $m_t$ measurements~\cite{:1900yx}. The separation between the 
two regions originates from the direct Tevatron limit on $M_H$. The lower (upper)
region corresponds to Higgs masses around 130\:\gev (190\:\gev).
Figure~\ref{fig:topscan} shows the $\DeltaChi$ profile versus $m_t$ for 
the standard fit (green band) and the complete fit (blue band). Also shown is
the world average of the direct $m_t$ measurements as well as the pole top mass 
derived from the inclusive $\ttbar$ cross section (dots with error bars).
Similar to the indirect $M_W$ determination, the results from the direct 
Higgs searches allow to significantly increase the precision of the indirect 
constraint. The grey-shaded band shows the constraint on $m_t$ one would obtain 
for a hypothetical Higgs discovery at 120\:\gev with negligible error on $M_H$. 
The precision of the indirect determination would reach $3.0\:\gev$.

\begin{figure}[!t]
  \centerline{\epsfig{file=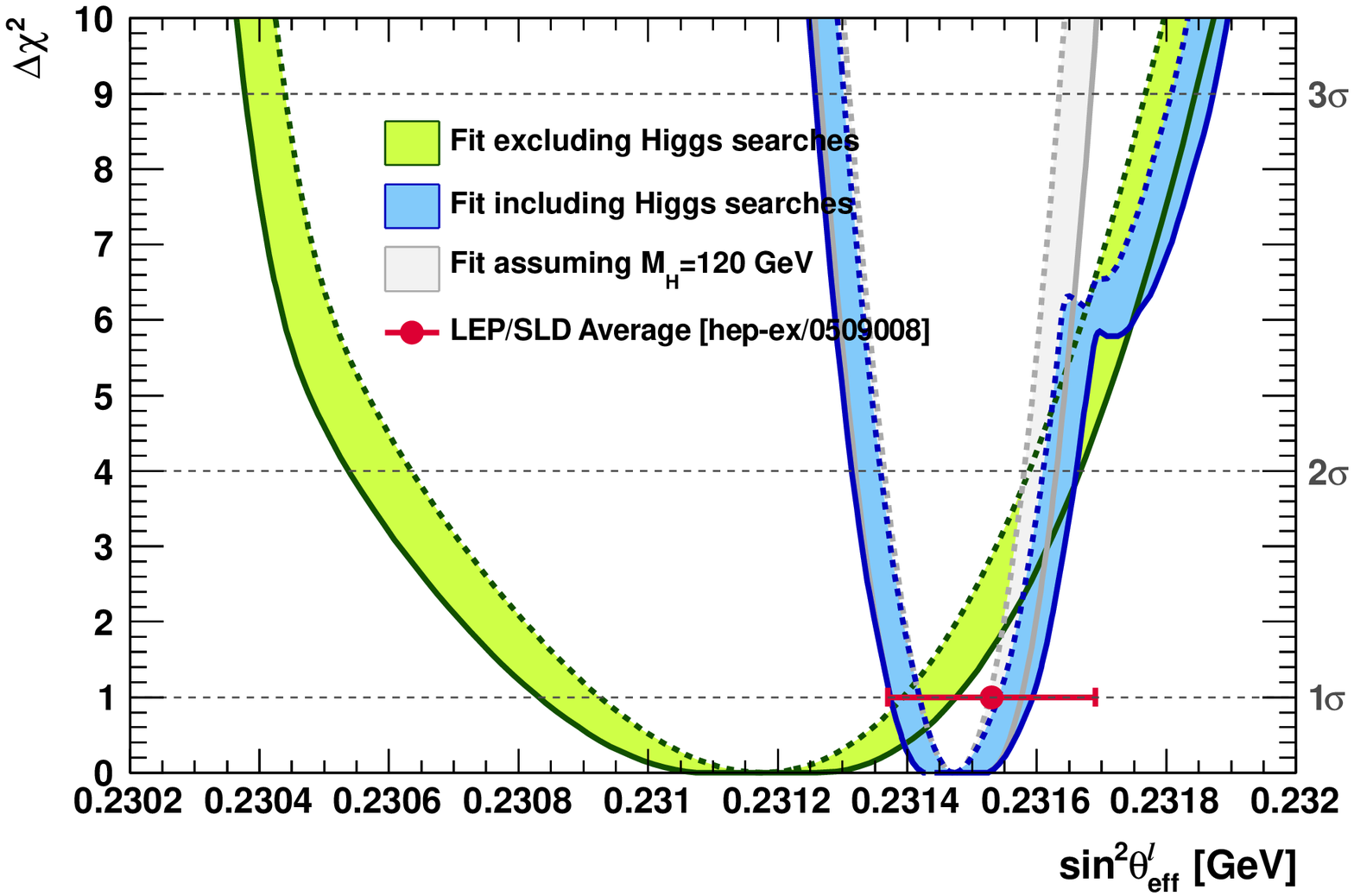, scale=\defaultFigureScale}}
  \vspace{0.1cm}
  \caption[]{Indirect determination of the effective weak mixing angle: 
             profile of $\DeltaChi$ versus $\sinleff$ for the complete fit 
             (blue shaded curve) and the standard fit (green shaded curve). 
             In both fits all measurements with a direct relationship to 
             $\sinleff$ (\eg the asymmetry parameters) are not included, and only 
             the measured or determined values of $m_t$, $M_W$, $\mc$, $\mb$, 
             $\dahadZf$ and $\asZ$ are used. 
             The widths of the bands indicate the size of 
             the cumulative theoretical uncertainty in the fit. 
             The LEP/SLD average for $\sinleff$ as derived directly from the asymmetry measurements is 
             indicated by the dot with $1\sigma$ error 
             bar. The light 
             shaded curve shows the constraint one would obtain for a hypothetical 
             Higgs discovery at 120\:\gev (with negligible error on $M_H$).
}
  \label{fig:sin2scan}
\end{figure}

\item For the indirect determination of the effective weak mixing angle $\sinleff$ we
      ignore all measurements of observables that are related to $\sinleff$ (\eg the
      asymmetry parameters), and instead only use experimental results for $m_t$, $M_W$, $\mc$,
      $\mb$, $\dahadZf$ and $\asZ$ (we use the result of Eq.~(\ref{eq:asZresult}) for the
      latter parameter) and the direct Higgs searches in the fit. From this we obtain
      the SM prediction
      \beq 
        \sinleff = 0.23148 \pm 0.00011\,, 
        \eeq 
        which is compatible with and more precise than the experimental average
        directly derived from the asymmetry measurements at LEP and SLD:
        $\sinleff = 0.23153\pm 0.00016$~\cite{:2005ema}.
        Figure~\ref{fig:sin2scan} shows the $\DeltaChi$ profile versus
        $\sinleff$ for the standard fit (green band) and the complete fit
        (blue band). Also
        shown is the LEP/SLD average from the direct determination.
        Similar to the indirect $M_W$ and $m_t$ determinations, the results from the direct 
        Higgs searches allow to significantly increase the precision of the indirect 
        constraint. The grey shaded band shows the constraint on $\sinleff$ one would obtain 
        for a hypothetical Higgs discovery at 120\:\gev with negligible error on $M_H$. 

\item Two-dimensional 68\%, 95\% and 99\%~CL allowed regions 
     obtained from scans of fits with fixed variable pairs $M_W$
     vs. $M_H$ are shown in Fig.~\ref{fig:MW_vs_MH}. The allowed
     region obtained without the $M_W$ measurement and the direct
     Higgs searches (largest/blue) agrees with the world average of
     the direct $M_W$ measurements (horizontal/green band). Inclusion
     of the $M_W$ measurements (narrow/purple) reduces significantly
     the allowed ranges in $M_H$ highlighting again the importance of
     the $M_W$ measurements for an accurate $M_H$ determination.
     After the inclusion of the direct Higgs searches two separate
     small regions (narrowest/green) in the parameter space remain as a result of the
     prominent maximum around $M_H=160$\,GeV in Fig.~\ref{fig:greenband} (bottom).

\item Two-dimensional 68\%, 95\% and 99\%~CL allowed regions 
     obtained from scans of fits with fixed variable pairs $M_W$
     vs. $m_t$ are shown in Fig.~\ref{fig:MW_vs_mtop}. The indirect
     determination (largest/blue) without the $M_W$ and $m_t$
     measurements and without the direct Higgs searches shows 
     agreement with the direct $M_W$ and $m_t$ measurements
     (horizontal and vertical green bands).  The inclusion of the
     results of the direct Higgs searches reduces significantly the
     allowed region (narrow/yellow), which is still in agreement with
     the direct results. For illustration, isolines for various values
     of $M_H$ representing the SM prediction in an indicative way are
     also shown. As these isolines do not include the theoretical
     uncertainties (e.g. on $M_W$), the allowed region of the fit
     including the direct Higgs searches (narrow/yellow region) as shown 
     in this plot extend slightly into regions of $M_H$ values smaller 
     than the strict exclusion bound from LEP ($M_H<114.4$\,GeV). For
     the same reason the 2-dimensional region allowed from the fit is
     not split into two separate region as naively expected from the
     Higgs boson mass limits around $M_H=160$\:GeV.

\end{itemize}
\begin{figure}[!p]
  \centerline{\epsfig{file=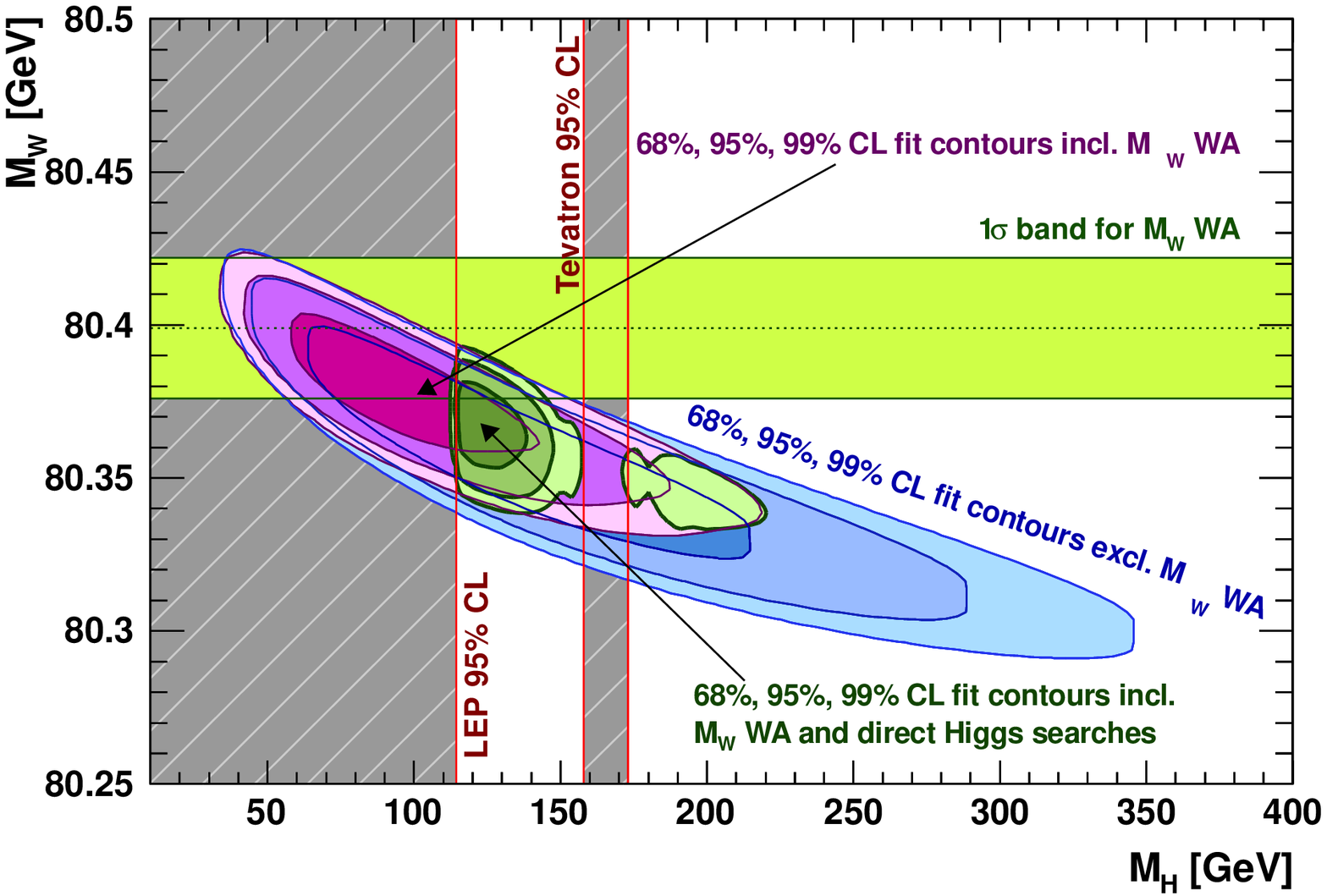, scale=\defaultFigureScale}}
  \vspace{0.1cm}
  \caption[]{Contours of 68\%, 95\% and 99\%~CL obtained from scans of
     fits with fixed variable pairs $M_W$ vs. $M_H$. The largest/blue
     allowed regions are the results of the fit excluding the $M_W$ measurement
     and any direct Higgs search information. The narrow/purple (narrowest/green)
     contours indicate the constraints obtained for a fit including 
     the $M_W$ measurement and the direct Higgs search results. The horizontal 
     band represents the $M_W$ world average experimental value with its $1\sigma$ 
     uncertainty.}
   \label{fig:MW_vs_MH}
  \vspace{0.4cm}
  \centerline{\epsfig{file=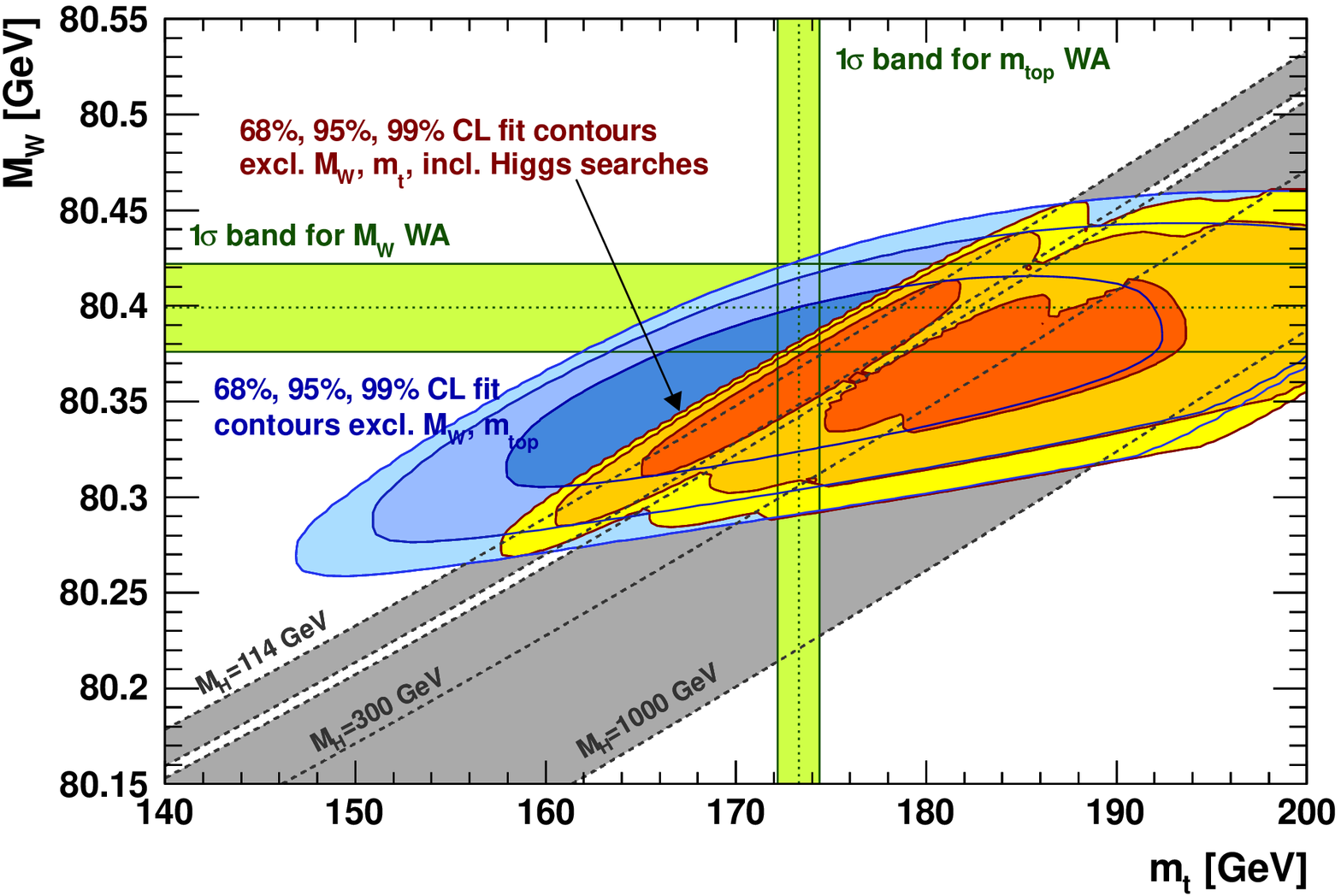, scale=\defaultFigureScale}}
  \vspace{0.1cm}
  \caption[]{Contours of 68\%, 95\% and 99\%~CL obtained from scans of
    fits with fixed variable pairs $M_W$ vs. $\mt$. The largest/blue
    allowed regions are the results of the standard fit
    excluding the measurements of $M_W$ and $\mt$. The narrow/orange
    areas indicate the corresponding constraints obtained for the
    complete fit. The horizontal bands indicate the $1\sigma$ regions
    of the measurements (world averages). The grey regions and
    isolines for various values of $M_H$ represent the SM prediction
    in an indicative way as the theory uncertainties (e.g. on $M_W$)
    are not included.  }
  \label{fig:MW_vs_mtop}
\end{figure}

More results and plots for the global SM fit are available on the Gfitter web 
site~\cite{GfitterWeb}.

%% file: SMtable_para_new.tex
\begin{table}
\setlength{\tabcolsep}{0.0pc}
{\small
\newcommand{\bm}{\boldmath}
\begin{tabular*}{\textwidth}{@{\extracolsep{\fill}}lc|cc|cc} 
\hline\noalign{\smallskip}
& & \multic{2}{c|}{Results from global EW fits:} & \multic{2}{c}{Fits w/o exp. input in given line:}   \\[-0.1cm]
\rs{Parameter} & \rs{Input value} & \multic{1}{c}{Standard fit} & \multic{1}{c|}{Complete fit} & \multic{1}{c}{Complete fit} & \multic{1}{c}{$M_H\equiv120$\:GeV} \\
\noalign{\smallskip}\hline\noalign{\smallskip}
\bm$M_{Z}$ {\ft [GeV]} &  $91.1875\pm0.0021$  &  $91.1874\pm0.0021$ &  $91.1877\pm0.0021$ &  $91.1959^{\,+0.0150}_{\,-0.0148}$ &   $91.1956^{\,+0.0141}_{\,-0.0136}$\\
$\Gamma_{Z}$ {\ft [GeV]} &  $2.4952\pm0.0023$ &  $2.4959\pm0.0015$ &  $2.4955\pm0.0014$ &  $2.4952\pm0.0017$ &   $2.4952\pm0.0017$\\
$\sigma_{\rm had}^{0}$ {\ft [nb]} &  $41.540\pm0.037$ &  $41.478\pm0.014$ &  $41.478\pm0.014$ &  $41.469\pm0.015$ &   $41.469\pm0.015$\\
$R^{0}_{\l}$ &  $20.767\pm0.025$ &  $20.743\pm0.018$ &  $20.741\pm0.018$ &  $20.719^{\,+0.025}_{\,-0.028}$ &   $20.717^{\,+0.027}_{\,-0.026}$\\
$A_{\rm FB}^{0,\l}$ &  $0.0171\pm0.0010$ &  $0.01640\pm0.0002$ &  $0.01624^{\,+0.0002}_{\,-0.0001}$ &  $0.01620^{\,+0.0002}_{\,-0.0001}$ &   $0.01620^{\,+0.0002}_{\,-0.0001}$\\
$A_\ell$ $^{(\star)}$  & $0.1499\pm0.0018$ & $0.1479\pm0.0010$ & $0.1472^{+0.0009}_{-0.0007}$ & -- &  --\\
$A_{c}$ &  $0.670\pm0.027$ &  $0.6683^{\,+0.00044}_{\,-0.00043}$ &  $0.6680^{\,+0.00040}_{\,-0.00028}$ &  $0.6679^{\,+0.00038}_{\,-0.00027}$ &   $0.6680^{\,+0.00038}_{\,-0.00026}$\\
$A_{b}$ &  $0.923\pm0.020$ &  $0.93469^{\,+0.00009}_{\,-0.00008}$ &  $0.93463^{\,+0.00007}_{\,-0.00005}$ &  $0.93462^{\,+0.00008}_{\,-0.00005}$ &   $0.93462^{\,+0.00008}_{\,-0.00003}$\\
$A_{\rm FB}^{0,c}$ &  $0.0707\pm0.0035$ &  $0.0741^{\,+0.0006}_{\,-0.0005}$ &  $0.0737^{\,+0.0005}_{\,-0.0004}$ &  $0.0738^{\,+0.0005}_{\,-0.0004}$ &   $0.0738^{\,+0.0005}_{\,-0.0004}$\\
$A_{\rm FB}^{0,b}$ &  $0.0992\pm0.0016$ &  $0.1037\pm0.0007$ &  $0.1032^{\,+0.0006}_{\,-0.0005}$ &  $0.1037^{\,+0.0003}_{\,-0.0005}$ &   $0.1037^{\,+0.0003}_{\,-0.0005}$\\
$R^{0}_{c}$ [$10^{-4}$] &  $1721\pm30$ &  $1722.9^{\,+0.7}_{\,-0.6}$ &  $1722.9\pm0.6$ &  $1722.9\pm0.6$ &   $1722.9\pm0.6$\\
$R^{0}_{b}$ [$10^{-4}$] &  $2162.9\pm6.6$ &  $2157.6^{\,+0.5}_{\,-0.8}$ &  $2157.5^{\,+0.5}_{\,-0.8}$  &  $2157.5^{\,+0.5}_{\,-0.8}$  &  $2157.5^{\,+0.5}_{\,-0.8}$ \\
$\sinleff(Q_{\rm FB})$ &  $0.2324\pm0.0012$ &  $0.23141^{\,+0.00012}_{\,-0.00013}$ &  $0.23150^{\,+0.00008}_{\,-0.00010}$ &  $0.23148^{\,+0.00010}_{\,-0.00009}$ &   $0.23149^{\,+0.00009}_{\,-0.00010}$\\
\noalign{\smallskip}\hline\noalign{\smallskip}
\bm$M_{H}$ {\ft [GeV]} $^{(\circ)}$ & ${\rm CL}_{\rm s+b}$ & $91^{+ 30[+ 74]}_{- 23[- 42]}$ & $120^{+ 12[+ 23]}_{-  5[-  6]}$ & $91^{+ 30[+ 74]}_{- 23[- 42]}$ &  $120$ (fixed) \\
\noalign{\smallskip}\hline\noalign{\smallskip}
$M_{W}$ {\ft [GeV]} &  $80.399\pm0.023$ &  $80.383^{\,+0.014}_{\,-0.015}$ &  $80.370^{\,+0.007}_{\,-0.009}$ &  $80.360^{\,+0.014}_{\,-0.013}$ &   $80.359^{\,+0.015}_{\,-0.008}$\\
$\Gamma_{W}$ {\ft [GeV]} &  $2.085\pm0.042$ &  $2.093\pm0.001$ &  $2.092\pm0.001$ &  $2.092\pm0.001$ &   $2.092\pm0.001$\\
\noalign{\smallskip}\hline\noalign{\smallskip}
\bm$\mc$ {\ft [GeV]} &  $1.27^{\,+0.07}_{\,-0.11}$ &  $1.27^{\,+0.07}_{\,-0.11}$ &  $1.27^{\,+0.07}_{\,-0.11}$ & --  &  -- \\
\bm$\mb$ {\ft [GeV]} &  $4.20^{\,+0.17}_{\,-0.07}$ &  $4.20^{\,+0.16}_{\,-0.07}$ &  $4.20^{\,+0.16}_{\,-0.07}$ & --  &  -- \\
\bm$m_{t}$ {\ft [GeV]} &  $173.3\pm1.1$ &  $173.4\pm1.1$ &  $173.7\pm1.1$ &  $177.2\pm3.4$$^{(\bigtriangledown)}$ &   $176.8^{+  3.1}_{-  3.0}$\\
\bm$\dalphaHadMZ$ $^{(\dag\bigtriangleup)}$ &  $2757\pm  10$ & $2758\pm  11$ & $2756\pm  11$ & $2729^{+  57}_{-  50}$ &  $2730^{+  57}_{-  46}$\\
\bm$\alpha_{s}(M_{Z}^{2})$ & -- &  $0.1193\pm0.0028$ &  $0.1194\pm0.0028$ &  $0.1194\pm0.0028$ &   $0.1194\pm0.0028$\\
\noalign{\smallskip}\hline\noalign{\smallskip}
\bm$\deltatheo M_W$ {\ft [MeV]}  & $[-4,4]_{\rm theo}$ & $4$ & $4$ & --  &  -- \\
\bm$\deltatheo \sinleff$ $^{(\dag)}$  & $[-4.7,4.7]_{\rm theo}$ & $4.7$ & $4.7$ & --  &  -- \\
\noalign{\smallskip}\hline
\noalign{\smallskip}
\end{tabular*}
{\ft
$^{(\star)}$Average of LEP ($A_\ell=0.1465\pm0.0033$) and SLD ($A_\ell=0.1513\pm0.0021$) measurements.
The complete fit w/o the LEP (SLD) measurement gives $A_\ell=$ $0.1473^{\,+0.0010}_{\,-0.0006}$ 
($A_\ell=$ $0.1469^{\,+0.0007}_{\,-0.0005}$ 
).
$^{(\circ)}$In brackets the $2\sigma$. 
$^{(\dag)}$In units of $10^{-5}$.
$^{(\bigtriangleup)}$Rescaled due to $\alpha_s$ dependency.
$^{(\bigtriangledown)}$Ignoring a second less significant minimum, \cf Fig.~\ref{fig:topscan} and Eq.~(\ref{eq:mtop_completefit}).
}}
\caption[.]{Input values and fit results for the observables and parameters of the global 
         electroweak fit. The first and 
         second columns list respectively the observables/parameters used in the fit, and their 
         experimental values or phenomenological estimates (see text for references). 
         The subscript ``theo'' labels theoretical error ranges. 
         Boldface letters indicate that a parameter is floating in the fit.
         The third (fourth) column quotes the results of the standard (complete) fit
         not including (including) the constraints from the direct Higgs searches at LEP, Tevatron 
         and the LHC in the fit. In case of floating parameters the fit results are directly given,
         while for (non-floating) observables the central values and errors are obtained by 
         individual profile likelihood scans. The last two columns give the fit results for each 
         parameter without using the corresponding experimental or phenomenological 
         constraint in the fit (indirect determination), for the complete fit and when 
         assuming the Higgs mass to be known and precisely measured at 120\:GeV, respectively.
}         
\label{tab:SMresults}
\end{table}


\begin{table}[t]
\setlength{\tabcolsep}{0.0pc}
{\normalsize
\centering
\begin{tabular*}{\textwidth}{@{\extracolsep{\fill}}lrrrrrrr} 
    \hline\noalign{\smallskip}
      Parameter & $\ln M_{H}$ & \dalphaHadMZ & $M_{Z}$ & $\asZ $ & $\mt$ &  $\mc$ &  $\mb$ \\
    \noalign{\smallskip}\hline\noalign{\smallskip}
      $\ln M_{H}$   & 1    &  $-0.18$ &  0.13  &  0.02  & 0.32   &   $-0.00$ &  $-0.01$ \\
      \dalphaHadMZ   &      &  1     & $-0.01$ &   0.35 & $0.00$ &  0.00 & 0.01 \\
      $M_{Z}       $  &      &        &  1     & $-0.01$ &  $-0.01$ &  $-0.00$ &  $-0.00$ \\
      $\asZ $        &      &        &        &  1     &  0.03       &  0.01 & 0.04 \\
      $\mt       $ &      &        &        &        &  1          &     0.00 & $-0.00$ \\
       $\mc$         &      &        &        &        &             &  1     & 0.00 \\
    \noalign{\smallskip}\hline
\end{tabular*} 
}
\caption{Correlation coefficients between the free fit parameters in the standard fit. 
         The correlations with and between the varying theoretical error parameters \deltatheo 
         are negligible in all cases. }
\label{tab:SMcormat}
\end{table}

%% file: Oblparam.tex
\section{Oblique Corrections} 
\label{sec:oblique}

A common approach to constrain physics beyond the SM using the global
electroweak fit is through the formalism of {\it oblique parameters}.

\subsection{Concept of oblique parameters}

Provided that the new physics mass scale is high, beyond the scale of
direct production, and that it contributes only through virtual loops
to the electroweak precision observables, the dominant BSM effects 
can be parametrised by three gauge
boson self-energy parameters named {\it oblique parameters}.  In
this section we recall only the relevant parameter definitions.  A
more general introduction of the oblique formalism is given in the
appendix, page~\pageref{app:obliqueParams}.

The literature focuses on two different, but equivalent oblique
parameter sets: $\varepsilon_{1,2,3}$~\cite{epspar,epspar2} and 
\STU~\cite{Peskin:1991sw}.  Both sets are reparametrisations of 
the variables $\Delta \rho$, $\Delta \kappa$ and $\Delta r$, which 
absorb the radiative corrections to the total $Z$ coupling 
strength, the effective weak mixing angle, and the
$W$ mass, respectively.  It is assumed that the new physics 
contributing to the radiative corrections is 
flavour universal, while for the $Z\to \bbbar$ vertex, receiving large top-quark 
corrections, an extra oblique parameter is introduced. In this analysis we 
implement the additional corrections to the $Z\to\bbbar$ coupling as described
in Ref.~\cite{Burgess:1993vc}.  More parameters $(X, Y, V, W)$ are
required if the scale of new physics is not much larger than the weak
scale~\cite{Burgess:1993mg, Barbieri:2004qk}. They can only be independently 
determined when including data at higher centre-of-mass energies than the 
$Z$ pole, which is not carried out in the present analysis, so that these 
additional parameters are set to zero.


The $\varepsilon_{1,2,3}$ parameters~\cite{epspar,epspar2} include SM contributions
dominated by top quark and Higgs boson corrections. By construction
they vanish at Born level if the running of $\alpha$ is accounted for. 
Their typical size is hence of order $\alpha$. They are defined by
\begin{eqnarray}
\label{eq:epsilon1}
  \varepsilon_1 &\!\! = \!\!& \Delta\rho\,, \\
\label{eq:epsilon2}
  \varepsilon_2 &\!\! = \!\!& \cos^2\! \theta_W\Delta\rho + \frac{\sin^2\! \theta_G}{\cos^2\! \theta_W-\sin^2\! \theta_G}\Delta r-2\sin^2\! \theta_G\Delta\kappa^\prime\,, \\
\label{eq:epsilon3}
  \varepsilon_3 &\!\! = \!\!& \cos^2\! \theta_W\Delta\rho + (\cos^2\! \theta_W-\sin^2\! \theta_G)\Delta\kappa^\prime\,,
\end{eqnarray}
with 
$2\sin^2\! \theta_G=1-\sqrt{1-\sqrt{8}\pi\alpha(M_Z^2)/(G_F M_Z^2)}$ 
and where $\Delta \kappa^\prime$ relates $\sinfeff$ to
$\sin^2\! \theta_G$ instead of $\sin^2\! \theta_W$.  The quadratic top
mass dependence present in all form factors has been removed
explicitly from the parameters $\varepsilon_2$ and $\varepsilon_3$. 


In the definition of the \STU parameters~\cite{Peskin:1991sw} the predicted
SM contributions are subtracted from the measured $\varepsilon$ parameters, so
that the \STU vanish in the SM.  
Due to the dominant virtual top quark and Higgs boson corrections, the subtracted 
SM terms depend on $M_H$ and $m_t$, which take fixed reference values. Thus, by 
construction, the \STU parameters depend on a (somewhat arbitrary) SM reference 
point, while the physically relevant difference between the experimental \STU parameters 
and a model prediction is independent of the reference. The \STU parameter  are 
normalised so that the expected BSM contributions are of order ${\cal O}(1)$. 
The so subtracted and normalised parameters are related to the $\varepsilon$ parameters by
\begin{eqnarray}
S        &\!\! = \!\!&           +  \varepsilon_3\frac{4\sin^2\! \theta_G}{\alpha(M_Z^2)} - d_S\,, \\
T        &\!\! = \!\!& \phantom{\pm}\varepsilon_1\frac{1      }{\alpha(M_Z^2)} - d_T\,, \\
U        &\!\! = \!\!&           -  \varepsilon_2\frac{4\sin^2\!\theta_G}{\alpha(M_Z^2)} - d_U\,, 
\end{eqnarray}
where $d_i$ are the SM predictions for the chosen $M_H$ and $m_t$
reference. Throughout this paper we use the reference values $M_{H,\rm ref}=120$\:GeV 
and $m_{t,\rm ref}=173$\:GeV that are chosen to agree with the experimental constraints.

Physics beyond the SM can also contribute to the $Z\to b\bbar$ vertex, 
which receives significant top quark corrections in the SM. Following  
Ref.~\cite{Burgess:1993vc} these additional vertex corrections are 
implemented via two new parameters, $\delta g_L^{b\bbar}$ and 
$\delta g_R^{b\bbar}$. These parameters are set to zero in the fits determining 
the experimental values of the \STU parameters.
The parameters are included for new physics models that significantly contribute
to the $Z\to b\bbar$ vertex, which, for the models studied in this paper,
is only the case for the Littlest Higgs model (\cf Section~\ref{sec:constraintsLH}).

The advantage of the \STU parametrisation lies in the convenience with
which it permits to compare model predictions with the electroweak data. 
It is therefore adopted in most parts of this paper. For a given model,
the prediction of any electroweak observable $O$ is given by
\begin{equation}\label{eq:fittingformula}
   O=O_{\rm SM,ref}(M_{H,\rm ref}, m_{t,\rm ref})+c_S S+c_T T+c_U U\, ,
\end{equation}
where $O_{\rm SM, ref}(M_{H,\rm ref}, m_{t,\rm ref})$ is the SM prediction of the observable
in the reference SM, including all known two-loop and beyond two-loop 
electroweak corrections. The linear terms ($c_SS$, $c_TT$, $c_UU$) 
parametrise the additional contribution from the BSM
model. The coefficients $c_S,c_T,c_U$ are available in the literature
for the full set of electroweak precision observables. This report
uses the values from Ref.~\cite{Burgess:1994zp}.  The precise
measurements of the electroweak observables thus allow to constrain \STU,
and hence parameters of specific BSM physics models whose contributions 
to the oblique parameters have been calculated. 

The BSM effects on the \STU parameters can be summarised as follows.
\begin{itemize}
\item The $T$ parameter measures the difference between the new
      physics contributions of neutral and charged current processes at
      low energies, i.e. it is sensitive to weak isospin violation.  $T$
      ($\varepsilon_1$) is proportional to $\Delta \rho$.
\item The $S$ $(S+U)$ parameter describes new physics contributions to
      neutral (charged) current processes at different energy scales. The $S$
      parameter ($\varepsilon_3$) takes the remaining part of $\Delta
      \kappa$, which is then free from quadratic top quark contributions
      due to weak-isospin breaking.
\item The third parameter, $U$, is only constrained by the $W$ boson
      mass and width. $U$ ($\varepsilon_2$) describes the
      remaining corrections to $\Delta r$ and is predicted to be small
      in most new physics models.
\end{itemize} 

Updated experimental results for the oblique parameters as obtained from
the global electroweak fit are given in the following section. Similar studies 
have been performed by the LEP Electroweak Working Group~\cite{LEPEWWG} and for 
the electroweak review of the Particle Data Group~\cite{ErlerNakamura:2010zzi}.

\subsection{Experimental constraints on the oblique parameters}

The \STU parameters are determined from the fit by
comparing the measured electroweak precision observables with the
respective theory predictions of Eq.~(\ref{eq:fittingformula}). Except 
for the fixed $M_{H,\rm ref}$ and $m_{t,\rm ref}$ all other SM fit
parameters, including $S$, $T$ and $U$, are free to vary in the fit 
(\cf bold quantities in Table~\ref{tab:SMresults}). After fit convergence 
we find
\begin{equation} 
    S=0.03\pm 0.10\,, \quad\quad\quad 
    T=0.05\pm 0.12\,, \quad\quad\quad 
    U=0.07\pm 0.11\,,
\label{eq:STU}
\end{equation}
and linear correlation coefficients of $+0.89$ between $S$ and $T$, 
and $-0.49$ ($-0.72$) between $S$ and $U$ ($T$ and $U$). Some BSM 
models predict a vanishing or negligible contribution to $U$, which 
allows to stronger constrain the remaining parameters. Fixing $U=0$ we
obtain
\begin{equation} 
    S|_{U=0}=0.06\pm 0.09,\quad\quad\quad T|_{U=0}=0.10\pm0.08,
\label{eq:STUfix}
\end{equation}
with a correlation coefficient of $+0.89$. The improved precision on $S$ 
and $T$ stems from the information of $M_W$ and $\Gamma_W$, which otherwise
is absorbed to determine the $U$ parameter. 

\begin{figure}[p]
  \newcommand\STUPlotSize{\defaultFigureScale}
  \centerline{\epsfig{file=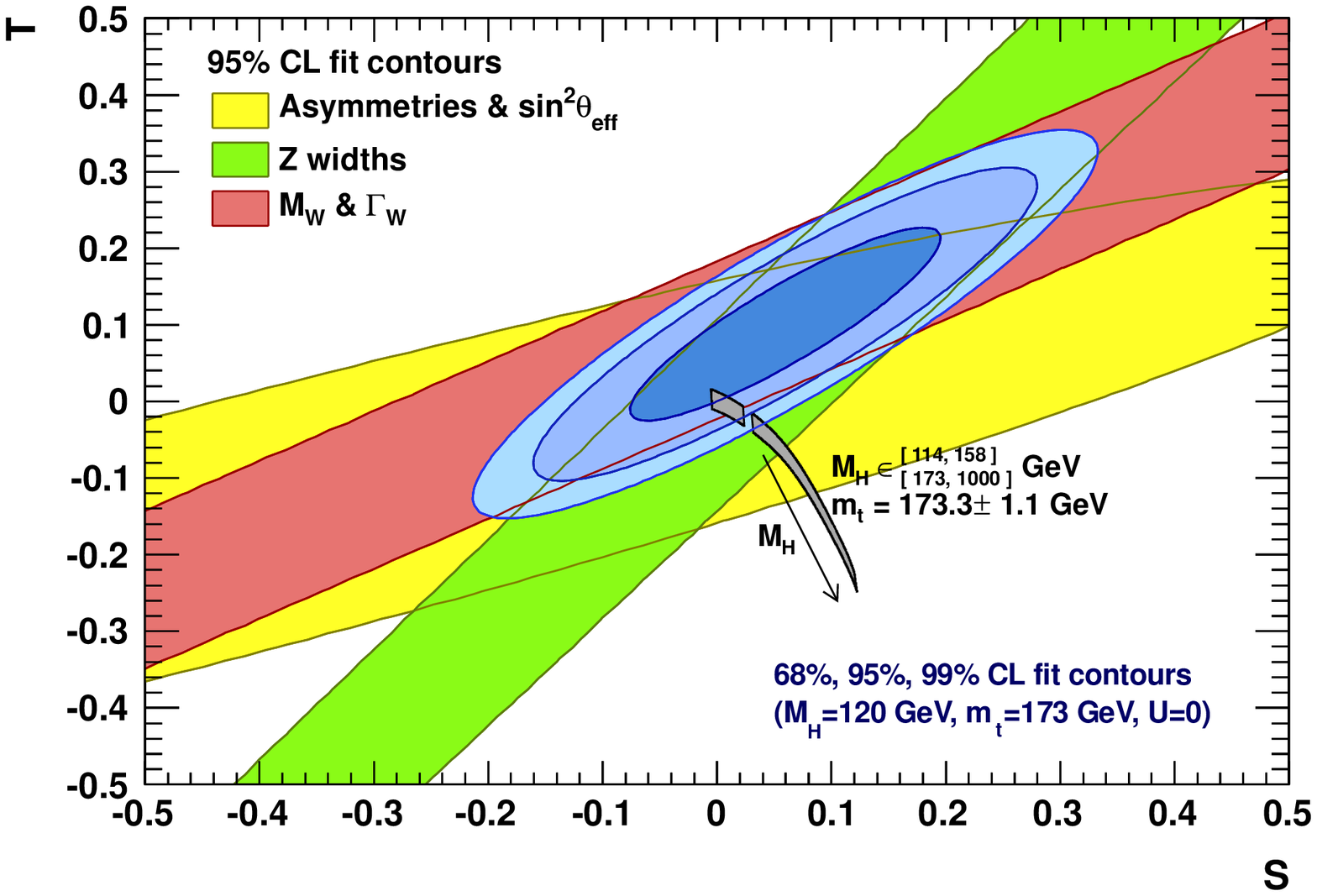, scale=\STUPlotSize}}
  \vspace{0.4cm}
  \centerline{\epsfig{file=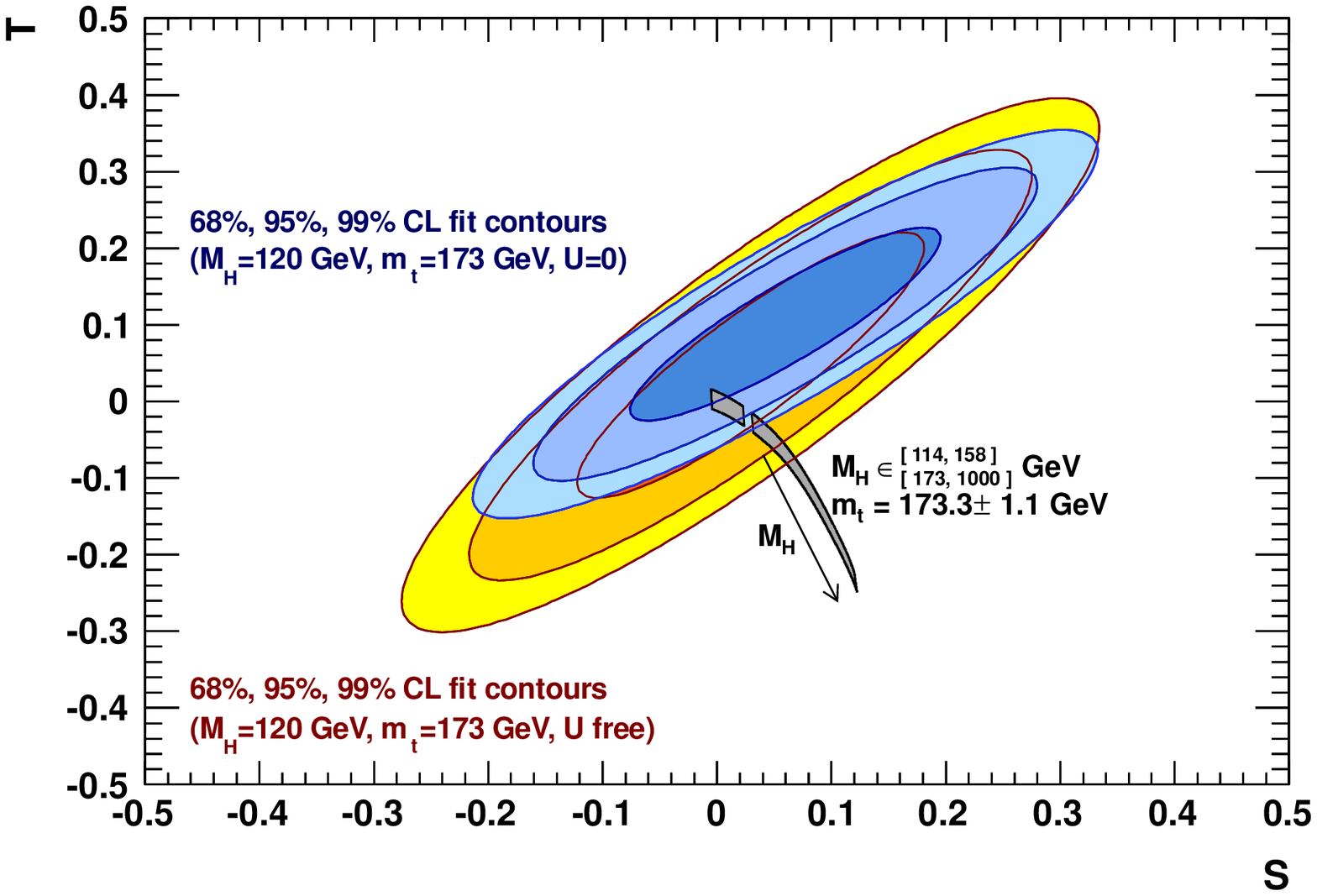, scale=\STUPlotSize}}
   \vspace{0.3cm}
   \caption[]{Experimental constraints on the $S$, $T$ parameters with respect to
     the SM reference represented by $M_{H,\rm ref}=120$\:GeV, $m_{t,\rm ref}=173$\:GeV 
     and the corresponding best fit values for the remaining SM parameters. 
     Shown are the 68\%, 95\% and 99\% CL allowed regions with the $U$ parameter
     fixed to zero (blue ellipses on top and bottom panels) or let free to vary 
     in the fit (orange ellipses on bottom panel). The top plot also shows for $U=0$
     the individual constraints from the asymmetry measurements (yellow), 
     the $Z$ partial and total widths (green), and the $W$ mass and width (orange). 
     The narrow dark grey bands illustrate the SM prediction for varying $M_H$ 
     and $m_t$ values (see figures for the ranges used).}
   \label{fig:SvsT}
\end{figure}

\begin{figure}[t]
  \newcommand\STUPlotSize{\HalfPageWidthScale}  
  \centerline{\epsfig{file=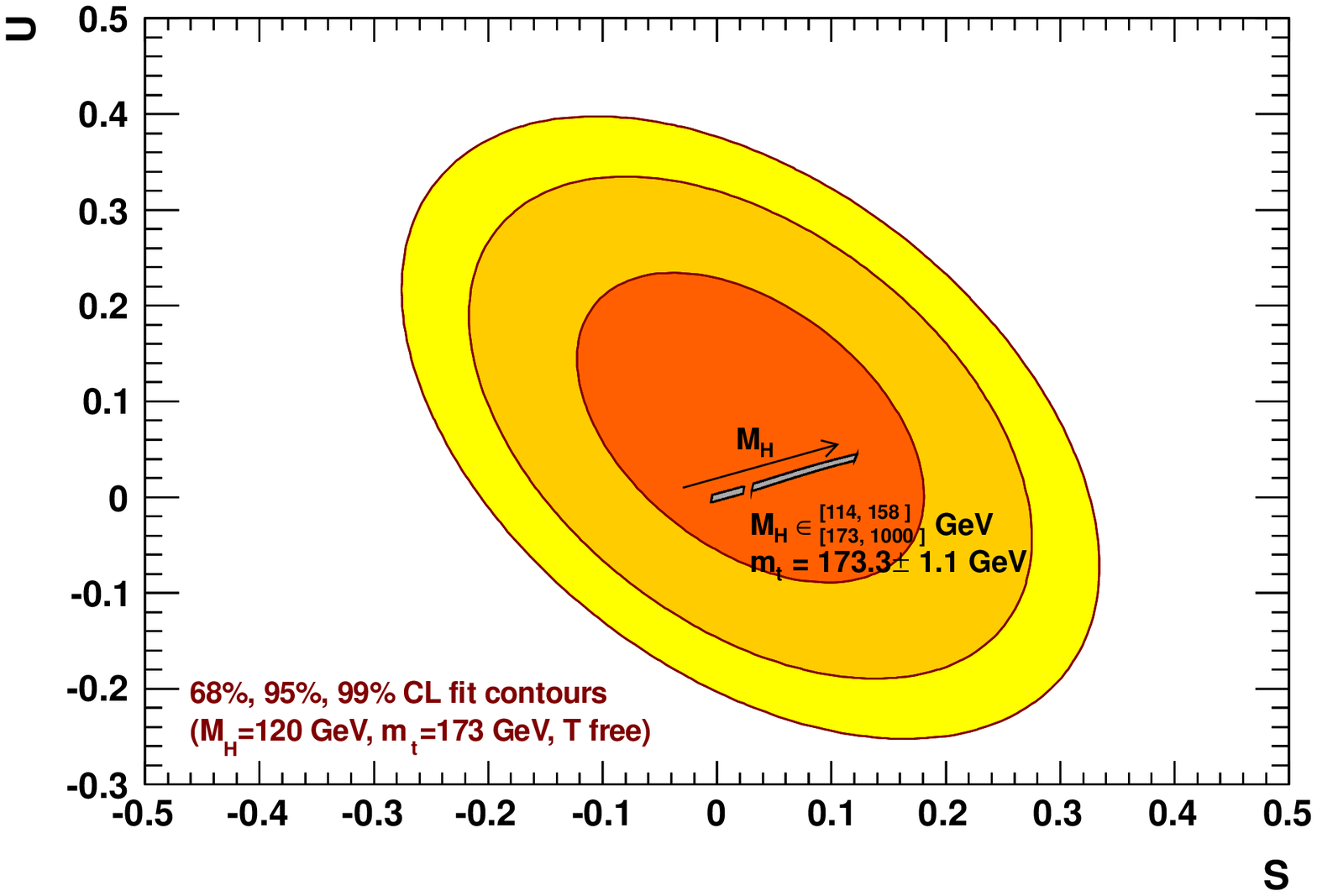, scale=\STUPlotSize}
              \epsfig{file=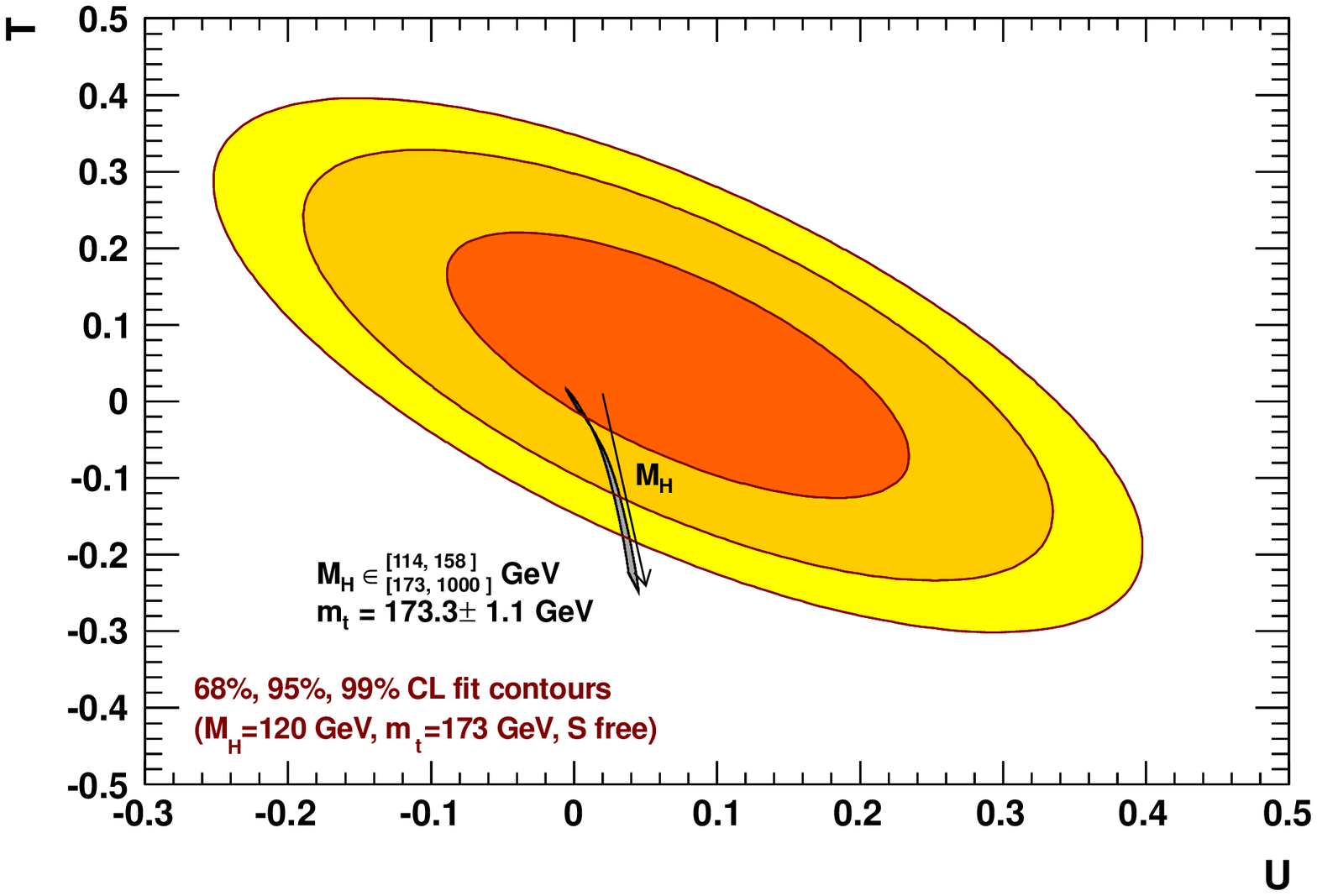, scale=\STUPlotSize}}
   \vspace{0.3cm}
   \caption[]{Experimental constraints on the $S$, $U$ (left) and $U$, $T$
     parameters (right)  with respect to the SM reference represented 
     by $M_{H,\rm ref}=120$\:GeV, $m_{t,\rm ref}=173$\:GeV and the 
     corresponding best fit values for the remaining SM parameters. 
     Shown are the 68\%, 95\% and 99\% CL allowed regions, where the 
     third parameter is left unconstrained. 
     The narrow dark grey bands illustrate the SM prediction for varying $M_H$ 
     and $m_t$ values (see figures for the ranges used).}
   \label{fig:U}
\end{figure}
As all the experimental \STU values are compatible with zero, the data are in 
agreement with our chosen SM reference. Figures~\ref{fig:SvsT} and \ref{fig:U}
show by the orange ellipses the 68\%, 95\% and 99\% confidence level (CL) allowed 
regions in the $(S,T)$, $(S,U)$ and $(U,T)$ planes. Figure~\ref{fig:SvsT} also 
gives the tighter constraints found when fixing $U=0$ (blue ellipses). 
The upper panel displays for $U=0$ the individual constraints from the asymmetry 
measurements, $Z$ partial and total widths, and $W$ mass and width. 
Leaving $U$ free would leave the former two constraints approximately unchanged, 
while the $W$ mass and width would then constrain $U$ rather than $S$ or $T$.
Also shown on all plots 
is the SM prediction for varying $M_H$ and $m_t$ values (ranges given on plots).
By construction, the SM prediction 
reproduces $S=T=U=0$ at the $\rm SM_{ref}$ bench mark.  While the variations 
of $S,T,U$ within the current $m_t$ uncertainty is small, $M_H$
values larger than the electroweak scale lead to larger (smaller)
values of $S$ $(T)$.  The $U$ parameter exhibits only a small
dependence on $M_H$, justifying the choice $U=0$ for the SM
interpretation.
All experimentally allowed ellipses show compatibility with the SM predictions 
for a light Higgs boson, reflecting the satisfactory goodness-of-fit obtained 
in the SM fit (\cf Section~\ref{sec:smfit-results}).\footnote
{
  Had we determined
  $M_H$ by confronting experimental and predicted oblique parameters,
  we would reproduce Fig.~\ref{fig:greenband} up to deviations due to
  the higher order and non-oblique corrections present in the standard
  electroweak fit.
} 

Many BSM models feature a similar agreement with the data as observed
for the SM.  The predictions of these models can cover large regions
in the $(S,T,U)$ space due to additional undetermined model parameters,
which in turn can be constrained via the oblique parameter formalism
from the data. Most (though not all) models decouple at high scales from 
the SM so that the oblique corrections reproduce the SM values. 
We will see in the following that models providing
additional weak isospin violation can readily accommodate large Higgs
boson masses, whose negative $T$ values are compensated by the
model-induced positive contributions.

%% file: Constraints.tex
\section{Constraints on New Physics Models }
\label{sec:constraints}

We proceed the analysis with confronting the \STU parameters determined 
in the previous section with predictions from SM extensions. For a given new physics
model, the \STU predictions consist of the sum of the BSM contributions and 
the non-vanishing SM remainders when the $M_H$ and $m_t$ values differ from 
those used for the SM reference. Numerous oblique parameter analyses have 
been performed in the past, usually following two separate steps: $(i)$ the 
determination of the \STU parameters by groups performing the electroweak fit, 
and $(ii)$ BSM studies using these \STU values in independent analyses. When 
fitting the BSM model parameters together with the top quark and Higgs boson 
masses, the dependence of \STU  on the latter two parameters is then usually
approximated by the one-loop terms~\cite{Peskin:1991sw}
\beqn 
    S &\!\!\approx\!\!& \frac{1}{12\pi}\ln\!\frac{M_H^2}{M_{H,\rm ref}^2} + 
                        \frac{1}{6\pi} \ln \frac{m_t^2}{m_{t,\rm ref}^2} \,, \\
    T &\!\!\approx\!\!& -\frac{3}{16\pi\cos^2\!\theta_W}\ln\!\frac{M_H^2}{M_{H,\rm ref}^2} + 
                         \frac{3}{16\pi\sin^2\!\theta_W \cos^2\!\theta_W } 
                         \ln \frac{m_t^2-m_{t,\rm ref}^2}{m_Z^2} \,,\\
    U &\!\!\approx\!\!&  \frac{1}{2\pi} \ln \frac{m_t^2}{m_{t,\rm ref}^2}\,.
\eeqn
The $m_t$ dependence is often neglected. 
The Gfitter software allows us to study the dependence of the oblique 
corrections on the BSM model parameters and  the SM parameters ($M_H$ and $m_t$)
taking into account the full two-loop and beyond-two-loop corrections of the SM.

In this section we revisit published \STU predictions for several prominent BSM models
and provide BSM constraints derived with consistent electroweak data 
(as used in Sect.~\ref{sec:smfit} for the {\it standard} fit), 
SM reference 
point and statistical procedure. At the beginning of each subsection we provide
a brief outline of the main model features and recall the available experimental 
search results. None of the direct searches is used to constrain the \STU predictions
or new physics parameter fits. 

\input FourthGen

\input 2HDM

\input InertHiggs

\input LittlestHiggs

\input LargeED

\input UniversalED

\input WarpedED

\input Technicolor

%% file: FourthGen.tex
\subsection{Models with a sequential fourth fermion generation}
\label{sec:constraints4th}

The fermion sector of the SM is composed of three generations of
leptons and quarks. Several SM
extensions suggest extra families of matter particles, which -- with
the dawn of the LHC -- have received increased attention in the
theoretical literature. As the new fermions would obtain their masses
via Yukawa couplings to the Higgs condensate they must be of order the
electroweak scale and hence should be experimentally accessible. The
phenomenological consequences of a fourth generation on the flavour
sector of neutrinos, charged leptons and quarks have been extensively
explored (\cf, \eg,
Refs.~\cite{Arhrib:2002md,Hou:2006mx,Hou:2005yb,Hou:2006jy,Soni:2008bc,Botella:2009zz,Eberhardt:2010bm,Bobrowski:2009ng,Soni:2010xh,Buras:2010pi,Lacker:2010zz, Erler:2010sk}
and the review~\cite{Frampton:1999xi}).  The impact on electroweak
precision data at the $Z$-pole and on Higgs physics has been studied
in
Refs.~\cite{Holdom:1996bn, Holdom:2006mr, He:2001tp,Kribs:2007nz,Novikov:2002tk,Chanowitz:2010bm,Chanowitz:2009mz,Eberhardt:2010bm}.
For the present analysis, we use the oblique corrections computed in
Ref.~\cite{He:2001tp}.

In a generic model with only one extra generation, two new fermions
$(\psi_1,\psi_2)$, with one left-handed weak isospin doublet 
$\psi_L=(\psi_{1},\psi_2)_L$ and two right-handed weak isospin singlet 
states $\psi_{1,R}$, $\psi_{2,R}$, and with charges equal to the 
three SM generations, are added to each of the quark and lepton sectors. 
The new unconstrained model parameters are the masses 
$m_{u_4}$, $m_{d_4}$, $m_{\nu_4}$, $m_{e_4}$ of the fourth generation quarks and leptons,
and \CP-conserving and \CP-violating neutrino and quark mixing parameters. 
The fourth generation neutrino must have a mass of at least $M_Z/2$ to 
not contribute to the invisible width of the $Z$. 

The most stringent experimental lower limits on sequential heavy
fourth generation quarks (SM4) stem from CMS~\cite{Chatrchyan:2011em}, 
excluding $d_4$ masses between 255 and 361\:GeV, and from the Tevatron
experiments where the newest analyses from CDF exclude $u_4$ quarks
below 358\:GeV and $d_4$ quarks below
372\:GeV~\cite{Gunion:1989we,Aaltonen:2011vr}. The $u_4$ searches
assume predominant decays into $W$ boson and SM quarks, requiring a small
$u_4$--$d_4$ mass splitting to inhibit the decay $u_4\to Wd_4$. The
CMS and Tevatron $d_4$ searches assume a $d_4\to Wt$ branching
fraction of one on the basis of the observed unitarity of the
three-generation CKM quark mixing matrix suggesting small
flavour-changing currents to light quarks. This also neglects the
possibility of an inverted fourth generation mass hierarchy. The CDF
limits have been reanalysed in
Refs.~\cite{Flacco:2011ym,Flacco:2010rg} under more general SM4 and
CKM4 quark mixing scenarios, leading to a weaker limit of 290\:GeV for
both quark flavours.  Fourth generation leptons have been best
constrained at LEP with a lower limit of about 101\:GeV for sequential
heavy leptons decaying to $W\nu$ and $Z\ell$, or to $W\ell$ and
$Z\nu$, depending on their electromagnetic
charge~\cite{Achard:2001qw}.

Assuming negligible mixing of the extra fermions with the SM
fermions,\footnote
{ 
  A detailed numerical SM4
  analysis~\cite{Eberhardt:2010bm} taking into account low-energy FCNC
  processes in the quark sector, electroweak oblique corrections, and
  lepton decays (but not lepton mixing) concludes that small mixing
  between the quarks of the first three and those of the fourth family
  is favoured. The value of $|V_{tb}|$ is found in this analysis to
  exceed 0.93.  The no-mixing assumption allows us to use the measured
  value of $G_F$, extracted from the muon lifetime under the SM3
  hypothesis, to its full precision~\cite{Lacker:2010zz}.  
}  
the one-loop fermionic contributions of a sequential fourth generation 
to the oblique corrections are given by~\cite{He:2001tp} 
\beqn
\label{eq:S4gen}
   S &=& \frac{\NC}{6\pi}\bigg\{(8Y+6)x_1 - (8Y-6)x_2 - 2Y\ln\frac{x_1}{x_2} \nonumber\\
     &&  \hspace{0.9cm} + \;\left[\left(\frac{3}{2}+2Y\right)x_1+Y\right]G(x_1)
                        +   \left[\left(\frac{3}{2}-2Y\right)x_2-Y\right]G(x_2)\bigg\}\,, \\[0.2cm]
\label{eq:T4gen}
   T &=& \frac{\NC}{8\pi s^2_W c^2_W}F(x_1,x_2)\,, \\[0.2cm]
\label{eq:U4gen}
   U &=& -\frac{\NC}{2\pi}\bigg\{\frac{1}{2}(x_1+x_2) - \frac{1}{3}(x_1-x_2)^2
                        + \left[\frac{1}{6}(x_1-x_2)^3 -
                                \frac{1}{2}\frac{x_1^2+x_2^2}{x_1-x_2}\right]\ln\frac{x_1}{x_2} \nonumber\\
     &&  \hspace{1.4cm} + \;\frac{1}{6}(x_1-1)f(x_1,x_1)+\frac{1}{6}(x_2-1)f(x_2,x_2) \nonumber\\
     &&  \hspace{1.4cm} + \;\left[\frac{1}{3}-\frac{1}{6}(x_1+x_2)-\frac{1}{6}(x_1-x_2)^2\right]f(x_1,x_2)
                          \bigg\}\,, 
\eeqn 
where $Y=1/6\,(-1/2)$ is the weak hypercharge for quarks (leptons), $\NC=3\,(1)$ 
for quarks (leptons), $x_i=(m_{i_4}/M_Z)^2$ with $i=1,2$ for the up-type and
down-type fourth generation fermions, respectively.\footnote
{\label{footnote:formula}
  The functions in Eqs.~(\ref{eq:S4gen}--\ref{eq:U4gen}) are defined as follows.
  $F(x_1,x_2)=(x_1+x_2)/2-x_1x_2/(x_1-x_2)\cdot\ln(x_1/x_2)$,
  $G(x)=-4y\arctan(1/y)$, $y=\sqrt{4x-1}$, and
  $f(x_1,x_2)=-2\sqrt{\Delta}[\arctan((x_1-x_2+1)/\sqrt{\Delta})-\arctan((x_1-x_2-1)/\sqrt{\Delta})]$
  for $\Delta>0$, $f(x_1,x_2)=0$ for $\Delta=0$, and
  $f(x_1,x_2)=\sqrt{-\Delta}\cdot\ln((X+\sqrt{-\Delta})/(X-\sqrt{-\Delta}))$
  with $X=x_1+x_2-1$ for $\Delta<0$, and where
  $\Delta=2(x_1+x_2)-(x_1-x_2)^2-1$.
}

In the limit of large and degenerate up and down-type fermion masses, the $S$ 
parameter in Eq.~(\ref{eq:S4gen}) reduces to $2/(3\pi) \simeq0.21$, exhibiting
the non-decoupling property of fourth generation models. Small 
fourth generation quark and lepton masses lead to larger positive $S$ values that, 
with increasing up$\,>\,$down mass splitting decrease (increase) for quarks (leptons). 
Negative contributions to $S$ are possible for a heavier up-type than down-type quark,
or for a heavier charged lepton than neutrino (e.g., for $m_{\nu_4}=400$\:\gev
and $m_{e_4}=660$\:\gev one has $\Delta S_\ell\simeq-0$). 

The $T$ parameter~(\ref{eq:T4gen}), sensitive to weak isospin violation, is 
always positive or zero, owing to $F(x_1,x_2)\ge0$, $\forall x_1,x_2>0$. In 
case of approximate mass degeneracy, $T$ is proportional to the difference 
between up and down-type mass-squared relative to $M_Z^2$.

The $U$ parameter~(\ref{eq:U4gen}) is positively defined and vanishes for 
degenerate fourth generation up-type and down-type fermion masses. For 
freely varying masses within the range $[100,1000]$\:GeV, the maximum 
value, obtained at maximum mass splitting, reads: $U\simeq0.49_{q}+0.16_\ell\simeq0.66$.

\begin{figure}[!t]
  \centering \epsfig{file=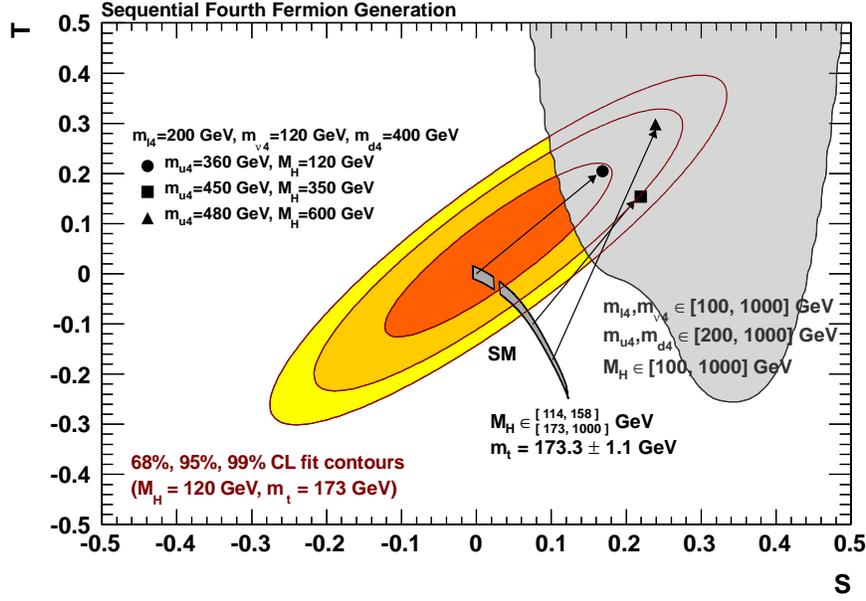,scale=\defaultFigureScale} 
  \vspace{0.2cm}
  \caption[]{Oblique parameters in a model with a fourth fermion generation.
             Shown are the $S$, $T$ fit results (leaving $U$ free) compared 
             with the prediction from the SM (dark grey) and the sequential
             fourth generation model with vanishing flavour mixing (light grey). 
             The symbols illustrate the predictions for three example settings 
             of the parameters $m_{U_4}$, $m_{d_4}$, $m_{{\nu}_4}$, $m_{l_4}$ and $M_H$. 
             The light grey area is obtained by varying the free mass parameters
             in the ranges indicated in the figure.}
   \label{fig:4thSvsT}
\end{figure}
\begin{figure}[!t]
  \epsfig{file=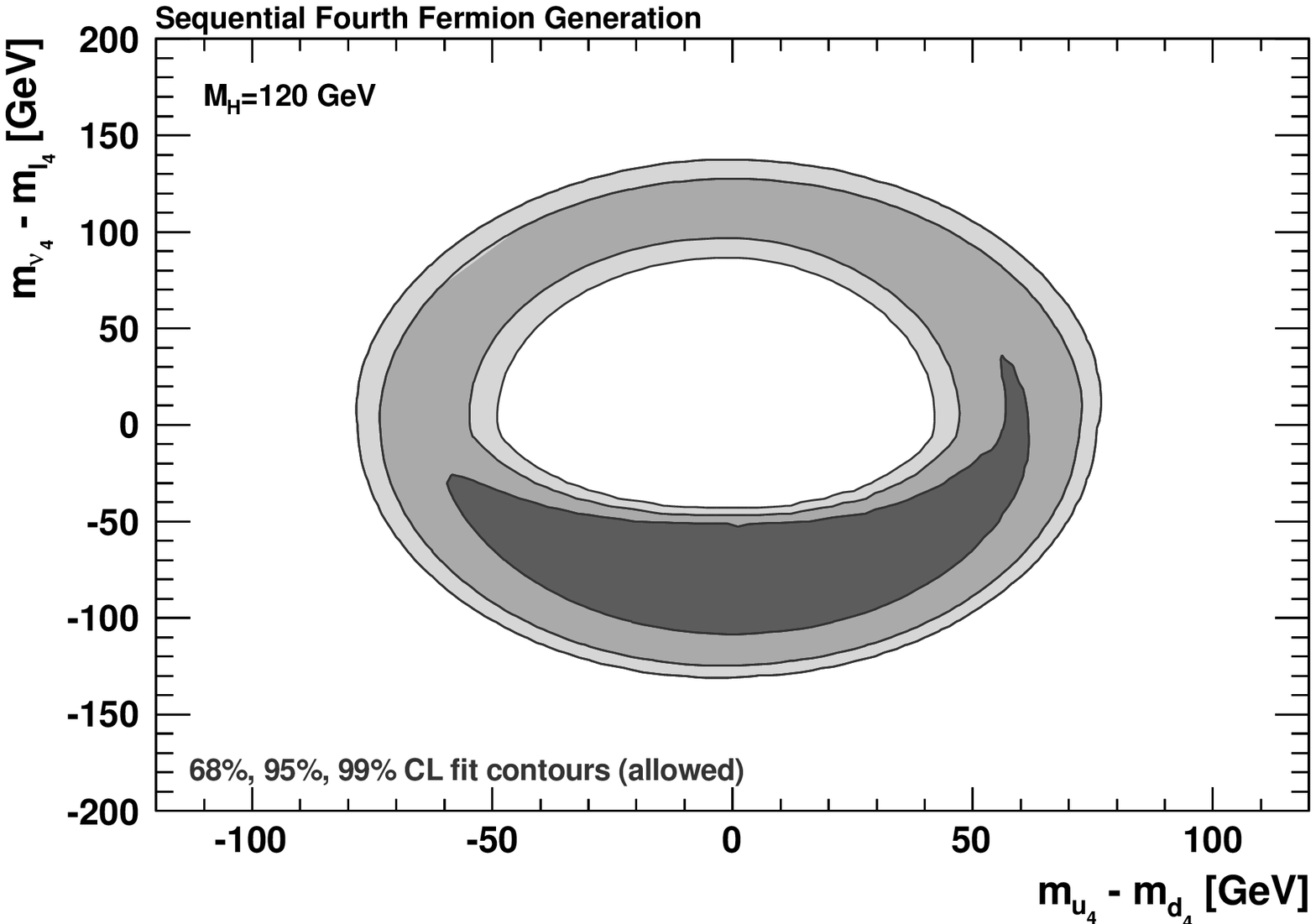, scale=0.4}
  \epsfig{file=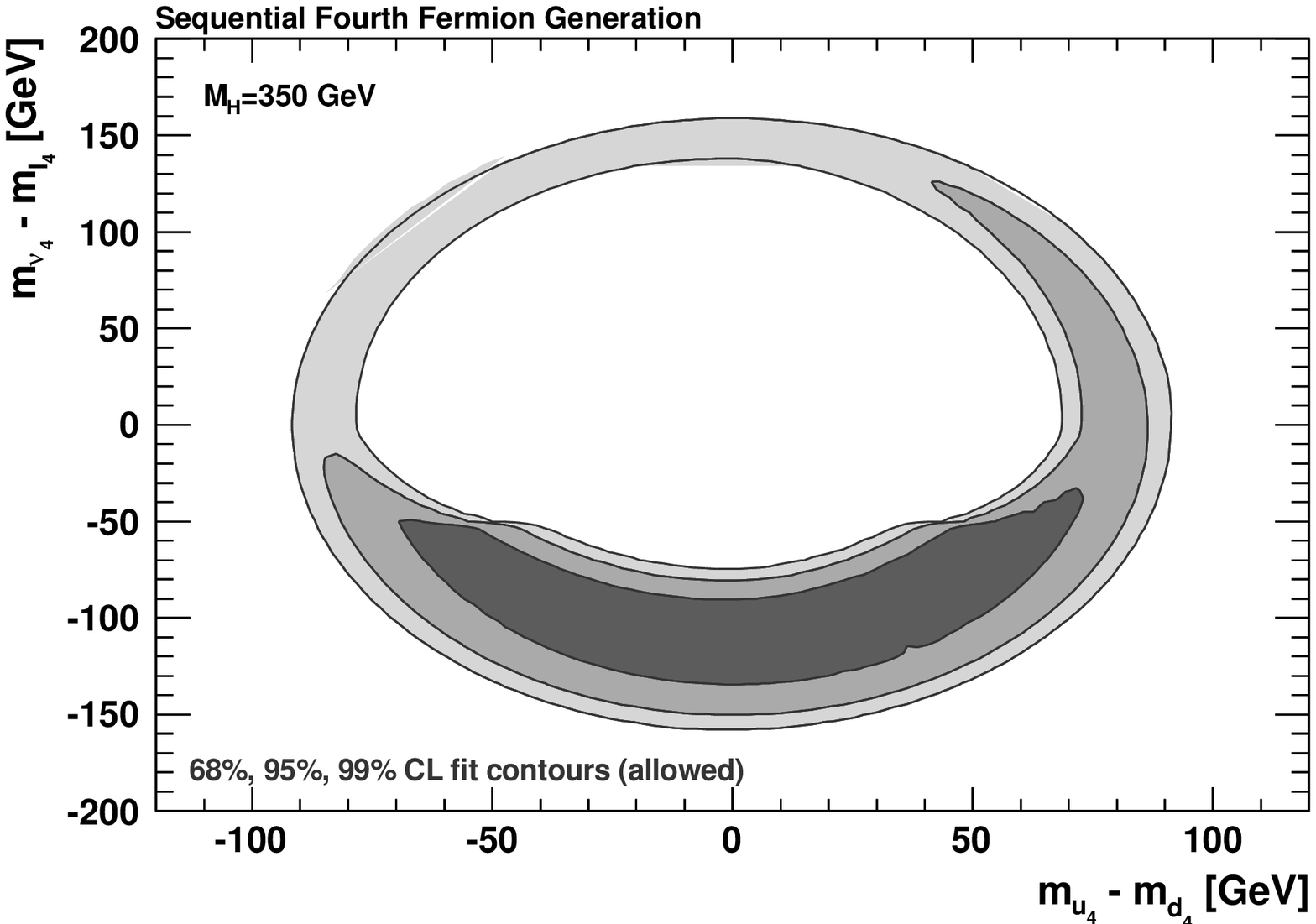, scale=0.4}\\
  \epsfig{file=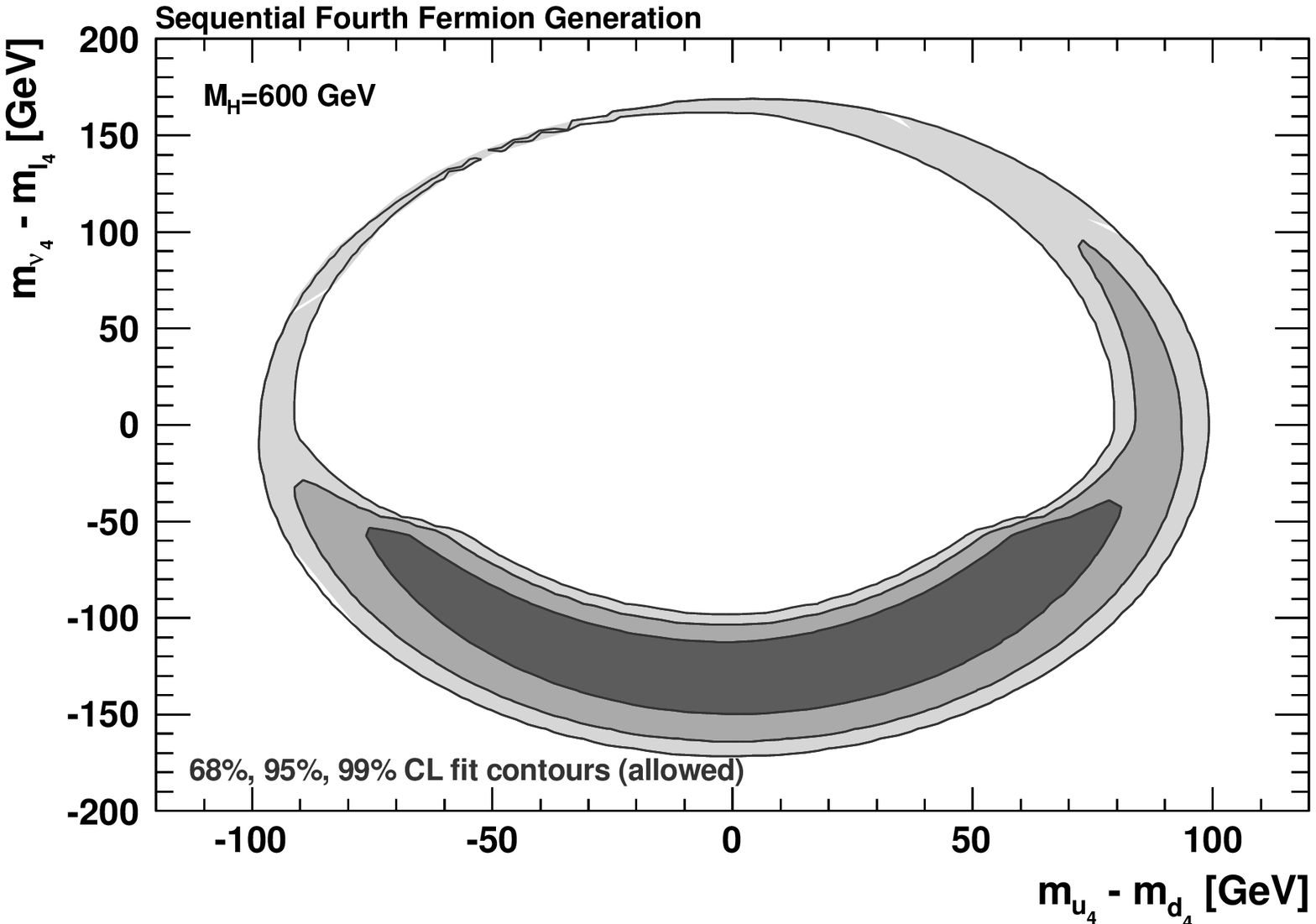, scale=0.4}
  \epsfig{file=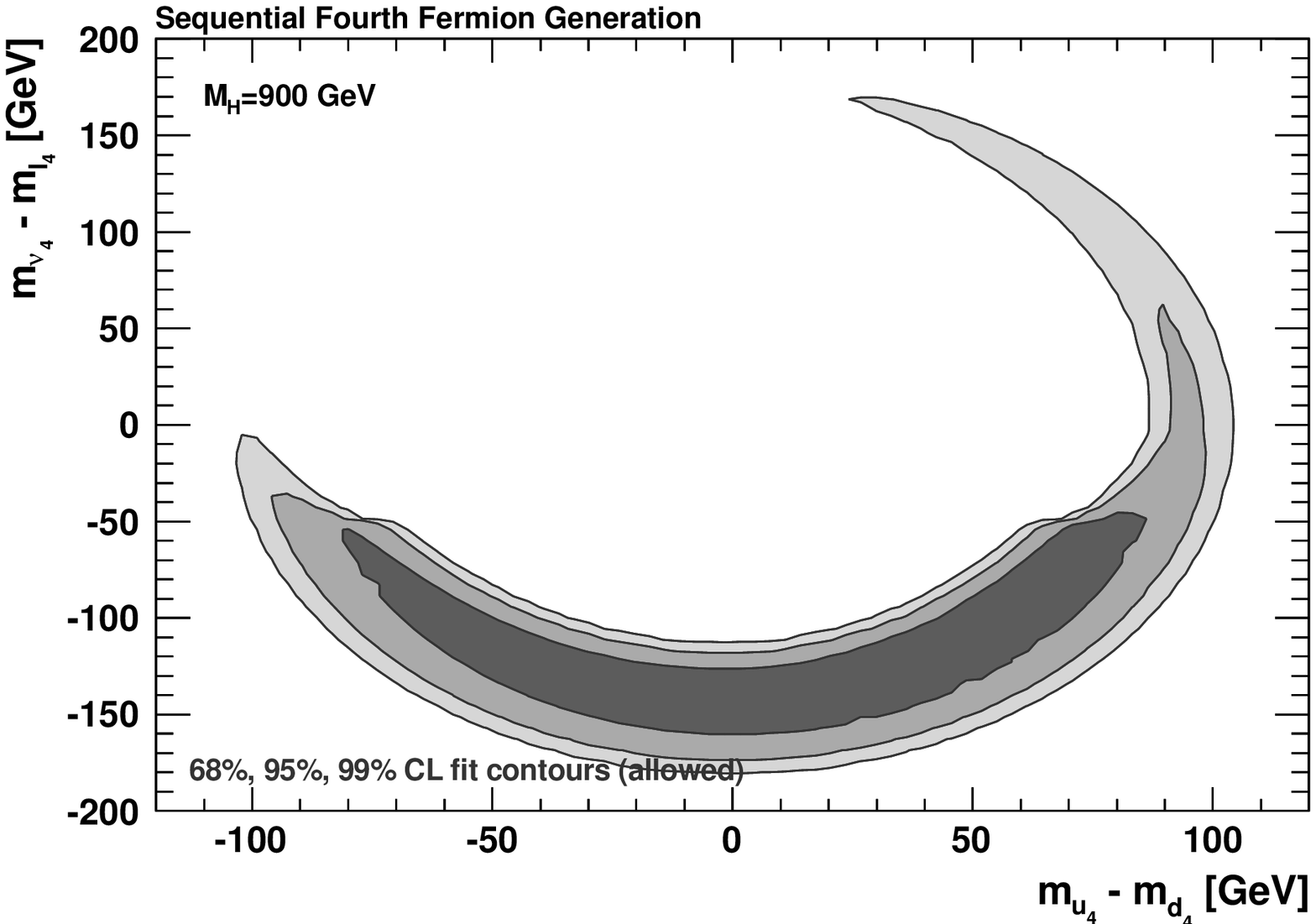, scale=0.4}
  \vspace{0.25cm}
  \caption[]{Constraints in a model with a fourth fermion generation. 
             Shown are the 68\%, 95\% and 99\% CL allowed fit countours 
             in the $(m_{u_4}-m_{d_4}, m_{l_4}-m_{{\nu}_4})$ plane as derived from the
             fit for $M_H=120,\,350,\,600,\,900\:\gev$ (top left to bottom 
             right).}
  \label{fig:4thscans}
\end{figure}
Figure~\ref{fig:4thSvsT} shows the experimental fit result in the $(S,T)$ plane 
for free $U$ together with the prediction from a fourth fermion generation with 
vanishing mixing. The markers indicate special model settings corresponding to the 
fixed masses $m_{\nu_4}=120\:\gev$, $m_{e_4}=200\:\gev$, $m_{d_4}=400\:\gev$, and 
various choices for $m_{u_4}$ and $M_H$.\footnote
{
   Ignoring flavour mixing between the fourth and the SM generations,
   it was found in Ref.~\cite{Hashimoto:2010at} that absolute vacuum 
   stability of the running Higgs self coupling approximately requires 
   the mass hierarchy $M_H \gtrsim m_{u_4}$. This strong lower bound on 
   the Higgs boson mass may possibly be weakened by looser stability 
   requirements. For example, in SM3 the absolute stability lower
   bound on $M_H$ is significantly reduced by allowing the minimum potential to 
   be metastable with finite probability not to have tunnelled into another, 
   deeper minimum during 
   the lifetime of the universe~\cite{Espinosa:2007qp,Ellis:2009tp}. 
   In the following discussion we will ignore the stability bound on $M_H$.    
} 
The $T$ parameter grows with the amount
of the up and down-type fermion mass splitting, while the $S$ parameter 
logarithmically grows with $M_H$ from the SM contribution, prevailing over the 
opposite trend from the increasing $m_{u_4}$. 
The shaded area in Fig.~\ref{fig:4thSvsT} depicts the allowed region when letting 
the fourth generation quark (lepton) masses free to vary within the interval 
$[200,1000]\,\gev$ ($[100,1000]\,\gev$), and $M_H$ within $[100,1000]\,\gev$. 
For specific parameter settings the fourth generation model is in agreement 
with the experimental data, and large values of $M_H$ are allowed. 

Because the oblique parameters are mainly sensitive to the mass differences
between the up-type and down-type fermions instead of their
absolute mass values, we have derived in Fig.~\ref{fig:4thscans} the 68\%, 95\% 
and 99\% CL allowed regions in the $(m_{u_4}-m_{d_4},m_{l_4}-m_{{\nu}_4})$ plane. Shown  
are the constraints obtained for, from the top left to the bottom right panel, 
increasing values of $M_H$. Large $M_H$ values of up to $1\:\tev$ 
can be accommodated by the data if the negative $T$ shift induced by $M_H$ is 
cancelled by a corresponding positive shift from a large fermion mass splitting. 
The data prefer a heavier charged lepton to counterweight the $S$ increase from 
the increasing $M_H$.

A sequential fourth generation of heavy quarks would increase the gluon fusion 
to Higgs production cross section, dominantly mediated by a triangular top loop, 
by approximately a factor of nine, hence increasing the experimental Higgs boson
discovery and exclusion potential. The Tevatron experiments~\cite{Aaltonen:2010sv}, 
ATLAS~\cite{Collaboration:2011qi} and CMS~\cite{Chatrchyan:2011tz} have reinterpreted 
their negative Higgs boson search results in the channel $H\to WW$ in terms of four generations
obtaining the 95\% CL exclusion bounds $131<M_H<204\:$GeV, $140<M_H<185\:$GeV and $144<M_H<207\:$GeV, 
respectively. Inserting these bounds into Fig.~\ref{fig:4thSvsT} does not alter
the allowed $(S,T)$ region of the fourth generation model. It also does not affect
the allowed fermion mass parameters shown in Fig.~\ref{fig:4thSvsT}, which were 
chosen to escape the excluded $M_H$ region.

%% file: 2HDM.tex
\subsection{Two-Higgs Doublet Model}
\label{sec:2hdm}

Two-Higgs doublet models (2HDM)~\cite{Haber:1978jt} are simple
extensions to the SM Higgs sector, which introduce one additional
$SU(2)_L\times U(1)_Y$ Higgs doublet with hypercharge $Y=1$. Two Higgs
doublets lead to five physical Higgs boson states of which three, 
$h^0$, $H^0$, $A^0$, are electrically neutral and the two remaining ones,
$H^{\pm}$, are electrically charged. Of the neutral states, $h^0$ and $H^0$
are scalars and $A^0$ is pseudoscalar. The free parameters of the
2HDM are the Higgs boson masses $M_{h^{0}}$, $M_{H^0}$, $M_{A^0}$ and
$\MHp$, the ratio of the vacuum expectation values of the two Higgs
doublets, $\tanb=v_2/v_1$, occurring in the mixing of charged and
neutral Higgs fields, and the angle $\alpha$ governing the mixing of
the neutral \CP-even Higgs fields.  In the
most general 2HDM $\tanb$ and, hence, the corresponding Higgs couplings
and mass matrix elements depend on the choice of basis for the Higgs
fields~\cite{Haber:2006ue,Davidson:2005cw}.

Models with two Higgs doublets intrinsically fulfil the empirical
equality $M_W^2\approx M_Z^2 \cos^2 \theta_W$. They also increase the
maximum allowed mass of the lightest neutral Higgs boson for
electroweak baryogenesis scenarios to values not yet excluded by LEP
(see \eg, Ref.~\cite{Cline:1996mga}) and they allow for \CP violation in
the Higgs sector. Flavour changing neutral currents can be suppressed
with an appropriate choice of the Higgs-to-fermion couplings (see \eg,
Ref.~\cite{Gunion:1989we,Gunion:1992hs}).  For example, in the
Type-I 2HDM this is achieved by letting only one Higgs doublet
couple to the fermion sector. In the Type-II
2HDM~\cite{Abbott:1979dt} one Higgs doublet couples to the up-type
quarks and leptons only, while the other one couples to the
down-type fermions. The Type-II 2HDM resembles the Higgs
sector of the Minimal Supersymmetric Standard Model. It fixes the
basis of the Higgs fields and promotes \tanb to a physical parameter.

Our previous analysis of the Type-II 2HDM
extension~\cite{Flacher:2008zq} was restricted to the \CP conserving
2HDM scalar potential, and only included observables sensitive to
corrections from the exchange of a charged Higgs boson. The most
constraining of these observables involve rare radiative or leptonic
decays of $B$ and $K$ mesons, where the charged current mediated by
the $W$ is replaced by a charged Higgs. The combination of the
constraints obtained excludes the high-\tanb, low-\MHp region spared
by the $B\to\tau\nu$ constraint, and leads to a 95\% CL charged-Higgs
exclusion below $240\:\gev$, irrespective of the value of $\tanb$.
This limit increases towards larger \tanb, \eg, $\MHp<780\:\gev$ are
excluded for $\tanb=70$ at 95\% CL. A similar analysis, which also
includes neutral \Bz meson mixing, has been reported in
Ref.~\cite{Deschamps:2009rh}.  There, a \tanb independent 95\% CL
lower limit of $316\:\gev$ was achieved.

Direct searches for the charged Higgs within the Type-II 2HDM
have been performed by the LEP collaborations. The main limitations were background from
diboson production and the kinematic limitation on the production cross 
section~\cite{Heister:2002ev,Abreu:2001fu,Achard:2003gt,Abbiendi:1998rd}.
The combined limit determined by the LEP Higgs Working Group is
$M_{H^\pm}>\rm 78.6\:GeV$~\cite{LHWG:2001xy}.

For the study of the 2HDM oblique corrections the type 
distinction between the models is irrelevant as they are defined according 
to the Yukawa couplings, which do not enter the oblique corrections at one-loop
order. For the prediction of the \STU parameters we use the formulas of 
Refs.~\cite{Haber:1993wf,Haber:1999zh,Froggatt:1991qw}\footnote
{
  The functions defined in Eqs.~(\ref{eq:S2hdm}--\ref{eq:U2hdm}) are 
  defined as follows. 
  $\mathcal{B}_{22}(q^2,m_1^2,m_2^2) = q^2/24\{2\ln q^2 +
  \ln(x_1x_2)+[(x_1-x_2)^3-3(x_1^2-x_2^2) +
  3(x_1-x_2)]\ln(x_1/x_2)-\left[2(x_1-x_2)^2-8(x_1+x_2)+10/3\right] -
  [(x_1-x_2)^2-2(x_1+x_2)+1]f(x_1,x_2)-6F(x_1,x_2)\}
  \overset{m_1=m_2}\Rightarrow q^2/24\left[2\ln q^2 + 2\ln x_1 +
    \left( 16x_1 - 10/3\right) +(4x_1-1)G(x_1)\right]$, where $x_i
  \equiv m_i^2/q^2$,
  $ \mathcal{B}_0(q^2,m_1^2,m_2^2) = 1 + 1/2\left[(x_1 + x_2)/(x_1 -
    x_2) - (x_1 - x_2)\right] \ln(x_1/x_2) + 1/2f(x_1, x_2)
  \overset{m_1 = m_2}\Rightarrow 2 - 2y\arctan(1/y), \quad y =
  \sqrt{4x_1 - 1}$,
  $\overline{B}_0(m_1^2,m_2^2,m_3^2) = (m_1^2\ln m_1^2 - m_3^2\ln
    m_3^2)/(m_1^2-m_3^2) - (m_1^2\ln m_1^2 - m_2^2 \ln
    m_2^2)/(m_1^2-m_2^2)$~\cite{He:2001tp}, see also Footnote~\ref{footnote:formula}
    on page~\pageref{footnote:formula}.
}.
\beqn
\label{eq:S2hdm}
S &= & \frac{1}{\pi M_{Z}^2} \bigg\{\sin^2(\beta-\alpha) \mathcal{B}_{22}(M_{Z}^2, M_{H^0}^2,M_{A^0}^2) - \mathcal{B}_{22}(M_{Z}^2, M_{H^{\pm}}^2,M_{H^{\pm}}^2) \nonumber \\
 && +\; \cos^2(\beta-\alpha)\bigg[\mathcal{B}_{22}(M_{Z}^2, M_{h^{0}}^2,M_{A^0}^2)+\mathcal{B}_{22}(M_{Z}^2, M_{Z}^2,M_{H^0}^2)-\mathcal{B}_{22}(M_{Z}^2, M_{Z}^2,M_{h^{0}}^2)\nonumber \\
 && -\; M_{Z}^2\mathcal{B}_0(M_{Z}^2, M_{Z}^2,M_{H^0}^2)+M_{Z}^2\mathcal{B}_0(M_{Z}^2, M_{Z}^2,M_{h^{0}}^2)\bigg] \bigg\}\,,\\
%
T &= & \frac{1}{16\pi M_{W}^2 \sin^2\theta_W}\bigg\{F(M_{H^{\pm}}^2,M_{A^0}^2)+\sin^2(\beta-\alpha)\bigg[F(M_{H^{\pm}}^2,M_{H^0}^2)-F(M_{A^0}^2,M_{H^0}^2)\bigg] \nonumber \\
 && +\; \cos^2(\beta-\alpha)\bigg[F(M_{H^{\pm}}^2,M_{h^{0}}^2)-F(M_{A^0}^2,M_{h^{0}}^2)+F(M_{W}^2,M_{H^0}^2)-F(M_{W}^2,M_{h^{0}}^2) \\
 && - F(M_{Z}^2,M_{H^0}^2) + F(M_{Z}^2,M_{h^{0}}^2) + 4 M_{Z}^2\overline{B}_0(M_{Z}^2, M_{H^0}^2,M_{h^{0}}^2) - 4 M_{W}^2 \overline{B}_0(M_{W}^2, M_{H^0}^2,M_{h^{0}}^2)\bigg]\bigg\}\,, \nonumber
\eeqn
\beqn
\label{eq:U2hdm}
U &= & -S + \frac{1}{\pi M_{Z}^2} \bigg\{ \mathcal{B}_{22}(M_{W}^2, M_{A^0}^2,M_{H^{\pm}}^2)-2\mathcal{B}_{22}(M_{W}^2, M_{H^{\pm}}^2,M_{H^{\pm}}^2) \nonumber \\
 && +\; \sin^2(\beta-\alpha)\mathcal{B}_{22}(M_{W}^2, M_{H^0}^2,M_{H^{\pm}}^2) \nonumber \\
 && +\; \cos^2(\beta-\alpha)\bigg[\mathcal{B}_{22}(M_{W}^2, M_{h^{0}}^2,M_{H^{\pm}}^2)+\mathcal{B}_{22}(M_{W}^2, M_{W}^2,M_{H^0}^2)-\mathcal{B}_{22}(M_{W}^2, M_{W}^2,M_{h^{0}}^2)\nonumber \\
 && -\;  M_{W}^2 \mathcal{B}_0(M_{W}^2, M_{W}^2,M_{H^0}^2)+M_{W}^2\mathcal{B}_0(M_{W}^2, M_{W}^2,M_{h^{0}}^2)\bigg] \bigg\}\,.
\eeqn

\begin{figure}[t]
  \centering \epsfig{file=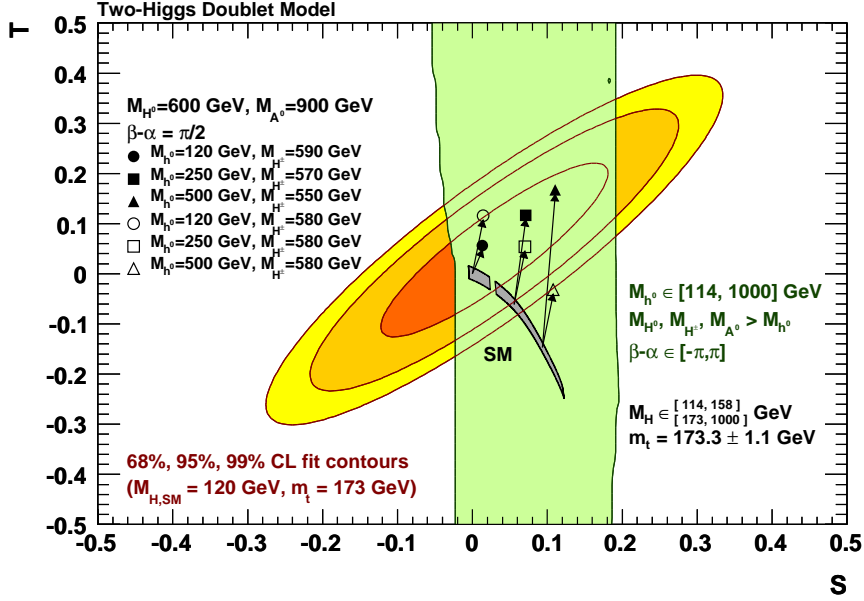, scale=\defaultFigureScale} 
  \vspace{0.2cm}
  \caption[]{Oblique parameters in the 2HDM. Shown are
             the $S$, $T$ fit results (leaving $U$ free) compared with
             predictions from the SM (grey) and 2HDM (light green).  The 2HDM
             area is obtained with the use of the mass and mixing parameter
             ranges given on the plot. The symbols illustrate the 2HDM
             predictions for six example settings, compared to the corresponding
             SM predictions via the arrows. }
   \label{fig:2HDMSvsT}
\end{figure}

Figure~\ref{fig:2HDMSvsT} shows the 68\%, 95\%, and 99\% CL allowed
contours in the ($S,T$)-plane (letting $U$ vary freely) as derived in
the electroweak fit together with the SM and 2HDM predictions (grey
and green areas, respectively). For the 2HDM prediction $M_{h^{0}}$ was
left free to vary within [114,1000]\:GeV and the masses of the other
Higgs bosons were allowed to vary between $M_{h^{0}}$ and $1000\:$GeV.
$S$ adopts relatively small and mainly positive values, whereas the
contribution to $T$ can take large positive and negative values. There
is a large overlap between the experimental fit and the 2HDM
prediction, so that a variety of model configurations exhibits
compatibility with the electroweak precision data. 
A few of these
configurations are shown for fixed values of $M_{h^{0}} = \rm 600\:GeV$,
$M_{A^0}=\rm 900\:GeV$, and $\beta-\alpha =\frac{\pi}{2}$ in
Fig.~\ref{fig:2HDMSvsT}. The open symbols depict the predictions for
three different masses of the lightest Higgs ($M_{h^{0}} = \rm
120,\,250,\,500\:GeV$) and a fixed charged Higgs mass of $580\:$GeV.
The arrows indicate the 2HDM-induced shifts in $S$ and $T$ with
respect to the SM prediction for the same $M_{h^{0}}$ values. Variations
of the charged Higgs mass (full symbols) induce strong effects on $T$.
By choosing adequate values ($M_{H^{\pm}} = \rm 590,\,570,\,550\:GeV$)
compatibility with the electroweak data can be achieved even for large
$M_{h^{0}}$.
\begin{figure}[t!]
   \epsfig{file=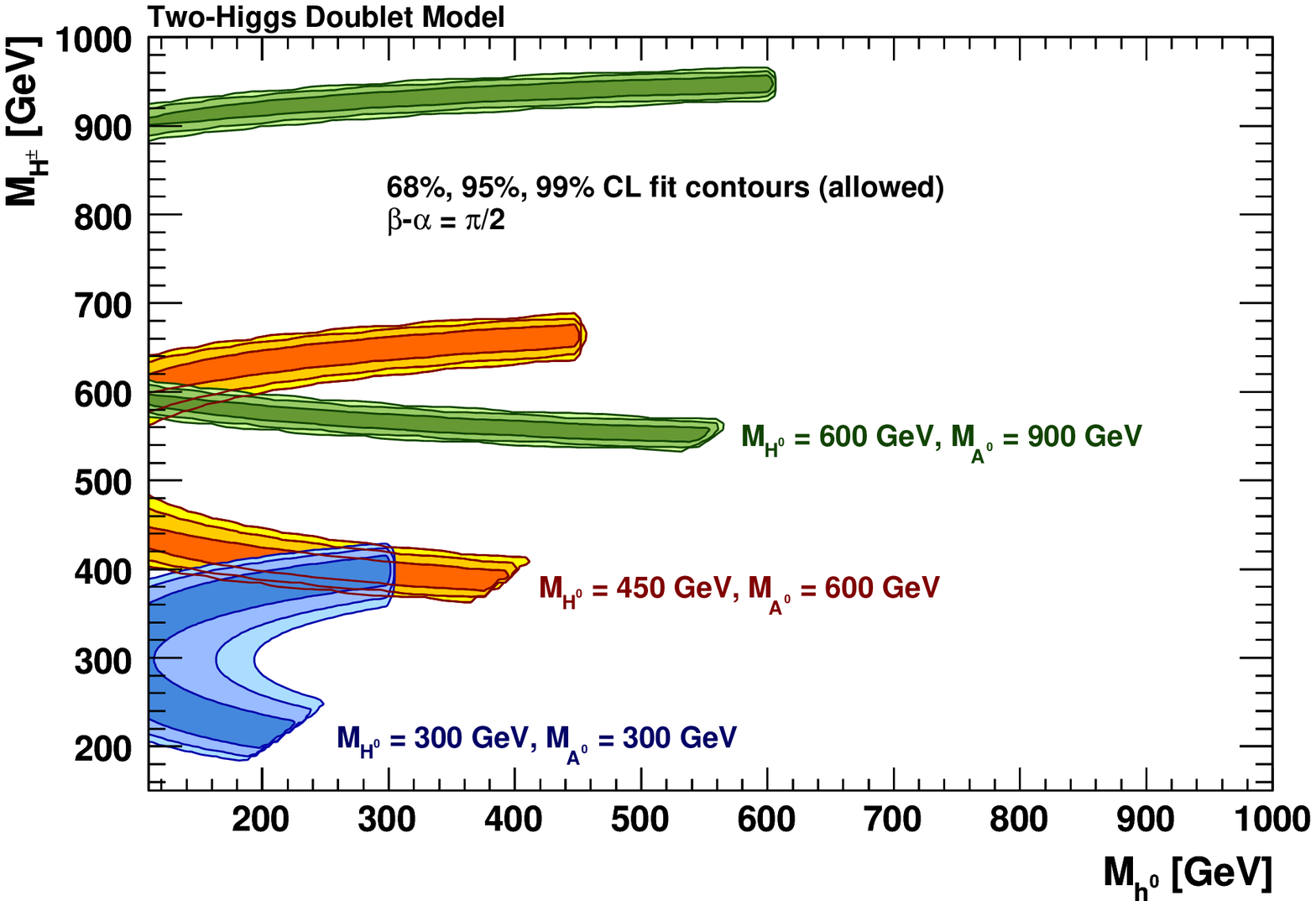, scale=\HalfPageWidthScale}
   \epsfig{file=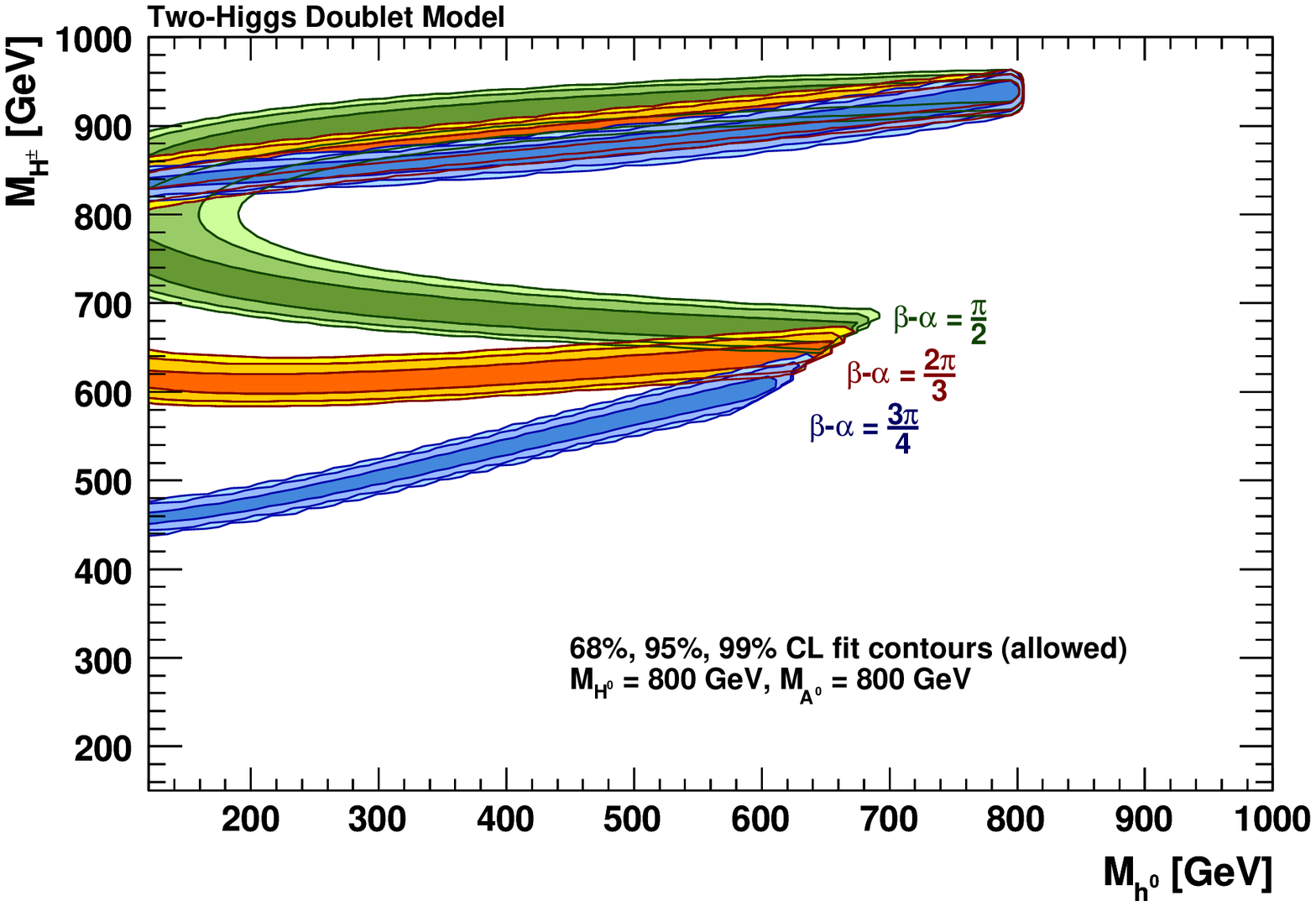, scale=\HalfPageWidthScale}

  \vspace{0.2cm}
   \epsfig{file=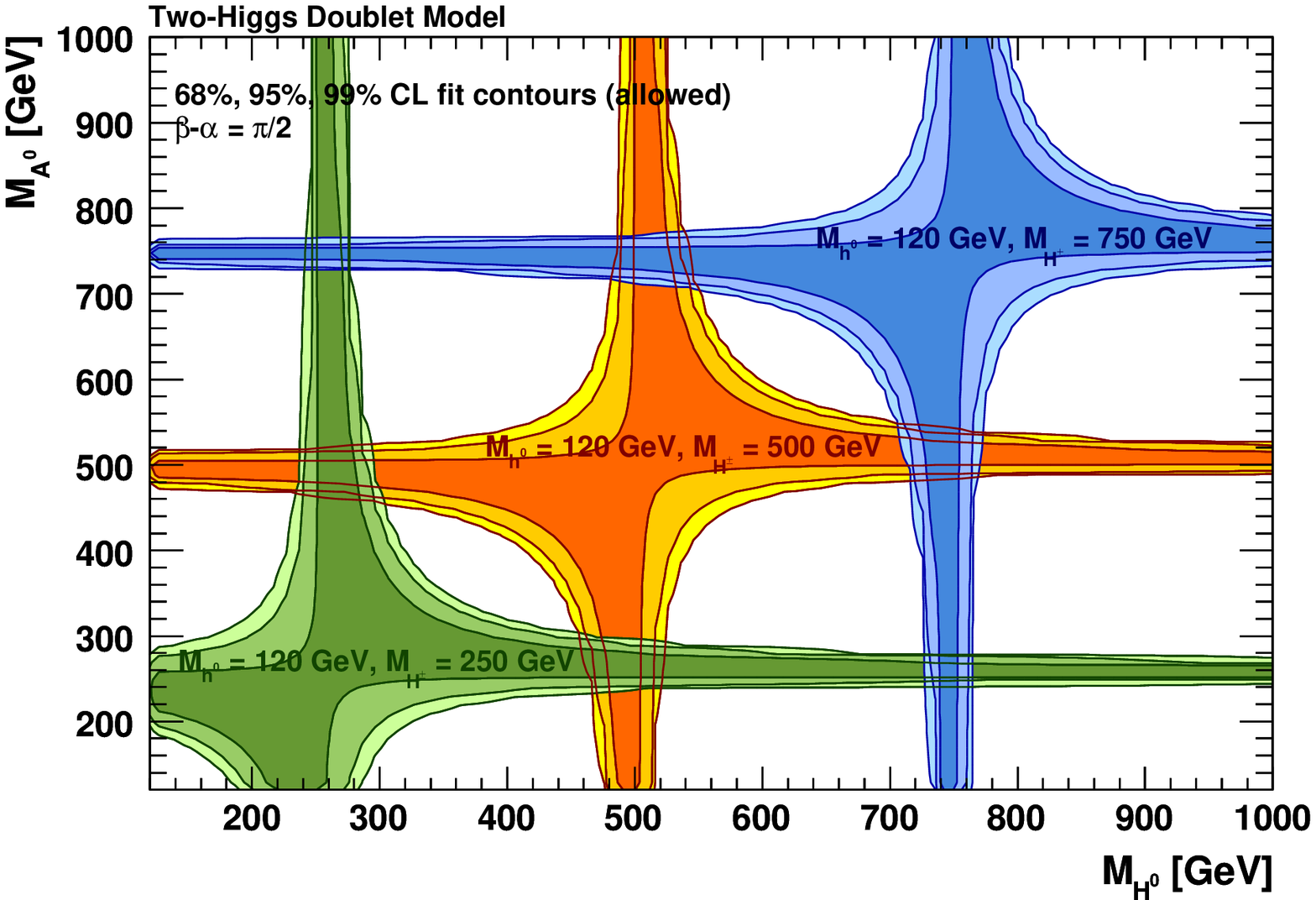, scale=\HalfPageWidthScale}
   \epsfig{file=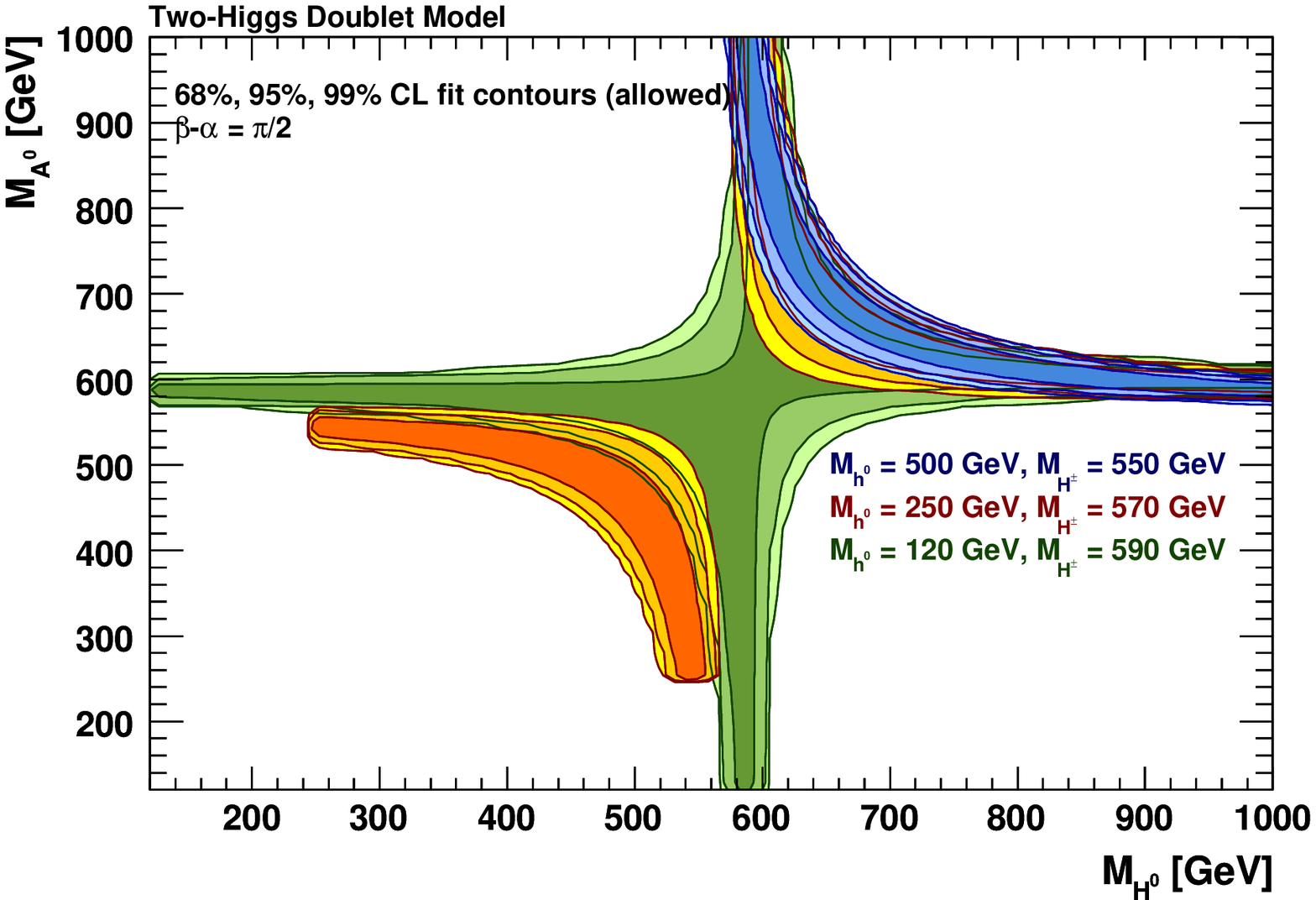, scale=\HalfPageWidthScale}
  \vspace{-0.3cm}
  \caption[]{Constraints in the 2HDM. Top panels: 68\%, 95\%, and
    99\% CL allowed fit contours in the ($M_{h^{0}},M_{H^{\pm}}$) plane
    as derived from the fit for $M_{H^0} = \rm 300,\,450,\,600\:GeV$
    and $M_{A^0} = \rm 300,\,600,\, 900\:GeV$ and
    $\beta-\alpha=\frac{\pi}{2}$ (left), and for $M_{H^0} = \rm
    800\:GeV$, $M_{A^0} = \rm 800\:GeV$, and $\beta-\alpha =
    \frac{\pi}{2}, \frac{2\pi}{3}, \frac{3\pi}{4}$, respectively (right).
    Bottom panels: 68\%, 95\%, and
    99\% CL allowed fit contours in the ($M_{H^0},M_{A^0}$) plane for
    $\beta-\alpha=\frac{\pi}{2}$ as derived from the fit for $M_{h^{0}}
    = \rm 120\:GeV$ and $M_{H^0} = \rm 250,\,500,\,750\:GeV$ (left),
    and for $M_{h^{0}} = \rm 120,\,250,\,500\:GeV$ and $M_{H^{\pm}} =
    \rm 590, 570, 550\:GeV$, respectively (right).}
       \label{fig:2HDM}
\end{figure}

Further 2HDM parameter configurations that are allowed by the electroweak
data are shown in Fig.~\ref{fig:2HDM}.  For fixed 
$M_{H^0}$, $M_{A^0}$, and $\beta-\alpha$, only two small 
bands of $M_{H^{\pm}}$ are allowed, namely masses very similar to either
$M_{H^0}$ or $M_{A^0}$, whereas $M_{h^{0}}$ cannot be constrained (see
Fig~\ref{fig:2HDM} (top left)) other than being the lightest Higgs boson. 
Towards closer $M_{H^0}$ and $M_{A^0}$ degeneracy the allowed bands for $M_{H^{\pm}}$
become broader. The widths of the bands also depend on the
error of $m_{t}$ and other relevant electroweak parameters. Varying
$\beta-\alpha$ (see Fig~\ref{fig:2HDM} (top right)) 
alters the preference of the charged Higgs to adopt
similar values as $M_{H^0}$ and $M_{A^0}$ slightly, preserving small
bands of allowed masses for $M_{H^{\pm}}$ but yielding an overall
wider range of masses. 

Figure~\ref{fig:2HDM} (bottom left) shows the ($M_{H^0},M_{A^0}$)-plane for
fixed $M_{h^{0}}=120$\:GeV and fixed $\beta-\alpha=\pi/2$ and varying 
$M_{H^{\pm}}$. Here, too, one notices that, for either $M_{H^0}$ or $M_{A^0}$,
similar values compared to $M_{H^{\pm}}$ are preferred, while the other
mass is hardly constrained.
This almost independent behaviour of $M_{H^0}$ and $M_{A^0}$ changes
slightly for heavier $M_{h^{0}}$ values, as illustrated in
Fig.~\ref{fig:2HDM} (bottom right). 
The larger $M_{h^{0}}$, the less freedom have $M_{H^0}$ and $M_{A^0}$ to 
adopt any value, whilst the other mass is fixed to a similar value of 
$M_{H^{\pm}}$. 
In these plots, the same values for $M_{h^{0}}$ and $M_{H^{\pm}}$ have been 
chosen as in Fig.~\ref{fig:2HDMSvsT}. 
The allowed fit contours clearly overlap for the above selected values of 
$M_{H^0} = \rm 600\:\gev$ and $M_{A^0}= \rm 900\:\gev$, indicating the 
compatibility of all three model configurations with the electroweak 
precision data.

Although the oblique parameter fits do not allow to determine any of the free 
2HDM parameters independently of the values of the other parameters, the electroweak 
precision constraints will become relevant in case of a discovery or the 
setting of significant 2HDM Higgs boson exclusion limits at the LHC.

%% file: InertHiggs.tex
%
%
%

\subsection{Inert-Higgs Doublet Model}
\label{sec:constraintsIH}

\newcommand{\Mh}{\ensuremath{M_{h^0}}\xspace}
\newcommand{\MA}{\ensuremath{M_{\!A^0}}\xspace}
\newcommand{\MH}{\ensuremath{M_{H^0}}\xspace}
\newcommand{\MHch}{\ensuremath{M_{H^\pm}}\xspace}

The inert-Higgs doublet model (IHDM) has recently been
re-introduced~\cite{Barbieri:2006dq} (see the original
paper~\cite{Deshpande:1977rw}) with the aim to accommodate a heavy
Higgs boson of mass between 400 and $600\:\gev$ that would lift the
divergence of the Higgs radiative corrections beyond the TeV scale,
where new physics is supposed to render the theory natural and the
Higgs quartic coupling perturbative. To respect the constraints from
the electroweak precision data, a second {\em inert} Higgs doublet,
$H_2$, is introduced. Although $H_2$ has weak and quartic interactions
just as in the ordinary 2HDM, it does not acquire a vacuum expectation
value (its minimum is at (0,0)), nor has it any other couplings to
matter. The IHDM therefore belongs to the class of Type-I 2HDMs. The
$H_2$ doublet transforms odd under a novel unbroken parity symmetry,
$Z_2$, while all the SM fields have even $Z_2$ parity.  As a
consequence, the lightest inert scalar (LIP) is stable and a suitable
dark matter candidate. To escape detection it should be electrically
neutral. The literature distinguishes three different LIP mass 
regions~\cite{Barbieri:2006dq,LopezHonorez:2006gr,Gustafsson:2007pc,Lundstrom:2008ai,Hambye:2009pw,Andreas:2009hj,Dolle:2009fn,Honorez:2010re,LopezHonorez:2010tb},  low mass 
(few\:\gev),  intermediate mass (40--160\:\gev), and  high
mass  (above 500\:\gev), a convention that we follow in the 
present analysis. 

Besides the SM-like Higgs, $h^0$, the remaining degree of freedom of the mass giving doublet $H_1$, 
the inert doublet $H_2$ enriches the scalar sector by two charged Higgs states of equal mass, $H^\pm$,  
and two neutral ones, $H^0$, $A^0$, where the lightest neutral state, which could be either $H^0$ 
or $A^0$, is typically assumed to be the LIP. The parameters of the extended sector are the three 
Higgs masses, $\MHch$, $\MH$, $\MA$, and two quartic couplings. One of the quartic 
couplings only affects the inert particles while the other one, involving both Higgs 
doublets, affects measurable observables~\cite{Barbieri:2006dq}. 

The oblique corrections induced by the IHDM have been computed in Ref.~\cite{Barbieri:2006dq}. 
They read
\beqn
\label{eq:ihdmS}
   S &=& \frac{1}{2\pi}
         \left(
                 \frac{1}{6}\ln\frac{\MH^2}{\MHch^2}
               - \frac{5}{36} + \frac{\MH^2\MA^2}{3(\MA^2-\MH^2)^2}
               + \frac{\MA^4 (\MA^2 - 3\MH^2)}{6(\MA^2-\MH^2)^3}
                 \ln\frac{\MA^2}{\MH^2}
         \right) \\[0.3cm]
\label{eq:ihdmT}
   T &=& \frac{1}{32\pi^2 \alpha v^2}
       \bigg(F(\MHch,\MH) + F(\MHch,\MA) - F(\MA,\MH)\bigg)\,,
\eeqn
where $F(m_1,m_2)=(m_1^2+m_2^2)/2 - m_1^2m_2^2/(m_1^2-m_2^2)\cdot\ln(m_1^2/m_2^2)$. 
The function $F$ is positive, symmetric with respect to an 
interchange of its arguments, and it vanishes for $m_1=m_2$. For approximate 
$H^0$, $A^0$ mass degeneracy one finds $T\propto(\MHch-\MH)(\MHch-\MA)$~\cite{Barbieri:2006dq}. 
Contributions to the $U$ oblique parameter are neglected.

\begin{figure}[t]
  \centerline{\epsfig{file=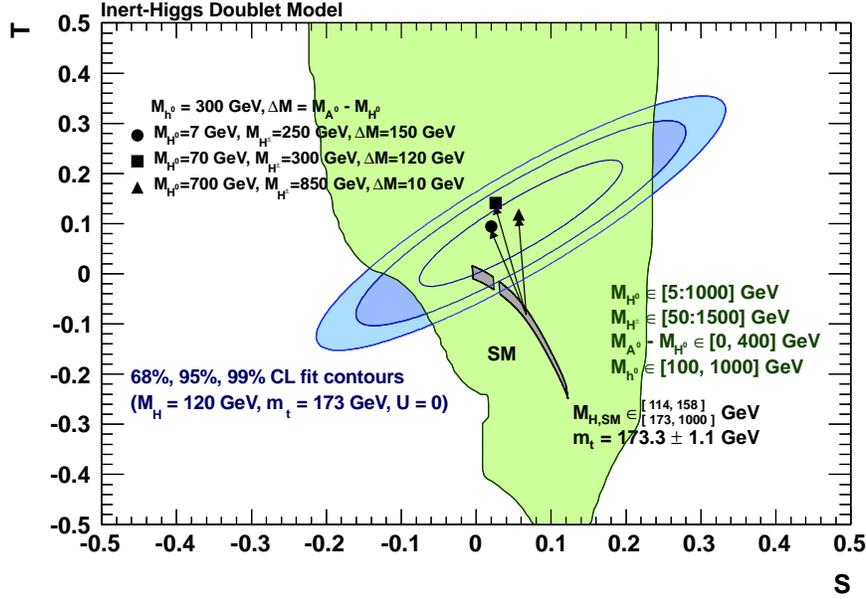, scale=\defaultFigureScale}}
  \vspace{0.2cm}
  \caption[]{Oblique parameters in the inert-Higgs doublet model. 
             Shown are the $S$, $T$ fit results (with $U=0$) compared
             with predictions from the SM and IHDM (grey and light green areas,
             respectively). The IHDM area is obtained with the use of the mass  
             parameter ranges given on the figure. The symbols illustrate 
             the IHDM predictions for three example settings, compared 
             to the corresponding SM predictions via the arrows.}
       \label{fig:IDM1}
\end{figure}

The IHDM predictions for $S$ and $T$ are shown in Fig.~\ref{fig:IDM1}. The 
solid circle, square and triangle indicate oblique corrections for three
representative $H_2$ mass parameter settings. The light shaded (green) area 
depicts the allowed region found for freely varying masses within the 
bounds: 
$100\:\gev<\Mh<1000\:\gev$, 
$50\:\gev<\MHch<1500\:\gev$, 
$5\:\gev<\MH<1000\:\gev$, and
$0<\MA-\MH<400\:\gev$,
assuming $\MA>\MH$ and $\MHch>\MH$. By construction, the IHDM grants
large $h^0$ masses. 

\begin{figure}[t]
   \epsfig{file=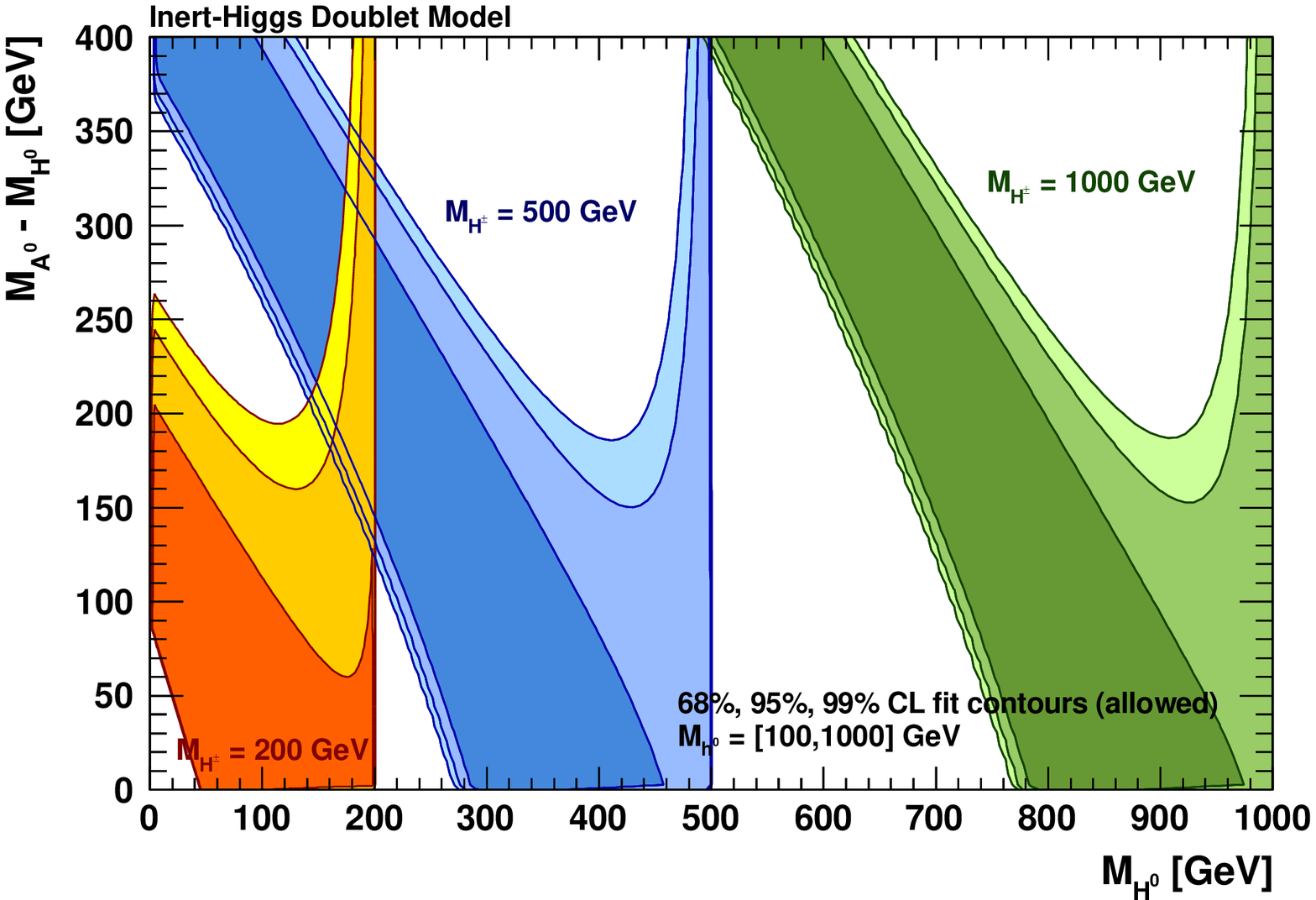, scale=\HalfPageWidthScale}
   \epsfig{file=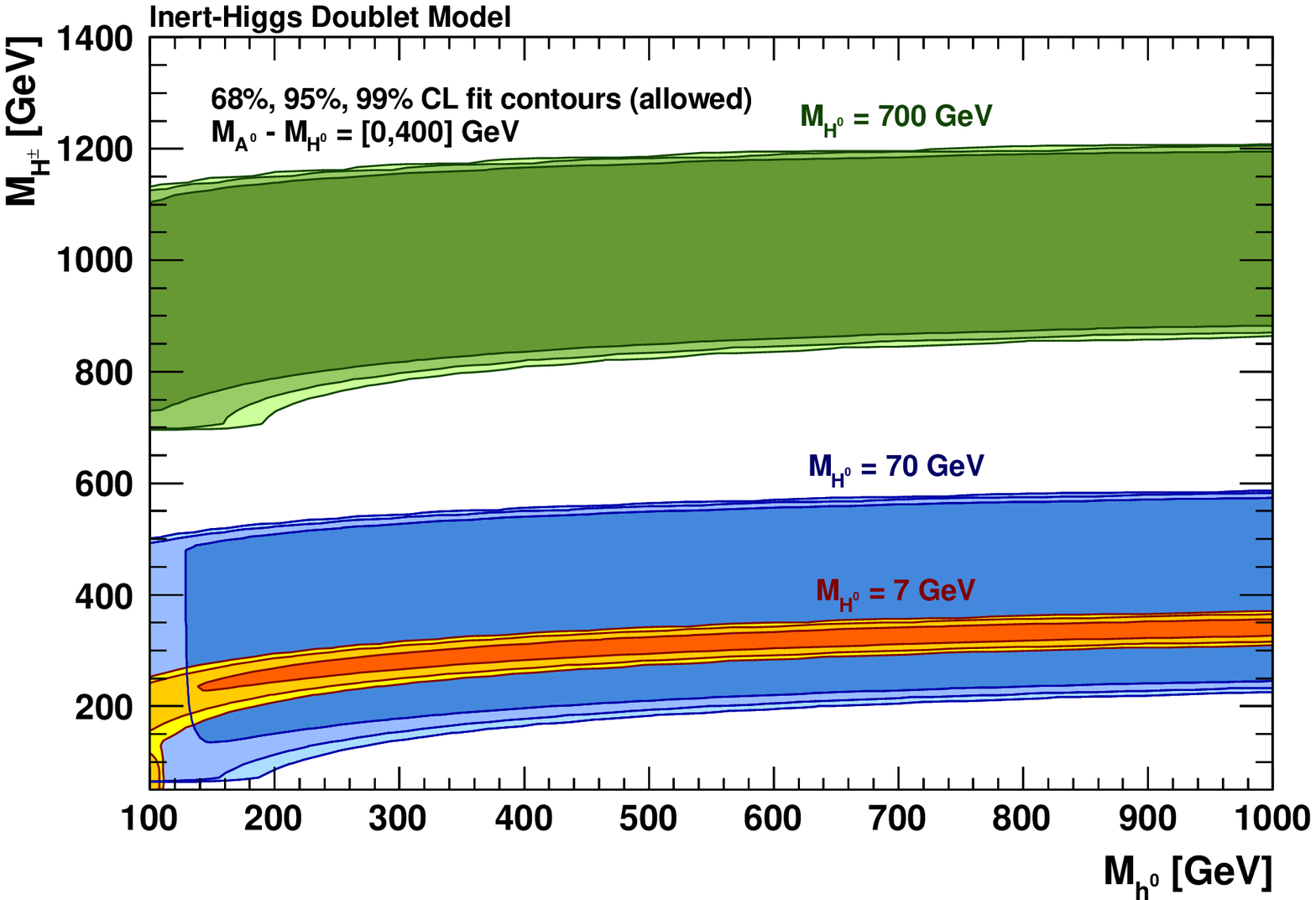, scale=\HalfPageWidthScale}
  \vspace{-0.25cm}
  \caption[]{Constraints in the inert-Higgs doublet model. Shown are the 
             68\%, 95\%, and 99\% CL allowed fit contours in the 
             ($M_{A^0}-M_{H^0},\,M_{H^0}$) (left) and ($M_{H^0},\,M_{h^{0}}$) (right)
             planes for the ranges of the other parameters given on the plots. }
       \label{fig:IDM2}
\end{figure}
Figure~\ref{fig:IDM2} translates the oblique parameter constraints from the 
electroweak precision data into constraints on the masses of the extended sector. 
There is a large freedom in the choice of the parameters. The neutral LIP 
requirement leads to the sharp vertical bound in the left hand plot of 
Fig.~\ref{fig:IDM2}. The constraint from $T$ puts bounds on the mass
splitting between the inert Higgs states. For the case of an almost mass 
degeneracy between charged Higgs and LIP, for which $\delta T_{\rm IHDM}$ 
approximately vanishes, $M_{A^0}$ is unconstrained. For large $M_{h^{0}}$ the 
allowed values for $M_{H^\pm}$ are approximately independent of $M_{h^{0}}$, but 
must rise along with $M_{H^0}$ and $M_{A^0}$.

%% file: LittlestHiggs.tex
\subsection{Littlest Higgs model with {\em T}-parity conservation}
\label{sec:constraintsLH}

An approach to realising a naturally light Higgs boson and tackle the
SM hierarchy problem are models in which the Higgs boson is a bound state of
more fundamental constituents interacting via a new strong
force~\cite{Kaplan:1983fs,Kaplan:1983sm,Georgi:1984ef,Georgi:1984af}.
Analogous to the pions in QCD, the Higgs is a pseudo-Goldstone boson in these 
models, generated by the spontaneous breaking of a global symmetry of the new
strong interaction. However, in these models the {\it little}
hierarchy between the symmetry breaking scale $f$ and the electroweak
scale cannot be realised without fine tuning. The new mechanism to
stabilise the little hierarchy is {\it collective} symmetry
breaking~\cite{ArkaniHamed:2001nc} of
several global symmetries. Under each symmetry alone the Higgs is a
Goldstone boson. However, the symmetries are only approximate;
they are broken explicitly by gauge, Yukawa and scalar couplings.
Quadratically divergent Higgs mass corrections can only occur if the
symmetries are broken at multi-loop level, featuring a light 
pseudo-Goldstone boson, denoted {\it little Higgs}~\cite{ArkaniHamed:2002pa}.
A common feature of little Higgs theories is a new global symmetry
broken at a scale $f\sim1$\:TeV where new gauge bosons, fermions and
scalars exist that cancel the one-loop quadratic divergences of $M_H$
in the SM. Evidence for the existence of these states can be searched for
directly at high-energy colliders and indirectly by exploiting their 
corrections to precisely measured observables such as the electroweak data.

The littlest Higgs (LH) model~\cite{ArkaniHamed:2002qy} is among the
simplest little Higgs realisations with a minimal
particle content. It is based on a non-linear 1$\sigma$ model
describing $SU(5)/SO(5)$ symmetry breaking at a scale $f$
of order TeV. The particle spectrum below this scale consists of the
SM states and a light Higgs boson, while at the TeV scale
a few new states are introduced. At an energy cut-off
$\Lambda=4\pi f\sim10\:$TeV the non-linear 1$\sigma$ model becomes strongly
coupled and the LH model needs to be replaced by a more fundamental
theory.  The originally proposed littlest Higgs models were found to 
provide large corrections to the precision electroweak observables, 
mainly due to the allowed tree-level exchange of the new heavy gauge
bosons~\cite{Csaki:2003si,Barbieri:2004qk,Perelstein:2003wd,Csaki:2002qg,Hewett:2002px,Han:2003wu}.
These problems were solved with the introduction of a conserved discrete 
symmetry, called {\it T-parity}~\cite{Cheng:2003ju,Cheng:2004yc}, featuring 
$T$-odd partners for all ($T$-even) 
SM particles, and a lightest $T$-odd particle that is stable.\footnote
{
   It has been shown~\cite{Hubisz:2004ft} that the $T$-odd partner of the 
   hypercharge gauge boson (the {\it heavy photon}) can give rise to the 
   observed relic density of the universe.
}
As a result tree-level contributions of the heavy gauge bosons to the 
electroweak precision observables are suppressed and corrections arise only
at loop level.

The study presented here follows the analysis 
of Ref.~\cite{Hubisz:2005tx}, where the dominant oblique 
corrections in the LH model with $T$-parity~\cite{Cheng:2004yc}
were calculated together with the $Zb\bbar$ vertex correction 
from the top sector. The largest oblique corrections result from 
one-loop diagrams of a new $T$-even top state $T_+$ which mixes 
with the SM top quark. In the limit $m_t \ll m_{T_+}$ these corrections 
are given by~\cite{Hubisz:2005tx}
\begin{eqnarray}
\label{eq:LHS}
S_{T_+} & \!\!=\!\! & \frac{1}{3 \pi} \left( \frac{1}{s_{\lambda}^2}-1 \right)
\frac{m_t^2}{m_{T_+}^2} \left(
  -\frac{5}{2} + \ln\! \frac{m_{T_+}^2}{m_t^2} \right)\,, \\
T_{T_+}  & \!\!=\!\! & \frac{3}{8 \pi} \frac{1}{\sin^2\!\theta_W \cos^2\!\theta_W}\,\left( \frac{1}{s_{\lambda}^2}-1 \right)\, \frac{m_t^4}{m_{T_+}^2 M_Z^2}
\,\left( \ln\! \frac{m_{T_+}^2}{m_t^2} - \frac{3}{2} +\frac{1}{2 s_{\lambda}^2} \right)\,, \\
\label{eq:LHU}
U_{T_+}  & \!\!=\!\! &
\frac{5}{6 \pi} \left( \frac{1}{s_{\lambda}^2}-1 \right)
\frac{m_t^2}{m_{T_+}^2}\,,
\end{eqnarray}
and
\beq
  m_{T_+}\ =\ m_t
  \sqrt{\frac{1}{s_{\lambda}^2\left(1-s_{\lambda}^2\right)}}\cdot 
  \frac{f}{v}\,,
\eeq
where $v$ is the Higgs vacuum expectation value and $f$ the symmetry
breaking scale. The parameter $s_{\lambda}$ is approximately the mass
ratio of the new $T$-odd and $T$-even top states, $s_{\lambda} \approx
m_{T^-}/m_{T^+}$,\footnote
{
   The parameter $s_{\lambda}$ is defined by 
   $s_{\lambda} = \lambda_2/\sqrt{\lambda_1^2+\lambda_2^2}$,
   where $\lambda_1$ and $\lambda_2$ are the Yukawa couplings of
   the new top states.
} 
which is restricted by the model to be
smaller than one. The $T$ parameter dominates over $S$ and $U$ by a 
factor of $\sim m_t^2/(\sin^2\!\theta_W \cos^2\!\theta_W M_Z^2)\sim 20$. 
Similar to the other new physics models discussed in this paper, the 
contribution to the $T$ parameter from $T_+$ loops in the LH model is 
positive and can thus cancel a negative SM correction due to a large $M_H$.

\begin{figure}[t]
  \centering \epsfig{file=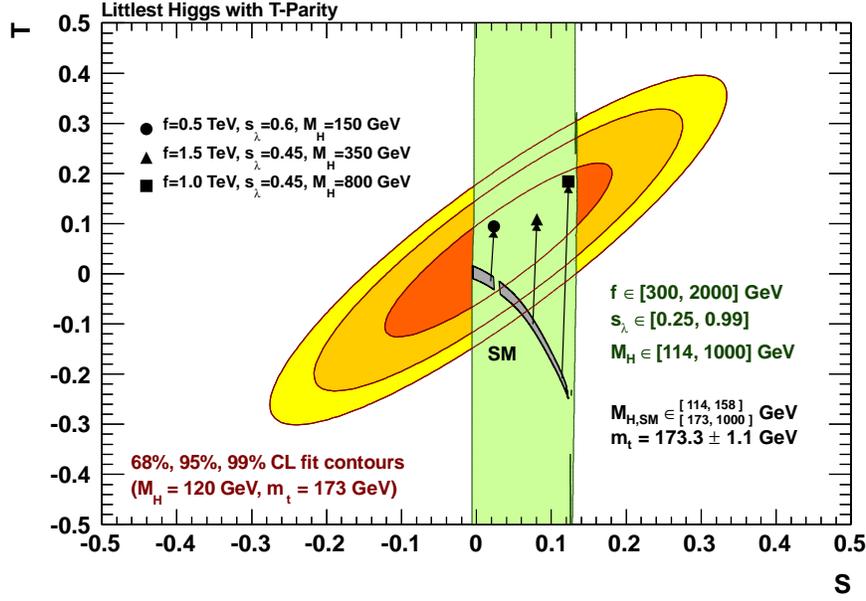,scale=\defaultFigureScale} 
  \vspace{0.2cm}
  \caption[]{Oblique parameters in the littlest Higgs model with $T$-parity
             conservation. 
             Shown are the $S$, $T$ fit results (without $U$ constraint) compared
             to predictions from the SM and the littlest Higgs model (grey and light green areas,
             respectively). The green area is obtained with the use of the 
             parameter ranges given on the figure. The symbols illustrate 
             the LH predictions for three example settings of the parameters $f$,
             $s_{\lambda}$ and $M_H$. The contribution from $T$-odd fermions is neglected.}
   \label{fig:LHSvsT}
\end{figure}

\begin{figure}[t]
  \centering \epsfig{file=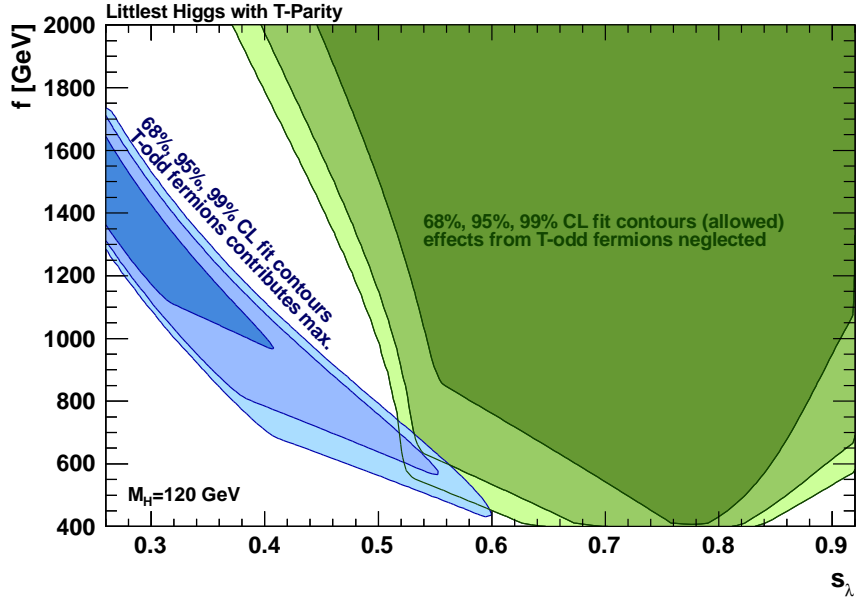, scale=\defaultFigureScale} 
  \vspace{0.2cm}
  \caption[]{Constraints in the littlest Higgs model with $T$-parity conservation. 
             Shown are the 68\%, 95\%, and 99\% CL allowed fit contours in the
             ($s_\lambda,f$) plane. The largest/green allowed regions are the results
             of a fit neglecting contributions from the $T$-odd partners of the light
             fermions to the $T$ oblique parameter. In the narrowest/blue allowed
             regions the $T$-odd fermion contribution is considered to have the
             maximal size consistent with the bound from four-fermion contact 
             interaction~(\ref{eq:toddf}). In both cases $M_H=120$\:GeV is assumed. }
  \label{fig:LHfvsSL}
\end{figure}
The oblique corrections~(\ref{eq:LHS}--\ref{eq:LHU}) vanish when $t$--$T_+$
mixing is suppressed (\ie for small values of $1/s_{\lambda}^2-1$). In this case 
additional contributions to the $T$ parameter arising from the gauge sector are 
non-negligible. They are given by~\cite{Hubisz:2005tx}\footnote
{
   A different result for the gauge sector contribution has been published in 
   Ref.~\cite{Asano:2006nr}, where the logarithmic term is found to cancel. 
   We thank Masaki Asano for pointing that out to us. 
   The numerical effect of this correction is 
   contained within the theoretical uncertainty of $\pm5$ assigned to the 
   $\delta_c$ coefficient.
}
\beq
   T_{\rm gauge} = -\frac{1}{4\pi\sin^2\!\theta_W}\frac{v^2}{f^2} 
                 \left(\delta_c+\frac{9}{4}\ln\! \frac{2\pi v}{M_W}\right)\,,
\eeq
where $v$ is the SM vacuum expectation value at the electroweak scale, 
$f$ is the ${\cal O}(\tev)$ symmetry breaking scale, 
and $\delta_c$ is a coefficient of order one whose exact value depends on
the details of the unknown UV physics~\cite{Hubisz:2005tx}.\footnote
{
    The $\delta_c$ parameter is treated as theory uncertainty varying in
    the range $[-5,5]$ in the fit.
}

The contribution to the $T$ oblique parameter from the $T$-odd
partners of the light SM fermions was found to increase with the masses 
of the partners~\cite{Hubisz:2005tx}. From LEP constraints on four-fermion 
contact interaction (the $ddee$ channel providing the most stringent lower bound 
on the contact interaction scale $\Lambda$), an upper bound on these masses can 
be derived leading to a maximum contribution to the $T$ parameter of~\cite{Hubisz:2005tx}
\beq
   T_{\mbox{\scriptsize $T$-odd fermions}}<0.05\,,
\label{eq:toddf}
\eeq
for each $T$-odd fermion partner of the twelve SM fermion doublets.

\begin{figure}[t]
  \epsfig{file=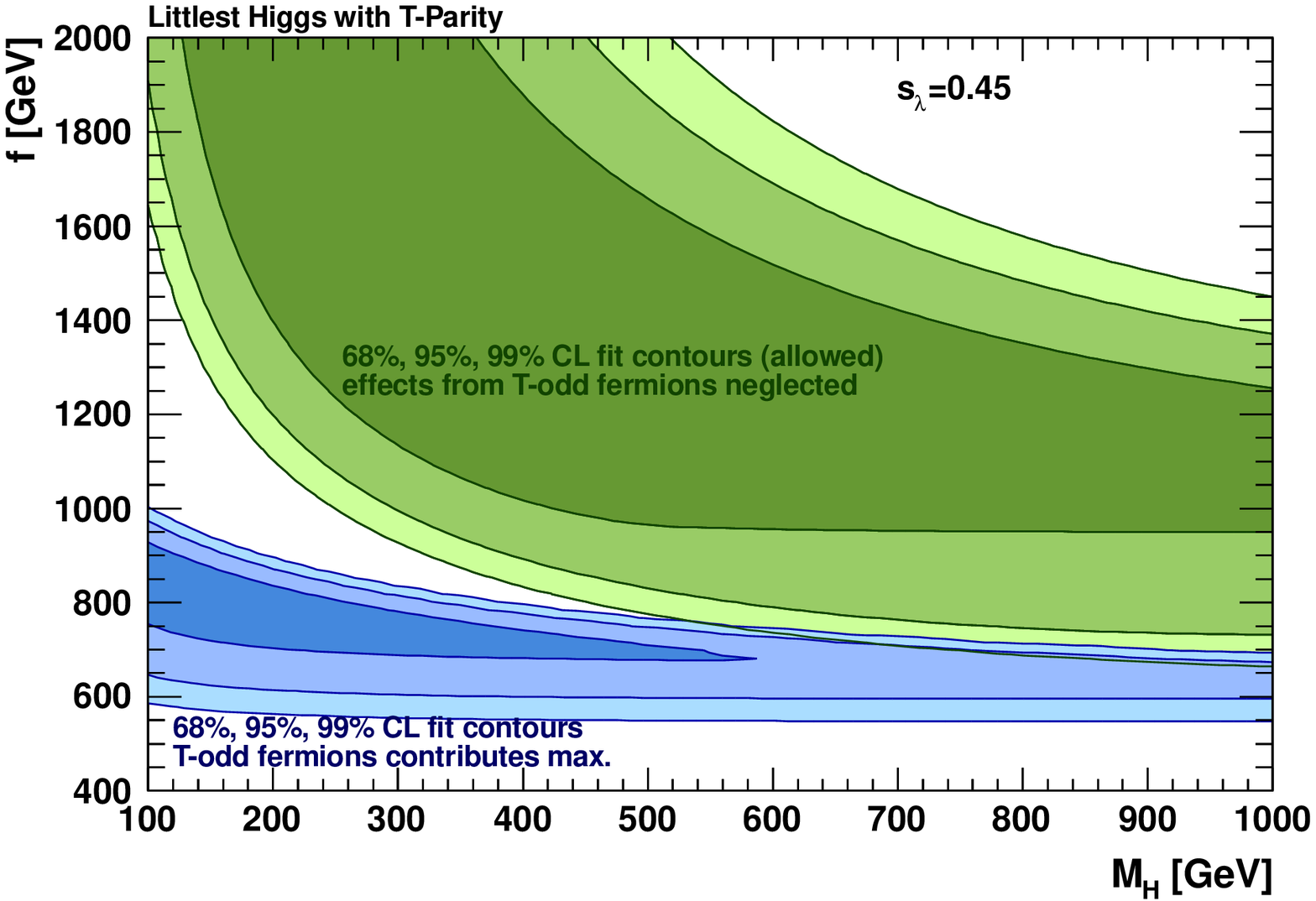, scale=\HalfPageWidthScale}
  \epsfig{file=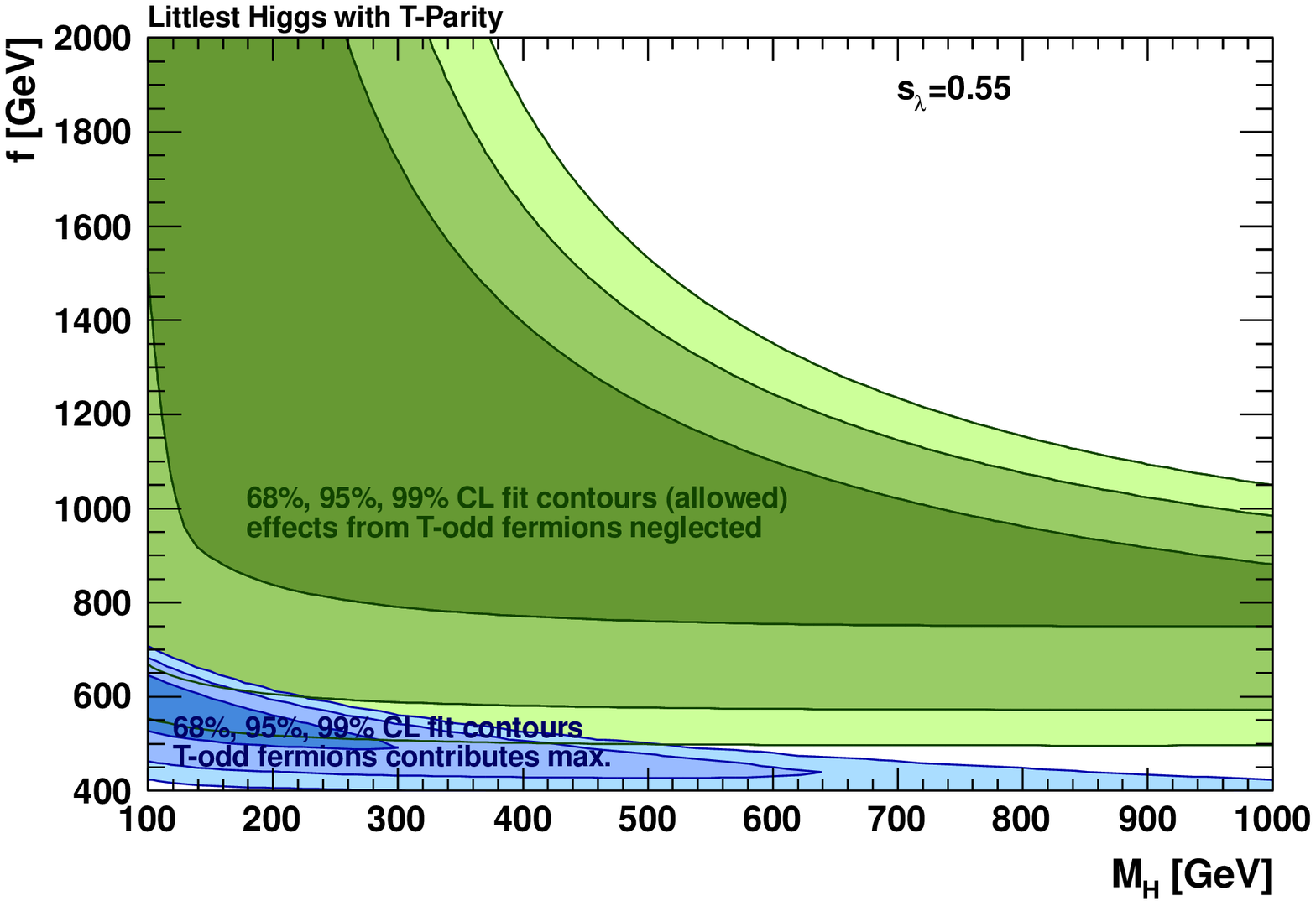, scale=\HalfPageWidthScale}
  \epsfig{file=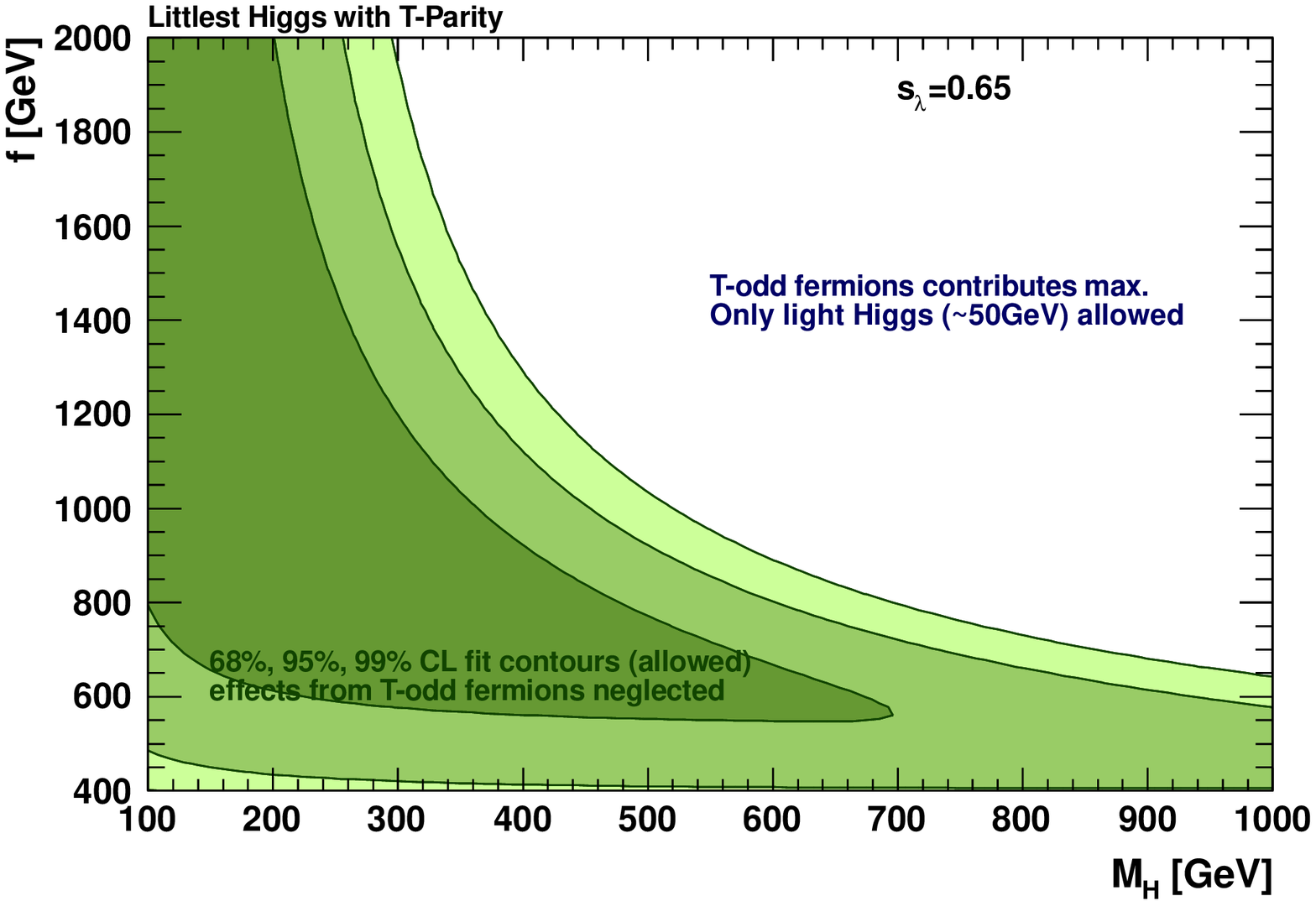, scale=\HalfPageWidthScale}
  \epsfig{file=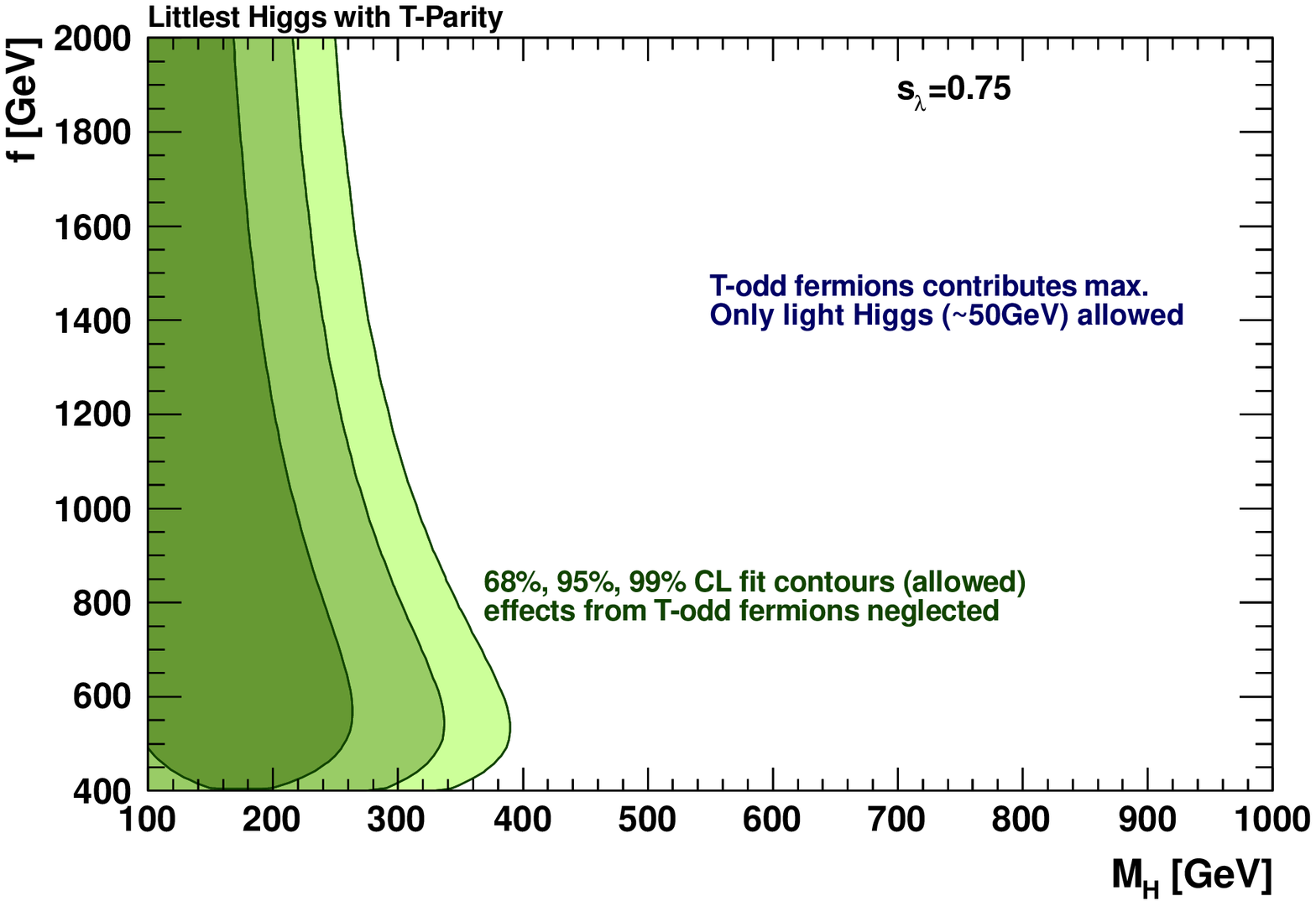, scale=\HalfPageWidthScale}
  \vspace{-0.3cm}
  \caption[]{Constraints in the littlest Higgs model with $T$-parity conservation. 
             Shown are the 68\%, 95\%, and 99\% CL allowed fit contours in the 
             ($M_H,f$) plane for fixed $s_\lambda$ values of 0.45, 0.55, 0.65 
             and 0.75 (top left to bottom right) when neglecting the effects from 
             the $T$-odd fermions (green), and when using the bound~(\ref{eq:toddf}) 
             (blue, only shown for $s_\lambda=$0.45 and 0.55).}
  \label{fig:LHfvsMH}
\end{figure}
Finally, the one-loop correction to the $Zb\bbar$ vertex in the LH model with 
$T$-parity conservation is dominated by diagrams involving Goldstone boson $\pi^{\pm}$
exchange. In the limit $m_{T_+} \gg m_t \gg M_W$ the additional
leading order correction reads~\cite{Hubisz:2005tx}
\begin{equation}
   \delta g_L^{b\bbar} = \frac{g}{c_w}\frac{\alpha}{8\pi \sin^2\!\theta_W} 
                        \frac{m_t^4}{M_W^2m_{T_+}^2}
                        \left( \frac{1}{s_{\lambda}^2}-1 \right) \ln\! \frac{m_{T_+}^2}{m_t^2}\,.
\end{equation}
The experimental fit result in the $(S,T)$ plane is compared in Fig.~\ref{fig:LHSvsT} 
to the LH prediction for example values of $f$, $s_{\lambda}$ and $M_H$, assuming 
that the $T$-odd fermions are sufficiently light ($\sim$300\:GeV)} to have a
negligible contribution to the $T$ parameter. Good overlap with the electroweak 
data is observed and, in particular, large $M_H$ values are allowed.

Figure~\ref{fig:LHfvsSL} shows, for a fixed value of $M_H=120\:\gev$, the 
68\%, 95\% and 99\% CL allowed regions in the ($s_\lambda,f$) plane 
when neglecting the effects from the $T$-odd fermions (green), and when 
assuming their maximum contribution to be consistent with the four-fermion 
contact interaction bound (blue), respectively. In both cases 
a large range of values for the breaking scale $f$ is allowed.
The green areas in the  panels of Fig.~\ref{fig:LHfvsMH}
illustrate the 68\%, 95\% and 99\% CL allowed regions
in the $(M_H,f)$ plane for the fixed values 
$s_{\lambda}=0.45,\,0.55,\,0.65,\,0.75$, and neglecting
the effects from $T$-odd fermions.  For large values of $f$ the $M_H$ 
constraint in the LH model approaches that of the SM, while for small $f$ 
significantly larger values of $M_H$ are allowed. Although the allowed  
$(M_H,f)$ regions strongly depend on $s_{\lambda}$ and no absolute exclusion 
limit on one of the parameters alone can be derived, the above statements 
are true for all values of $s_{\lambda}$.  For $s_{\lambda}=0.45$ (top left) and
$s_{\lambda}=0.55$ (top right) also the constraints obtained when including 
the maximum effect of $T$-odd fermions are shown. In that case, 
the allowed values for the breaking scale $f$ are largely reduced.

%% file: LargeED.tex
\subsection{Models with large extra dimensions}
\label{sec:constraintsLED}

Models with large flat extra spatial dimensions (ADD)~\cite{ArkaniHamed:1998rs,Antoniadis:1998ig} 
of compact size up to microns provide a possible solution to the hierarchy problem by reducing the 
size of the non-fundamental Planck scale, $M_D$, close to that of the (fundamental) electroweak 
scale. In these models only gravity propagates into the extra dimensions (the {\em bulk}),
while the SM fields are confined in the four-dimensional space-time where the gravitational 
flux is diluted. The larger the number of extra dimensions, $\delta$, the larger the amount
of the dilution. Reducing the $4+\delta$ dimensional Planck scale to TeV size requires at least 
$\delta=2$, where the size of the extra dimension would be of order $100\,\mu$m. For 
$\delta>2$ the required size would be $10^{-7}\,$cm or less. A direct search for a
violation of the Newtonian inverse-square law sets a 95\% CL upper bound of 
$R \le 44\,\mu$m on the size of the largest extra dimension~\cite{Kapner:2006si},
where $R$ is the radius of an extra dimension that is compactified on a torus. 
For $\delta=2$ the above result on $R$ is slightly tighter giving  
the lower bound $M_D>3.6\,\tev$~\cite{NakamuraGiudice:2010zzi}. 
Under certain model assumptions there exist strong astrophysical constraints 
on large extra dimensions, excluding Planck scales of up to $M_D>1700\,(60)\,\tev$ 
for $\delta=2\,(3)$~\cite{Hannestad:2001xi}. 

Gravitons propagating in the compact extra dimensions exhibit towers of 
Kaluza-Klein (KK) excitations with masses that are multiples of $\sim$$R^{-1}$. 
Due to the smallness of $R^{-1}$, the mass spectrum is quasi-continuous and cannot
be resolved in an accelerator experiment. In spite of the small gravitational 
coupling of each individual KK graviton to the SM particles, detectable scattering 
cross sections are achieved by summing over the large number of KK graviton states in 
a tower. However, this sum is ultraviolet divergent requiring a cut-off and
the modelling of the ultraviolet completion. The cut-off scale $\Lambda$ is related, 
but not necessarily equal to $M_D$~\cite{Giudice:1998ck,Han:1998sg,Hewett:1998sn,Giudice:2003tu}.
It should, however, not be chosen much larger than $M_D$ due to the unknown 
ultraviolet physics. Naive dimensional analysis, for example, sets upper limits
at which gravity becomes strongly interacting of $\Lambda/M_D\simeq5.4\,(2.7)$ for 
$\delta=1\,(2)$, and further decreasing limits for rising 
$\delta$~\cite{Giudice:2003tu,Contino:2001nj}. 

Direct accelerator-based searches for large extra dimensions have been
carried out at LEP, the Tevatron and LHC (see \eg the
review~\cite{Landsberg:2008ax}). The LEP experiments have searched for
direct graviton production and for virtual effects in fermion pair
and diboson production, leading to $M_D$ exclusion limits between
1.6\,TeV for $\delta=2$ and 0.66\,TeV for $\delta=6$. See Ref.~\cite{Ask:2004dv} 
for a review of the LEP results.
The Tevatron experiments have searched for large extra dimensions 
in dielectron, diphoton, monojet and monophoton channels 
(see~\cite{Landsberg:2008ax} and references therein). These 
searches lead to $M_D$ exclusion limits exceeding the LEP bounds 
for $\delta\ge 4$~\cite{Landsberg:2008ax}.
By searching for deviations in the diphoton invariant mass spectrum, 
CMS sets limits excluding $M_D$ values lower than $1.6$--$2.3\,\tev$ at
95\% CL, depending on the number of extra dimensions and on the
ultraviolet cut-off prescription used~\cite{Chatrchyan:2011jx}. 
In a recent analysis of the monojet channel ATLAS excludes $M_D$ values 
smaller than $2.3$\,TeV, $2.0\,\tev$ and $1.8\,\tev$ for $\delta=2$, 
$\delta=3$ and $\delta=4$, respectively~\cite{Collaboration:2011xw}.

For the implementation of the electroweak precision constraints on large extra 
dimensions we follow Ref.~\cite{Contino:2001nj}. The graviton corrections to the 
electroweak precision observables scale like $M_Z^2 \Lambda^\delta/M_D^{2+\delta}$ and
thus decrease with $\delta$ in the better 
controlled region $\Lambda < M_D$, while increasing with $\delta$ for $\Lambda > M_D$.
Graviton loop effects have been computed in Ref.~\cite{Contino:2001nj} for a 
simplifying combination of $\eps$ oblique parameters in which only 
the vacuum polarisation correction difference between $W$ and $Z$ loops appears
\beq
     \overline \eps = \eps_1 - \eps_2 - \eps_3\cdot\tan\!^2\theta_W\,.
\eeq
Using Eqs.~(\ref{eq:stu_eps}) (appendix) this combination can be readily 
transformed into \STU parameters giving 
$\overline \eps=\alpha(M_Z^2)(T + U/(4\sinthw) - S/(4\costhw))$. 
For the experimental value at $M_H=120\,\gev$ ($M_H=600\,\gev$) we find 
$\overline \eps=(8.8 \pm 6.1)\cdot10^{-4}$ ($\overline \eps=(25.4 \pm 6.1)\cdot10^{-4}$). 
In the limit of heavy graviton states and by choosing the renormalisation 
scale equal to $\Lambda$ and cutting off the KK tower at $n<R\cdot \Lambda$, 
the graviton loop gives~\cite{Contino:2001nj}
\beq
   \delta \overline \eps \simeq \sinthw \frac{M_Z^2}{M_D^2}
                                \left(\frac{\Lambda}{M_D}\right)^{\!\!\delta}
                                \frac{5(8+5\delta)}{48\,\Gamma(2+\delta/2)\pi^{2-\delta/2}}\,.
\eeq
By inverting this equation, one can use the measurement of $\overline \eps$ to constrain
$\Lambda/M_D$ versus $M_D$ as a function of $\delta$. 

\begin{figure}[t]
  \epsfig{file=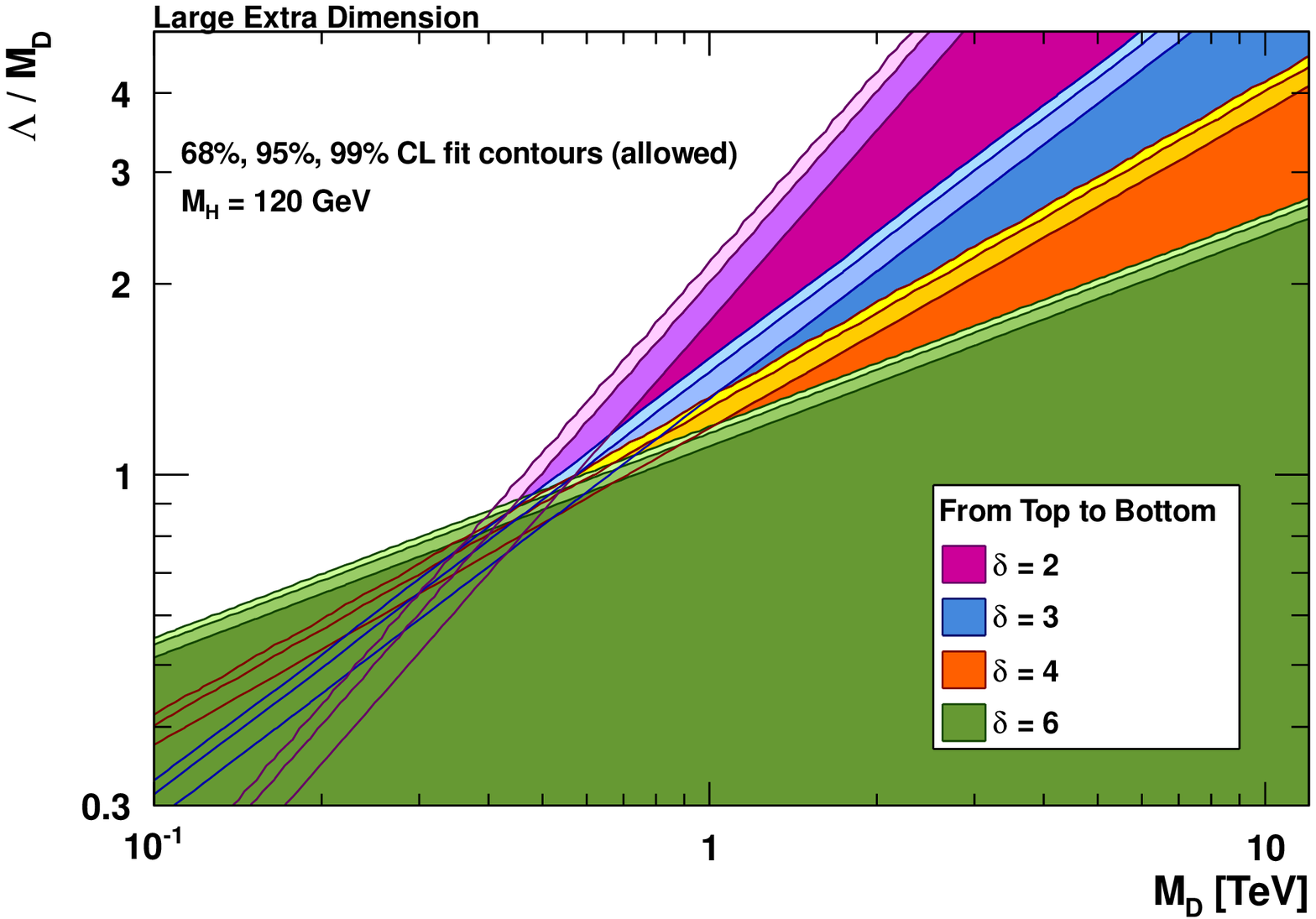, scale=0.4}
  \epsfig{file=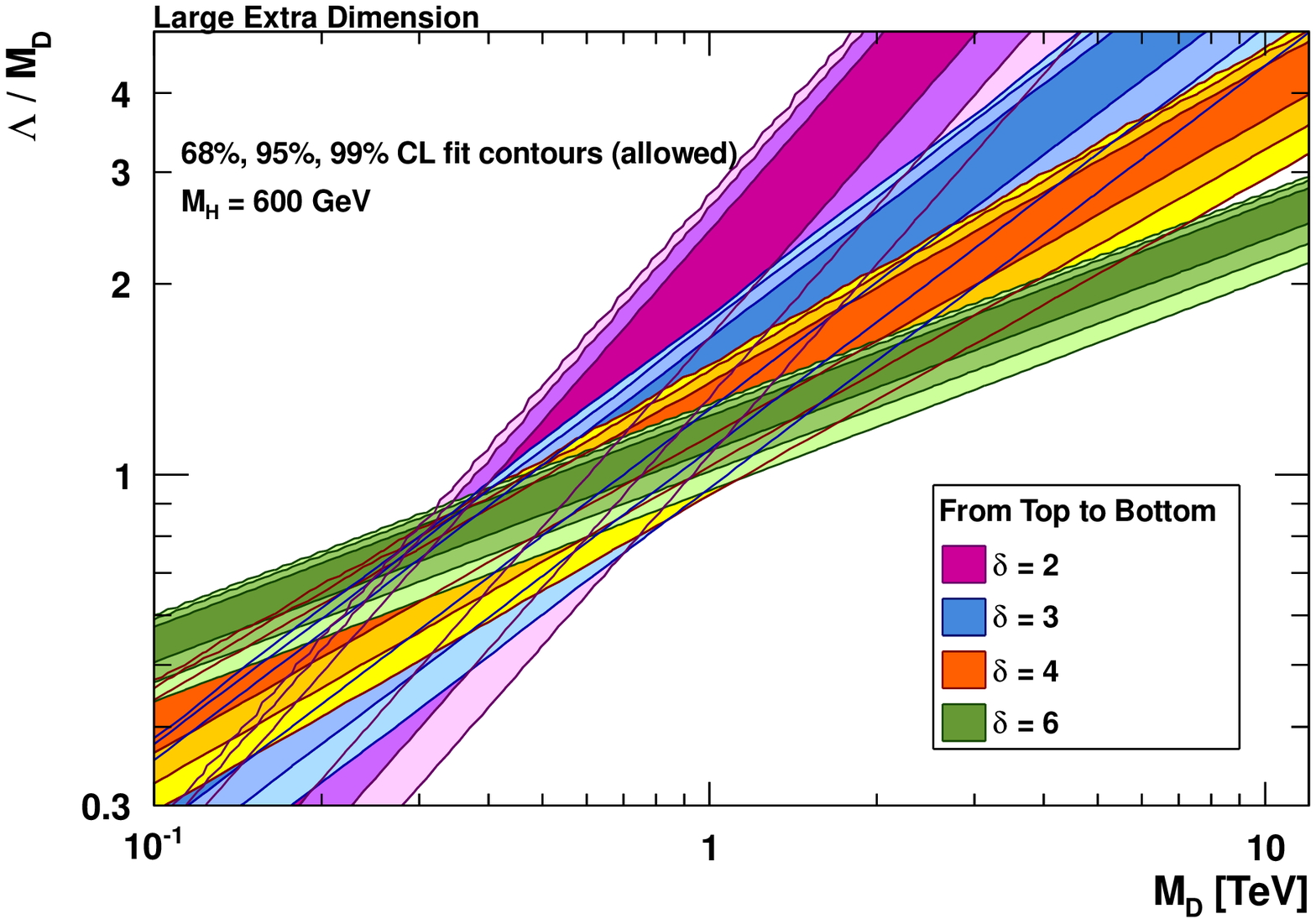, scale=0.4}
  \vspace{0.1cm}
  \caption[]{Constraints from the electroweak precision data on the ADD model 
             parameters. Shown are the 68\%, 95\% and 99\% CL 
             allowed fit contours in the ($M_D,\Lambda/M_D$) plane for various 
             numbers of extra dimensions $\delta$ and for Higgs masses of 
             120\,\gev (left) and 600\,\gev (right).}
       \label{fig:LED}
\vspace{0.6cm}
  \centerline{\epsfig{file=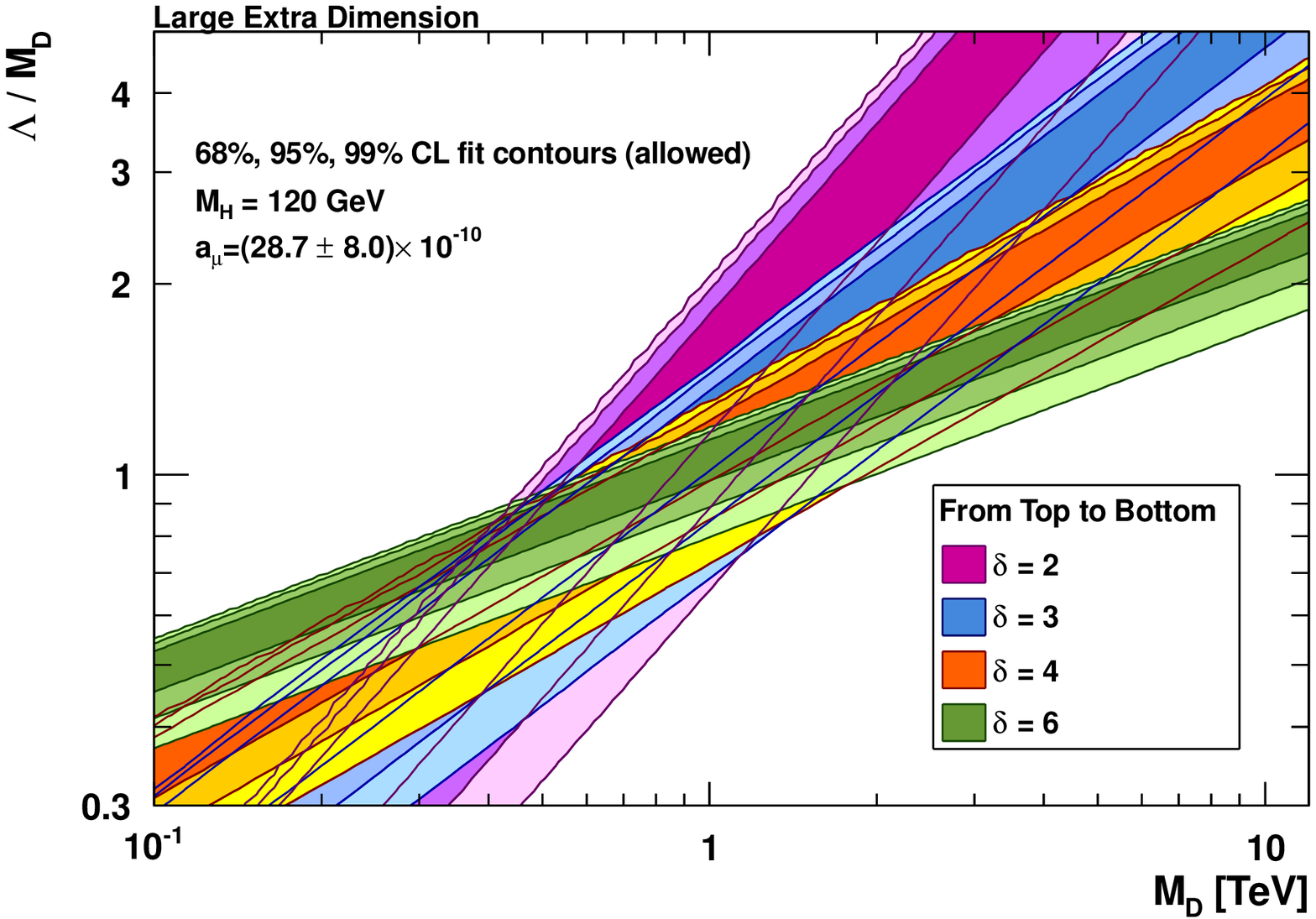, scale=0.4}}
  \vspace{0.1cm}
  \caption[]{Constraints on the ADD model parameters obtained by combining
             the electroweak precision data with the muon anomalous magnetic moment.
             Shown are the 68\%, 95\% and 99\% CL allowed fit contours in the 
             ($M_D,\Lambda/M_D$) plane for various numbers of extra dimensions 
             $\delta$ and for a Higgs mass of 120\,\gev.}
       \label{fig:LEDamu}
\end{figure}
The constraints obtained for various $\delta$ in terms of 68\%, 95\% and 99\% CL 
allowed regions in the ($M_D,\Lambda/M_D$) plane 
are drawn in Fig.~\ref{fig:LED} for hypothetical Higgs masses of 120\,\gev (left panel) 
and 600\,\gev (right panel). They show the expected behaviour of a weaker experimental 
constraint for rising $\delta$ where $\Lambda/M_D<1$, and the opposite effect for 
$\Lambda/M_D>1$. Owing to the significant deviation of $\overline \eps$ from zero for 
$M_H=600\,\gev$, contributions from large extra dimensions, which effectively counterweight
the large negative $T$ term in the SM, are required. These heavy-Higgs scenarios are 
already excluded by the direct searches for $\Lambda/M_D<1$ (see references above).

It is possible to enhance the electroweak constraint by also including
the difference between the measured~\cite{Bennett:2006fi} 
and predicted anomalous magnetic moment of the muon (\amu). We use a recent 
evaluation of this difference, $(28.7 \pm 8.0)\cdot10^{-1}$~\cite{Davier:2010nc}, 
which exhibits a $3.6\sigma$ deviation from zero (the corresponding $\tau$-data based 
deviation amounts to $2.6\sigma$). The contribution of the large extra dimensions 
model to \amu is also given in Ref.~\cite{Contino:2001nj}
\beq
    \delta\amu = \frac{m_\mu^2}{M_D^2}\left(\frac{\Lambda}{M_D}\right)^{\!\!\delta}
                 \frac{34 + 11\delta}{96\,\Gamma(2+\delta/2)\pi^{2 - \delta/2}}\,.
\eeq
Figure~\ref{fig:LEDamu} shows the constraints obtained for various $\delta$ 
from the combined usage of the electroweak precision data for $M_H=120\,\gev$ 
and \amu. The deviation of the latter quantity from the SM value can only
be accommodated by a low Planck scale, already excluded by direct experimental 
searches, or by a very large ultraviolet cut-off scale.

%% file: UniversalED.tex
\subsection{Models with universal extra dimensions}
\label{sec:constraintsUED}

Models with flat, compactified extra dimensions where all of the SM
fields are allowed to propagate into the bulk~\cite{Antoniadis:1990ew}
are referred to as universal extra dimensions
(UED)~\cite{Appelquist:2000nn} (see also the review on UED 
phenomenology in~\cite{Hooper:2007qk} and references therein). 
In its minimal version one
extra dimension is compactified on an $S^1/Z_2$ orbifold with two
fixed points at $y=0$ and $\pi$ to obtain the SM chiral fermions
from the corresponding extra dimensional fermion fields. The SM
fields appear as towers of Kaluza-Klein (KK) states with tree-level
masses
\beq
\label{eq:UEDmass}
   m^2_n = m^2_0+ \frac{n^2}{R^2}\,,
\eeq
where $m_n$ is the mass of the {\it n}th KK excitation of the SM
field, $m_0$ is the ordinary mass of the SM 
particle and $R\sim\tev^{-1}$ is the size of the extra dimension
with the compactification scale $M_{\KK}=R^{-1}$. 
Bulk loops and brane-localised kinetic terms can lead to corrections of the
KK masses of up to 20\% for the KK quark and KK gluon states and of
a few percent or less for the other states.

In UED models, momentum conservation in the higher dimensional space
leads to a conserved {\em KK-parity} $P=(-1)^n$. As a consequence, 
the lightest KK state is stable and could be a candidate particle for 
the cold dark matter in the universe. Indeed it has been
shown~\cite{Servant:2002aq,Burnell:2005hm,Kong:2005hn} that the first
excitation of the hypercharge gauge boson $B^{(1)}$ can account for the
relic dark matter abundance of the universe if its mass is approximately
600\:\gev.\footnote
{
   If the UED is embedded into large extra dimensions of size eV$^{-1}$ 
   accessible to gravity only, the lightest KK state could 
   decay via KK-number violating gravitational interaction into a
   photon and an eV-spaced graviton tower of mass equivalent 
   between zero and $R^{-1}$~\cite{Macesanu:2002ew}. Such a model 
   provides a clear collider signature with two isolated photons
   and missing transverse energy in the final state, which has 
   been searched for at ATLAS~\cite{Aad:2010qr} and D0~\cite{Abazov:2010us}.   
} 
The odd-level KK states can only be pair produced
at colliders and their couplings to even number KK modes are
loop suppressed. The LHC experiments should be able to detect
the new KK states up to $R^{-1}\sim1.5\:\tev$~\cite{Cheng:2002ab}.

The SM particles that are allowed to propagate into the bulk 
contribute with quantum corrections to the lower energy observables. 
In particular, extra dimension models where only the gauge bosons 
are allowed to propagate into the bulk, while all other particles 
are confined to the SM brane, are strongly constrained by the LEP 
data forcing the masses of the lowest KK excitations to several
TeV~\cite{Masip:1999mk,Rizzo:1999br}, beyond the reach of possible
direct detection at the LHC. KK-parity conservation in
UED models forbids a direct coupling of a single KK excitation to
the SM fermions and thus weakens the impact of the electroweak
data. The heavy KK states can only contribute to the self energies of
the gauge bosons parametrised in terms of the $S,T, U$ parameters.

The complete one-loop corrections of a given KK level $n$ of the SM
fields to the gauge-boson self energies have been calculated
in Ref.~\cite{Appelquist:2002wb,Gogoladze:2006br}. The corrections are proportional to 
$m_t^2/M_{\KK}^2$, $M_H^2/M_{\KK}^2$ and $M_W^2/M_{\KK}^2$ for the top quark,
Higgs, and gauge boson excitations, respectively.
The contributions from top (Higgs) excitations dominate for small (large) 
Higgs masses. The total UED contribution corresponds to an infinite sum 
over $n$, which is convergent for one extra dimension. For the leading order
terms of the oblique corrections for one extra dimension we follow 
Refs.~\cite{Appelquist:2002wb,Gogoladze:2006br} where results very
similar to the present study were presented. The terms read
\begin{align}
\label{eq:UEDS}
S ~& =\!\!\!\!\!\! & {4\sinthw \over \alpha}   & \ \left[{3 g^2 \over 4 (4 \pi)^2}
\left( {2\over 9} {m_t^2 \over M_{\KK}^2}  \right)\zeta(2) \ + \
{ g^2 \over 4 (4 \pi)^2}
\left( {1\over 6} {M_H^2 \over M_{\KK}^2} \right)\zeta(2)\right] \,,\\[0.2cm]
\label{eq:UEDT}
T ~& =\!\!\!\!\!\! & {1\over \alpha}   & \ \left[{3 g^2 \over 2 (4 \pi)^2} {m_t^2 \over M_W^2 } \left( {2\over 3}
{m_t^2 \over M_{\KK}^2}  \right)\zeta(2) \ + \
{ g^2 \sinthw \over (4 \pi)^2 \costhw} 
\left( -{5\over 12} {M_H^2 \over M_{\KK}^2} \right)\zeta(2)\right]\,, \\[0.2cm]
\label{eq:UEDU}
U ~& =\!\!\!\!\!\! & -{4\sinthw \over \alpha} & \ \left[{ g^2 \sinthw \over  (4 \pi)^2} {M_W^2 \over M_{\KK}^2}
\left( {1\over 6}  \zeta(2) - {1\over 15}
{M_H^2 \over M_{\KK}^2} \zeta(4) \right) \right] \, ,
\end{align}
where the $\zeta$-functions arise from the summation over the KK tower states.
Because $M_W^2\ll m_t^2 \ll M_{\KK}^2$, the oblique parameter $T$
will dominate the electroweak precision constraints for small values
of $M_H$, and $U$ is negligible compared to $T$ and $S$.
Top quark and Higgs loops contribute with opposite signs to the $T$ parameter.  
Cancellation between these contributions is  achieved for 
$M_H=\sqrt{12/5}\cdot\sqrtcotthw\cdot m_t^2/M_W\approx1.1\:\tev$. 
For smaller (larger) $M_H$, $T$ takes
positive (negative) values. The positive contribution to $T$ from the top 
loops also weakens the Higgs mass constraint from the global electroweak 
fit.\footnote
{
  This effect is similar to the cancellation of the negative SM Higgs 
  contribution to $T$ with the positive contribution from the top sector 
  in the littlest Higgs model (cf. Section~\ref{sec:constraintsLH}).
}  
On the other hand, the top quark and Higgs loop contributions to $S$ have 
the same sign. One notices the decoupling from the SM in 
Eqs.~(\ref{eq:UEDS})-(\ref{eq:UEDU}) for small extra dimensions.

\begin{figure}[p]
   \centerline{\epsfig{file=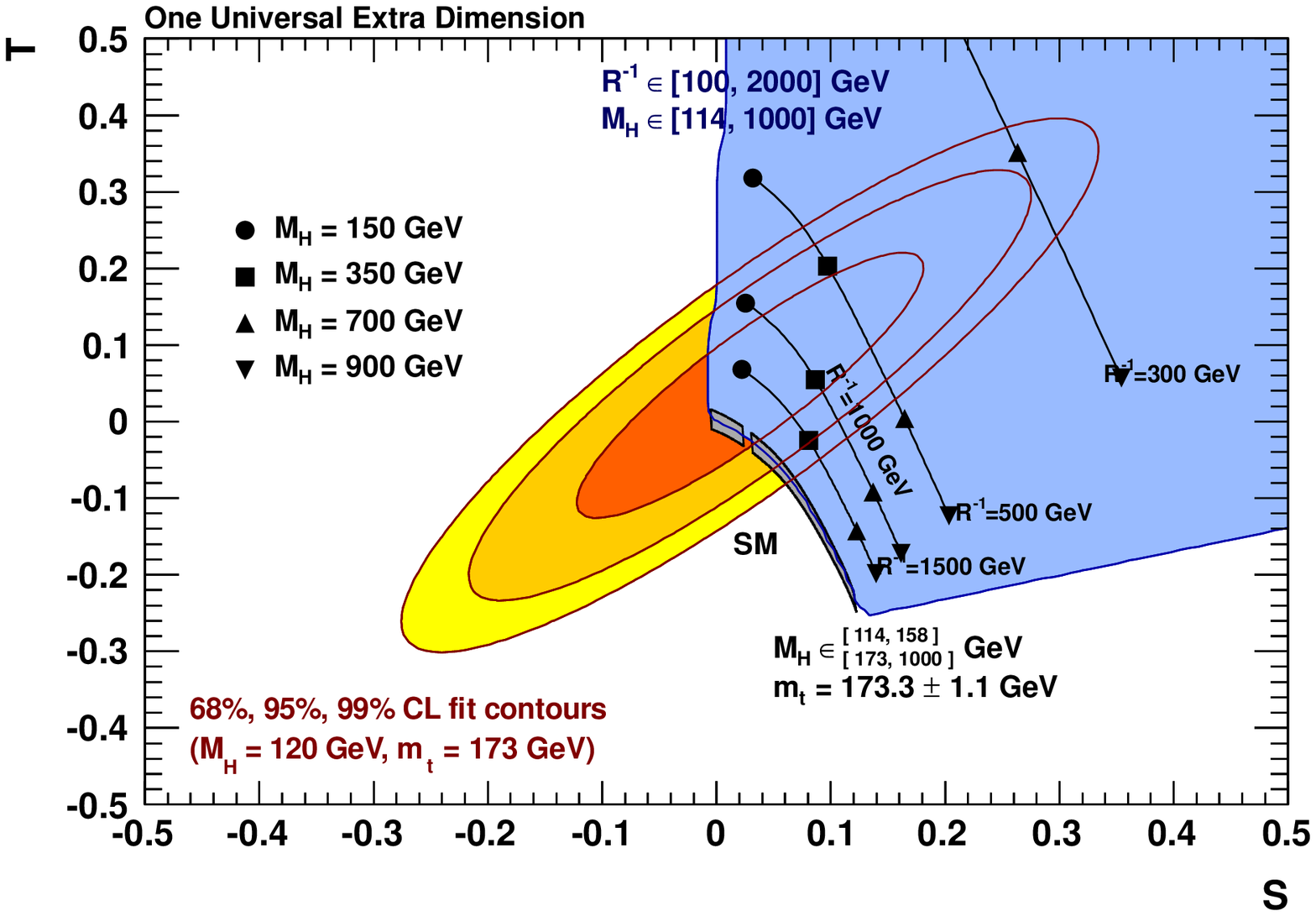, scale=\defaultFigureScale}}
  \vspace{0.2cm}
  \caption[]{Oblique parameters in a model with one universal extra dimension.
             Shown are the $S$, $T$ fit results (leaving $U$ free) compared
             to predictions from the SM and the UED model (grey and blue areas,
             respectively). The UED area is obtained with the use of the
             parameter ranges quoted on the figure. The solid lines and symbols illustrate
             the UED predictions for example models with varying values of $M_H$ and $R^{-1}$.}
       \label{fig:UEDSvsT}
  \vspace{0.7cm}
   \centerline{\epsfig{file=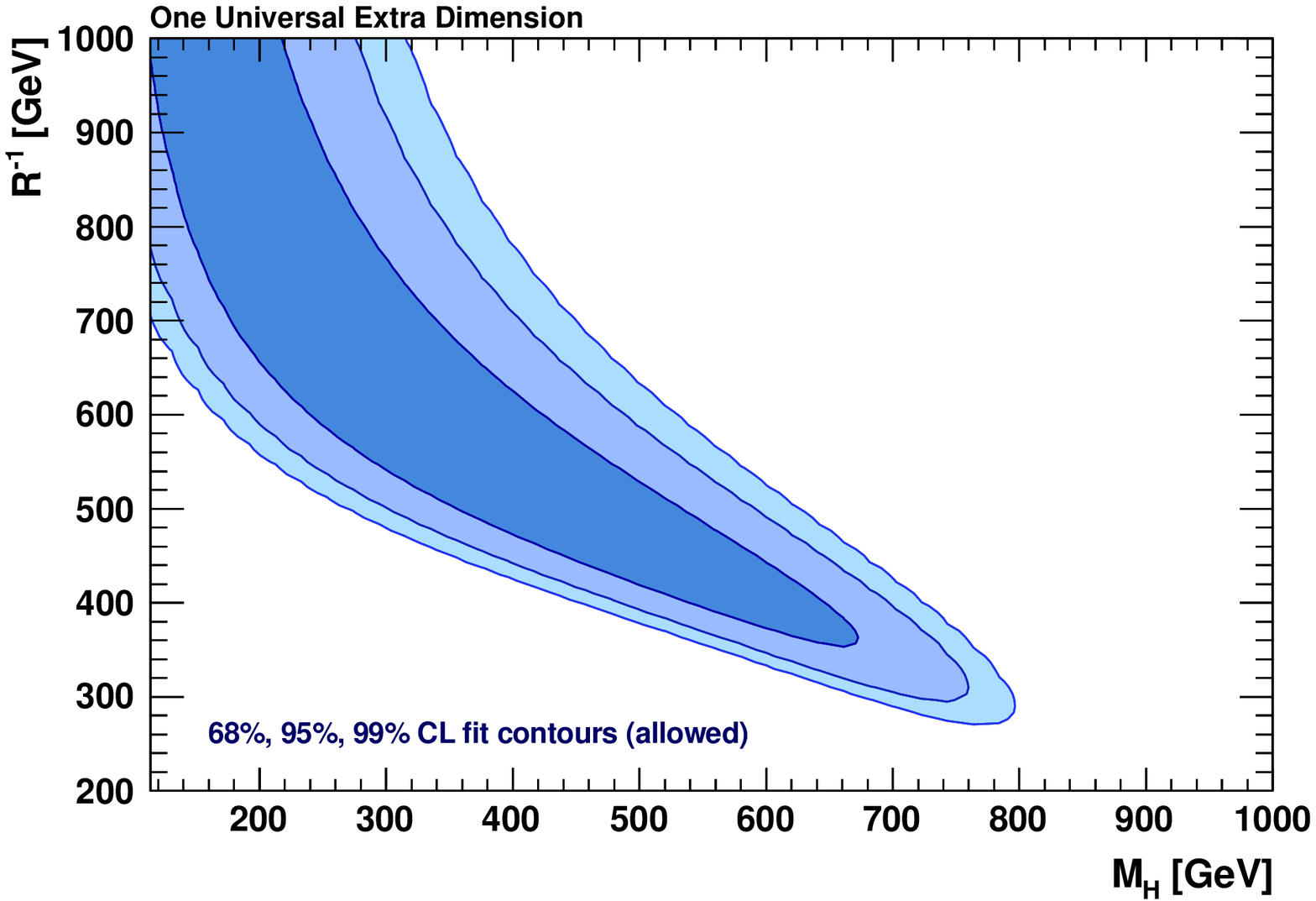, scale=\defaultFigureScale}}
  \vspace{0.2cm}
  \caption[]{Constraints in the model with one universal extra dimension. Shown are the
             68\%, 95\%, and 99\% CL allowed fit contours in the ($M_H,R^{-1}$) plane. }
       \label{fig:UED}
\end{figure}

Figure~\ref{fig:UEDSvsT} shows the UED prediction in the $(S,T)$ plane for 
various $R^{-1}$ and $M_H$ hypotheses. Constant values of $R^{-1}$ are depicted 
by the solid contour lines. The plot reproduces the
UED decoupling from the SM at large compactification scales, while for
small scales $T$ and $S$ can become large. The steepness of the
prediction for constant $M_H$ in the $(S,T)$ plan reduces with
increasing values of $M_H$ reflecting the negative (positive) sign of
the Higgs contribution to $T$ ($S$) in Eq.~(\ref{eq:UEDT})
(Eq.~(\ref{eq:UEDS})). For the $T=0$ cancellation value of
$M_H=1.1\:\tev$ a horizontal prediction is obtained as expected (not
drawn in the plot). By comparison with the electroweak data (ellipses)
one notices that for large UED scales $M_H$ must be small and vice versa.

This behaviour is emphasised in Fig.~\ref{fig:UED}, which shows the 68\%, 
95\% and 99\% CL allowed regions in the $(M_H,R^{-1})$ plane. For large $R^{-1}$ 
the constraint on $M_H$ approaches that of the SM, while for small 
compactification scales significantly larger Higgs masses are allowed. 
The region $R^{-1}<300\:\gev$ and $M_H>750\:\gev$ can be excluded 
at 95\% CL. These findings are in agreement with the results of previous
publications~\cite{Appelquist:2002wb,Gogoladze:2006br,Flacke:2005hb}.

It has been shown~\cite{Gogoladze:2006br} that constraints derived from fits 
including the subleading contribution from the additional oblique parameters 
$X,Y,V,W$ are very similar to the results of the $S,T$ analysis. Tighter 
constraints can be obtained~\cite{Gogoladze:2006br} when including $e^+e^-$ 
data from centre-of-mass energies beyond the $Z$ pole~\cite{Barbieri:2004qk}.
 
%
%

%% file: WarpedED.tex
\subsection{Models with warped extra dimensions}
\label{sec:constraintsRS}

To solve the hierarchy problem, Randall and Sundrum (RS) have proposed
a single, small and non-factorisable extra space dimension accessible to
gravity only~\cite{Randall:1999ee}. The geometry of this model is
determined by the extra dimension confined by two three-branes. The
model assumes only one fundamental mass scale, which is the ultraviolet
(UV) Planck scale.  The generation of the weak scale on the infrared
(IR) brane from the UV brane is achieved by introducing a
\textit{warp} factor altering the four-dimensional Minkowski metric.
The warp factor is an exponential function of the compactification
radius of the extra dimension, which is small and thus precludes the
extra dimension to be observed at low-scale gravity experiments. The warp factor is
considered to be the source of the observed large hierarchy between Planck and
weak scales in four space-time dimensions. The effective four-dimensional Planck 
scale is determined by a higher dimensional Planck scale and the geometry
of the extra dimension.

The RS model features fundamental spin-2 KK graviton excitations,
which strongly couple to the SM particles and would thus manifest
themselves in form of TeV scale resonances of pairs of jets, leptons,
photons, and gauge bosons in collider experiments. The scale can be
reduced if either a heavy Higgs is allowed or an ultraviolet cut-off
below the Planck scale is introduced.  The simplest RS models contain
only the SM particles and their KK excitations.  These models are
characterised by only two new parameters, of which one is the
order-one logarithm of the warp factor, $L=kr\pi$, where $k$ and $r$
are the dimensional curvature of the five-dimensional space-time and the
compactification radius, respectively. The inverse warp factor sets
the scale of the other free parameter, $M_\KK=k e^{-L}$.

In the minimal RS model, all SM fields are confined to one brane.
Since in this model the unification of the gauge couplings cannot be
described by an effective field theory~\cite{Agashe:2002pr} and the 
flavour hierarchy is not addressed
alternatives have been developed. In a first extension, the SM gauge
bosons are allowed to propagate into the bulk. However, $S$
and $T$ then adopt very large and negative values~\cite{Csaki:2002gy}.
In following variations also the SM fermions are let to propagate into
the bulk, which reduces the amount of the oblique corrections and shifting
them to small, positive values. $M_\KK$ then determines the lowest KK
excitations of the SM fields in  the bulk. The masses of the first 
KK gluon and photon excitations are approximately $2.5 \cdot M_\KK$.

The leading contributions to the $S$ and $T$ parameters for a model
with a brane-localised Higgs sector and bulk gauge and matter fields
are found to be~\cite{Delgado:2007ne,Casagrande:2008hr,Carena:2003fx}
\beqn
\label{eq:Swed}
  S & \!\!=\!\! & \frac{2 \pi v^2}{M_\KK^2}\left( 1 - \frac{1}{L} \right ) \,, \\
  T & \!\!=\!\! & \frac{\pi v^2}{2\cos^2\theta_W M_\KK^2} \left( L - \frac{1}{2L} \right )\,,
\eeqn
whereas there are no contributions to $U$. In the analysis presented here 
we follow the studies of Ref.~\cite{Casagrande:2008hr} where similar results 
have been obtained.

\begin{figure}[p]
  \centerline{\epsfig{file=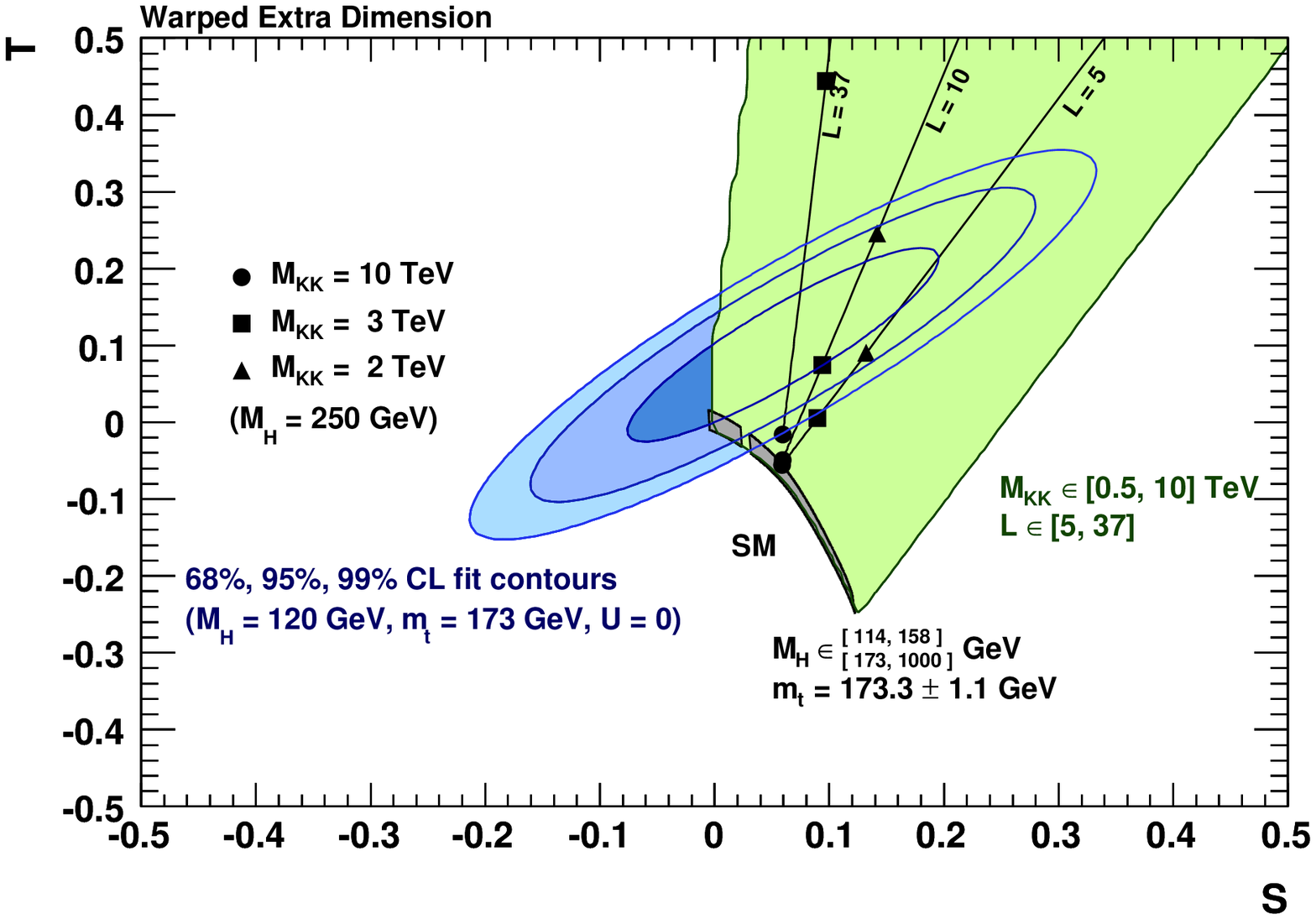,    scale=\defaultFigureScale}} 
  \vspace{0.4cm}
  \centerline{\epsfig{file=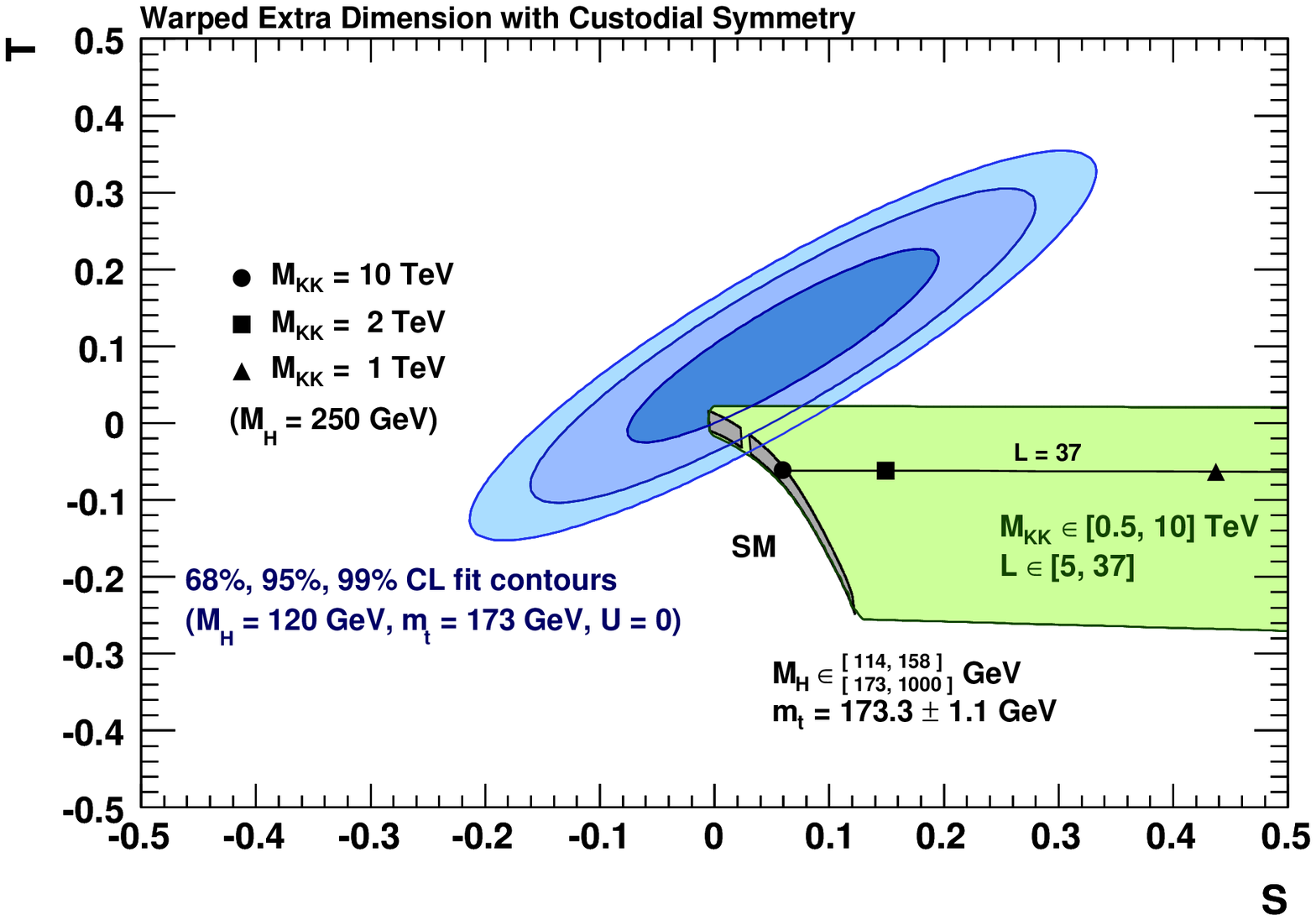, scale=\defaultFigureScale}}
  \vspace{0.3cm}
  \caption[]{Oblique parameters in the warped extra dimension model.
             Shown are the $S$, $T$ fit results (for $U=0$) compared
             to predictions from the SM and the RS model (grey and green areas,
             respectively) where gauge bosons and
             fermions are allowed to propagate into the bulk (top),
             and where in addition a custodial $SU(2)_L\times SU(2)_R$
             isospin gauge symmetry is introduced (bottom). 
             The predicted areas are obtained with the use of the $L$ and $M_\KK$ 
             parameter ranges given on the figures. The symbols and lines 
             illustrate example model settings.}
  \label{fig:RSSvsT}
\end{figure}

The predicted $S$ and $T$ regions for $0.5\le M_\KK\le 10\:\tev$
and $5\le L \le 37$ are shown by the shaded (green) region on the top panel of 
Fig.~\ref{fig:RSSvsT}. There is a large overlap with the electroweak data (ellipses). 
The figure also illustrates the decoupling of the RS model for large $M_\KK$. 

Specific constraints from the electroweak fit on the RS model
parameters in correlation with the Higgs mass are shown in the top and
middle panels of Fig.~\ref{fig:RSscans}. Large Higgs
masses can be accommodated for comparatively low $M_\KK$ values
counteracting on the strong constraint from $T$. A large Higgs mass is in
agreement with the Higgs field being localised on the TeV brane. Assuming 
new physics to stabilise the hierarchy problem at a UV scale 
of approximately $10^3\:\tev$ (corresponding to $L\approx9$) would 
relax the $M_\KK$ lower bound, cf. Fig.~\ref{fig:RSscans}. Vice versa, 
one finds that small $M_\KK$ values lead to an increased constraint on $L$.
Addressing the full hierarchy problem ($L\approx39$) requires the lightest 
KK modes to be heavy, albeit this constraint would be alleviated if the 
Higgs boson is heavy. 

\newcommand\HalfPageWidthScaleWED{0.406}

\begin{figure}[p]
\centering
  \epsfig{file=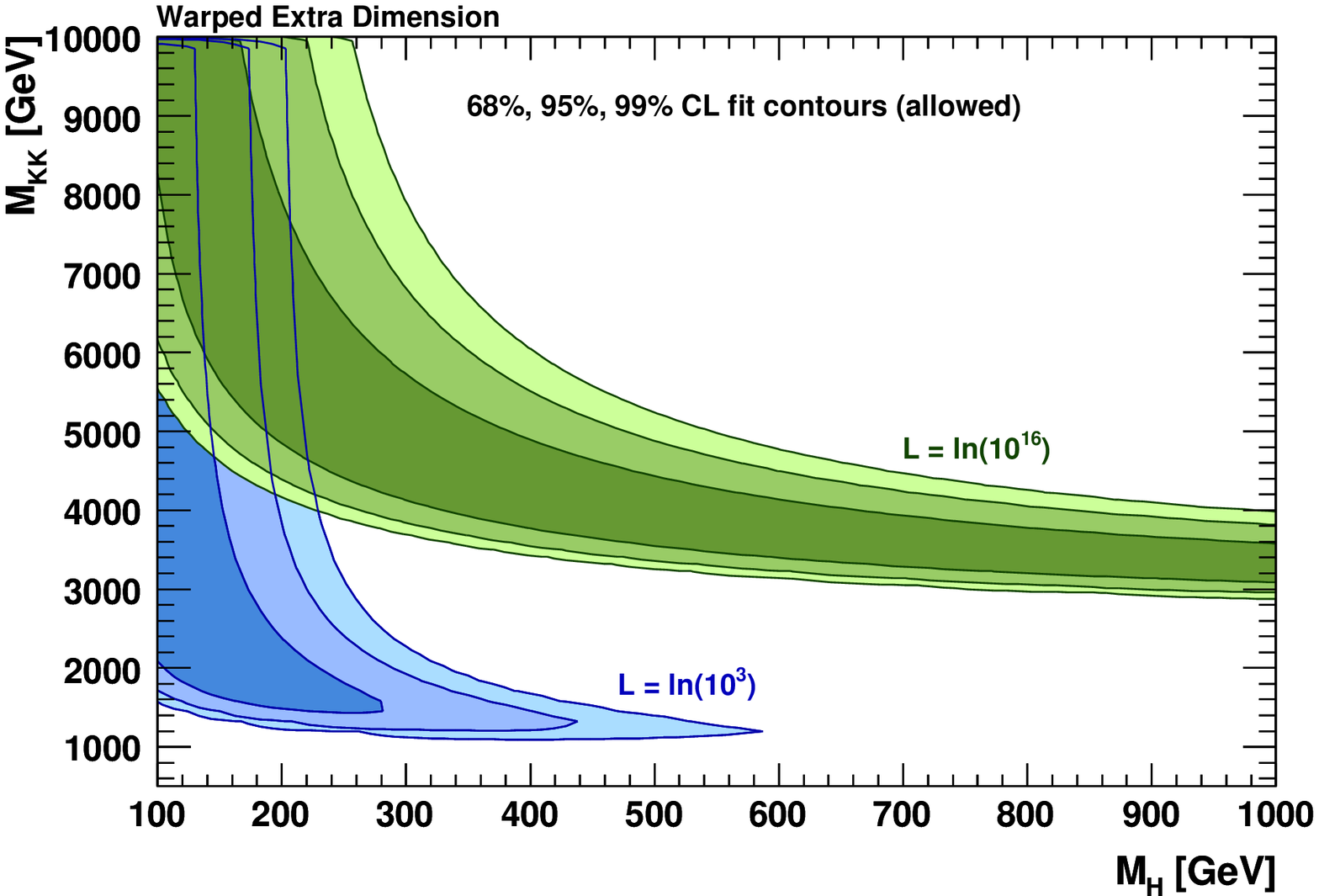, scale=\HalfPageWidthScaleWED}
  \epsfig{file=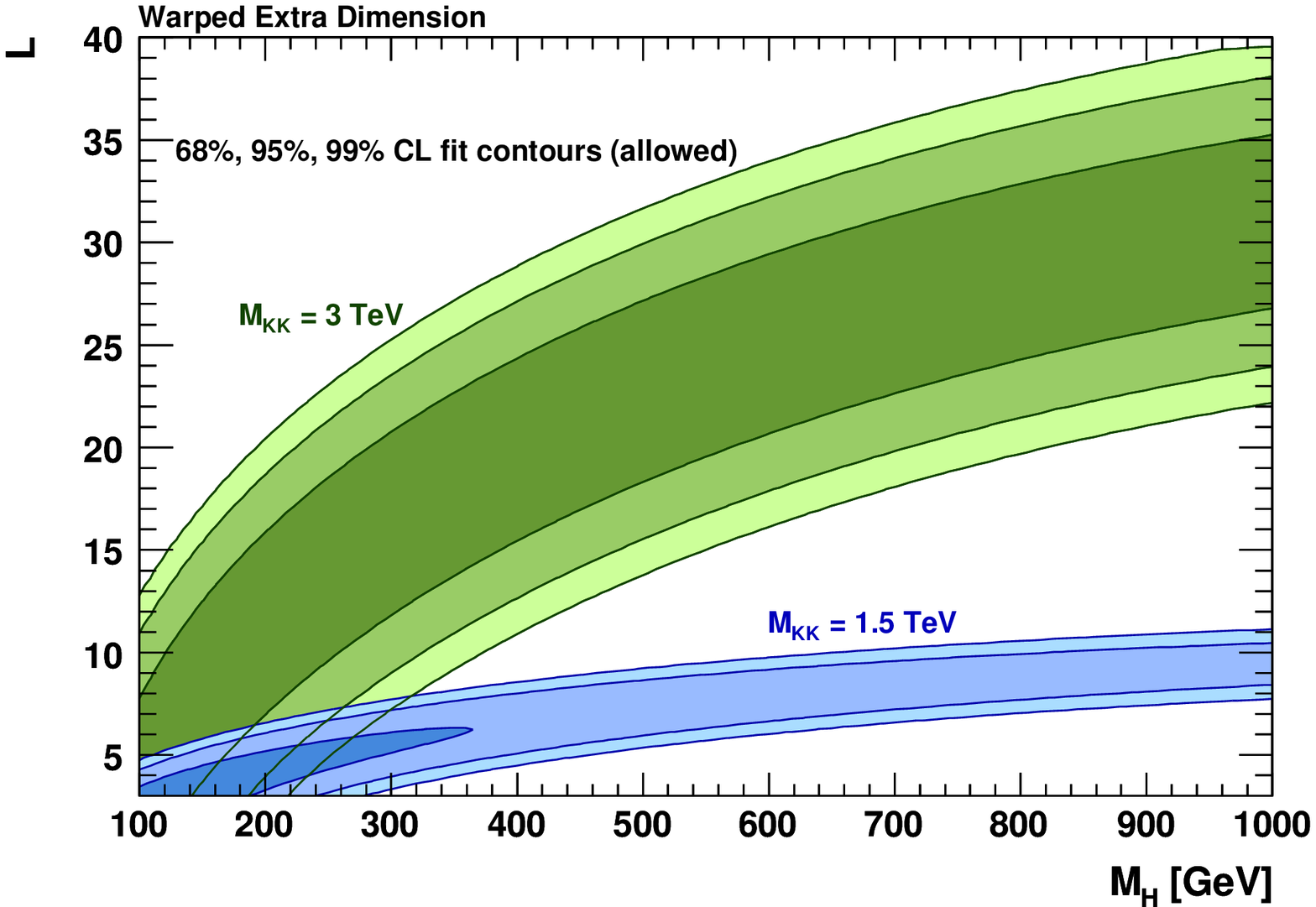, scale=\HalfPageWidthScaleWED} \\
  \vspace{0.5cm}
  \epsfig{file=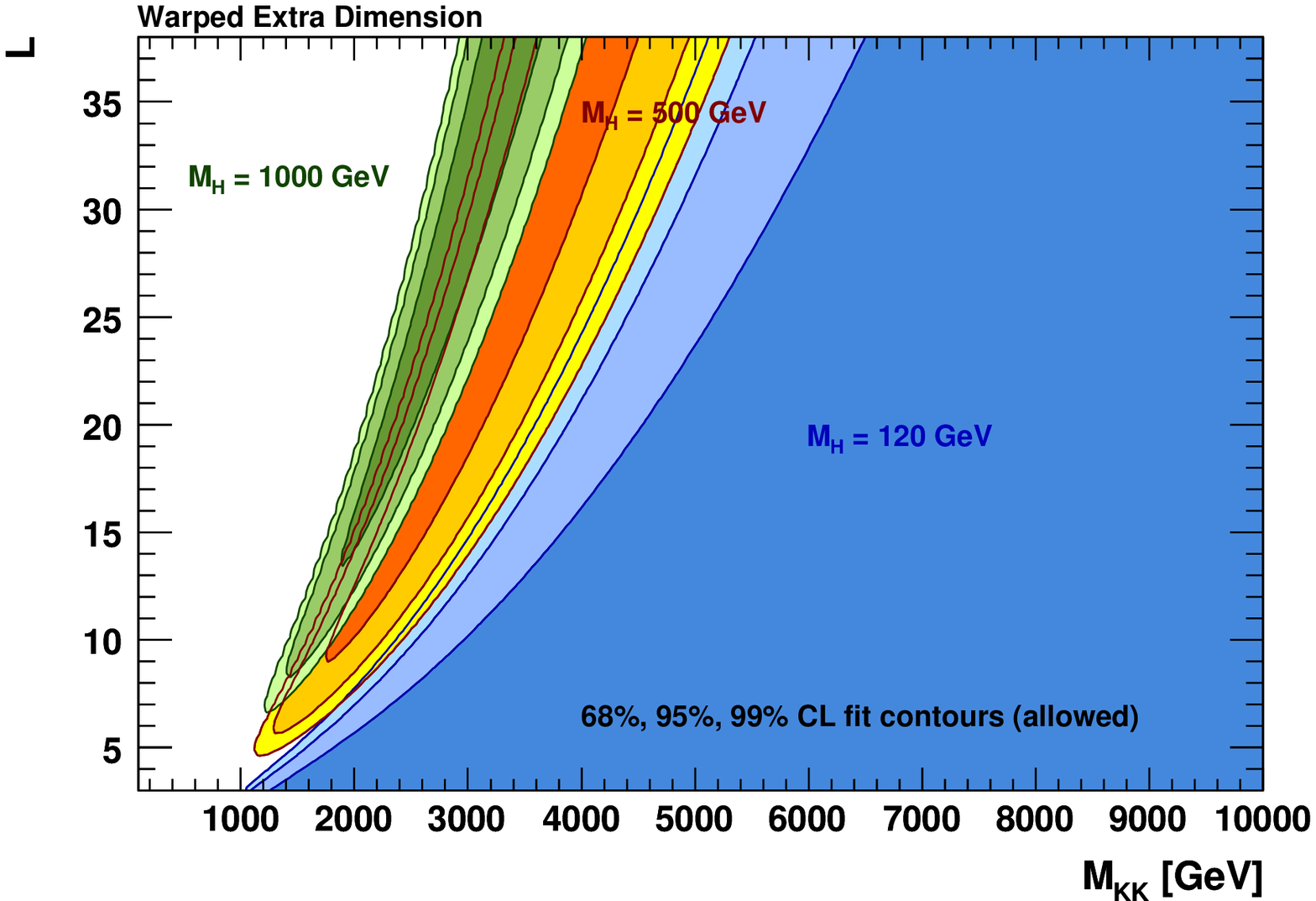, scale=\HalfPageWidthScaleWED} \\
  \vspace{0.5cm}
  \epsfig{file=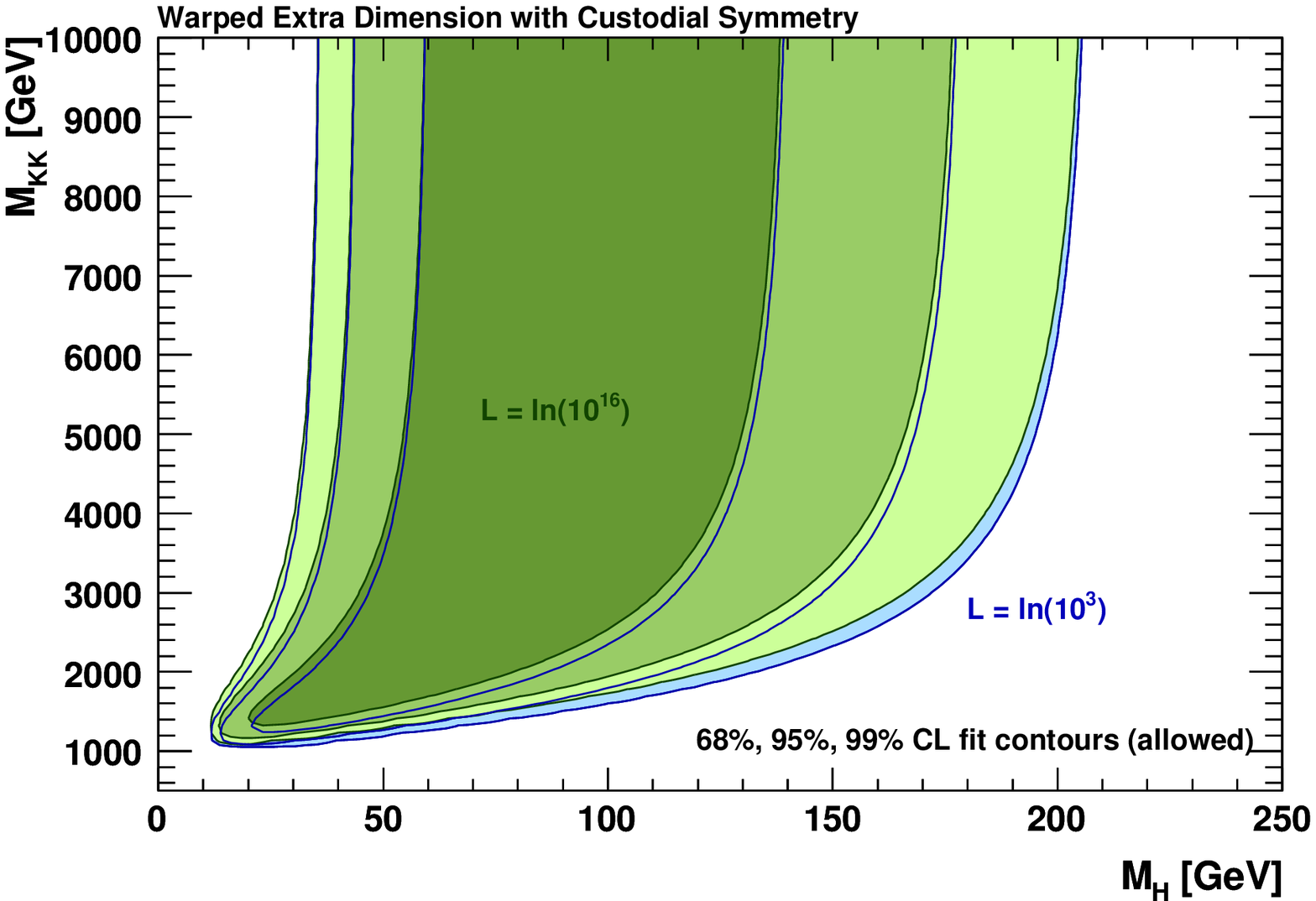, scale=\HalfPageWidthScaleWED}
  \epsfig{file=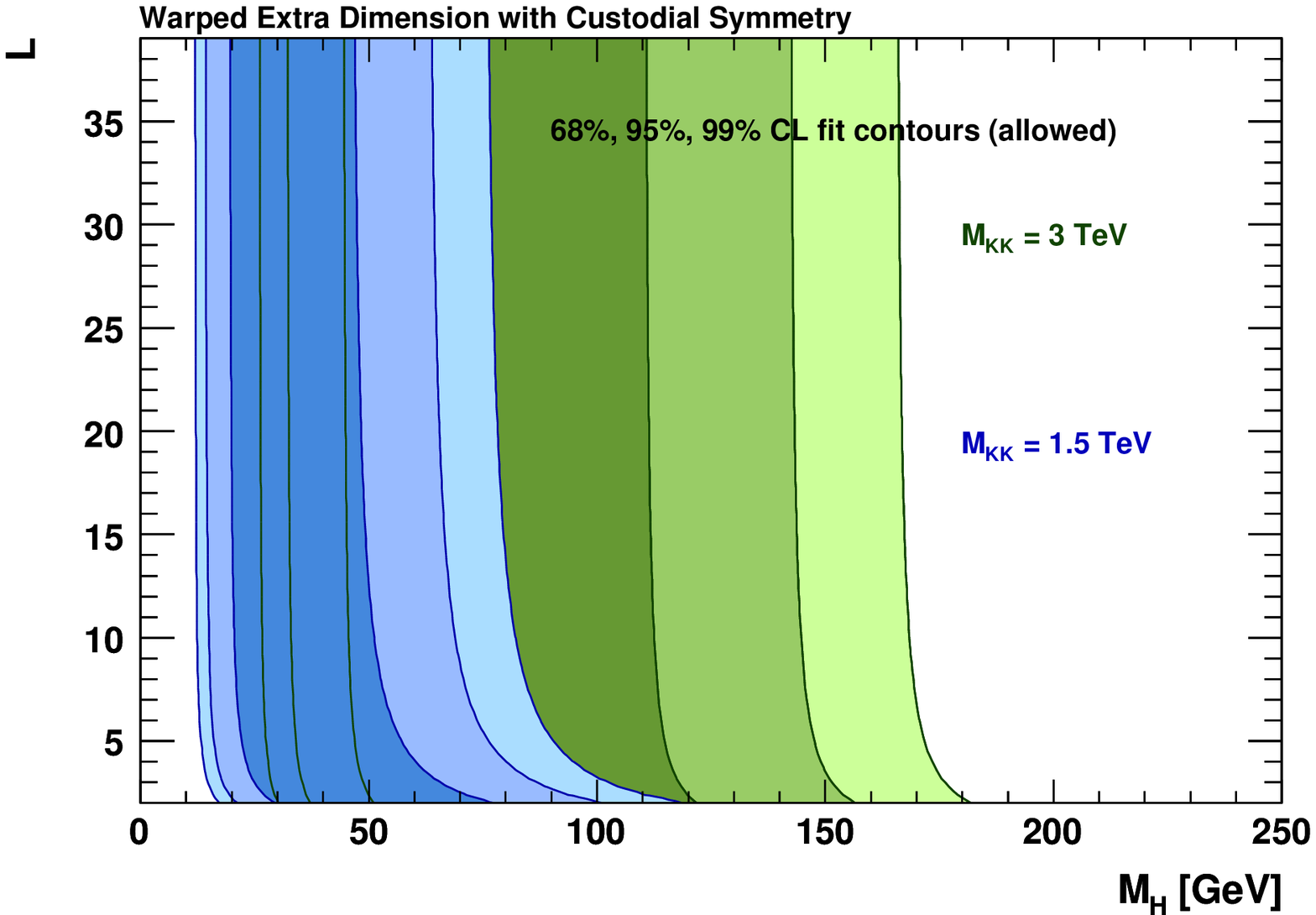, scale=\HalfPageWidthScaleWED}
  \vspace{-0.2cm}
  \caption[]{Constraints in the warped extra dimensions model. Shown 
             are the 68\%, 95\%, and 99\% CL allowed fit contours 
             in two-dimensional planes involving $M_H$, $L$ and $M_\KK$ as
             derived from the RS model fits to the electroweak data 
             without (top and middle) and 
             with (bottom) introducing custodial symmetry. }
  \label{fig:RSscans}
\end{figure}

In a different approach to lowering the constraint on $M_\KK$ from the
$T$ parameter, one introduces a so-called \emph{custodial isospin gauge
symmetry}~\cite{Agashe:2003zs}. The electroweak gauge
symmetry is thereby enhanced to $SU(2)_L \times SU(2)_R \times
U(1)_{B-L}$ yielding a $SU(3)_C\times SU(2)_L \times SU(2)_R \times
U(1)_{B-L}$ gauge symmetry in the bulk of the extra dimension. 
$SU(2)_R$ is then broken to $U(1)_R$ on the Planck brane resulting 
in a spontaneous breaking of $U(1)_R\times U(1)_{B_L}$ to
$U(1)_Y$. Consequently, the right-handed fermionic fields are promoted
to doublets of this symmetry.
Adding custodial isospin symmetry to the RS model leaves the $S$
parameter unchanged with respect to Eq.~(\ref{eq:Swed}), while 
$T$ becomes warp-factor suppressed~\cite{Agashe:2003zs,Casagrande:2008hr}
\beq
  T  =  - \frac{\pi v^2}{4\cos^2\theta_W M_\KK^2} \frac{1}{L}\,.
\eeq
The bottom plot of Fig.~\ref{fig:RSSvsT} shows the corresponding
allowed region for the same parameter ranges as in the top plot. The
negative $T$ oblique correction inherent in the custodial model adds
to that of the SM so that only small values of $M_H$ are allowed.

The bottom panel of Fig.~\ref{fig:RSscans} shows the dependence of the
two model parameters on the Higgs mass. Even though a light Higgs
cannot counteract on the new physics contributions the model parameters are
less constrained leading to a reduced lower bound on $M_\KK$.
However, very small $M_\KK$ lead to excluded Higgs masses. In
addition, the strong correlation between $M_\KK$ and $L$ is removed so that
the Higgs mass and $M_\KK$ are practically independent of $L$. Therefore, 
there is no need to introduce a cut-off at a specific scale.

As an alternative to custodial symmetry, it was proposed to reduce
the contribution to the $T$ parameter by also allowing the Higgs to
propagate into the bulk. This leads to a preferably heavy Higgs, which can
lower the bound on $M_\KK$ by several TeV and therefore shift the lightest
KK modes in the accessible range of the LHC~\cite{Cabrer:2011vu}.

In a bulk version of the Rattazzi-Zaffaroni
model~\cite{Rattazzi:2000hs}, it is assumed that the SM Yukawa
hierarchy is set by UV physics and that the fundamental 5D Yukawa
couplings are shined through the bulk by scalar flavor fields in
agreement with flavor and CP violation constraints. Thus, a bound on
the KK-scale as low as 2\:TeV is allowed for specific parameter
configurations~\cite{Delaunay:2010dw}. However, this alternative
description of flavor causes the new physics contributions to the EW
parameters to be not oblique and a description by the usual \STU
formalism would be incomplete. Nonetheless, the Higgs mass may adopt
values up to 200\:GeV~\cite{Delaunay:2011vv}.  Recent hints for new
physics from the Tevatron, e.g. the top quark
forward-backward-asymmetry can be easily accomodated in this Flavor
Triviality model.

There have been various experimental searches for high-mass graviton resonances 
decaying to, \eg, photon or electron pairs within the original RS model at 
the LHC and 
Tevatron~\cite{CMS-EXO-10-019,Abazov:2010xh,Aaltonen:2010cf,Aaltonen:2011xp}. 
In these analyses, the invariant mass of the two-particle final states is used
to set limits on the RS-graviton production cross section and lowest-level graviton 
mass scale. The latter one is found to be $M_G > \rm 1058\:GeV$ and 
$M_G > \rm 560\:GeV$ at 95\% CL for $\sqrt{8\pi}k/M_{\rm Pl} = 0.1$ and 
$\sqrt{8\pi}k/M_{\rm Pl} = 0.01$, respectively~\cite{Aaltonen:2011xp}.

%% file: Technicolor.tex
\subsection{Technicolour}

Elementary Higgs models provide no dynamical explanation for electroweak symmetry 
breaking and require a high degree of finetuning. One of the first attempts to 
address these shortcomings of the Standard Model were so-called technicolour (TC) 
models which were developed in the late 1970s~\cite{Weinberg:1979bn,Susskind:1978ms}. 
These models introduce a new QCD-like gauge 
interaction that is asymptotically free at high energies but confining at the 
electroweak scale. It is assumed that the technicolour gauge interaction is to be 
based on a $SU(N_{\TC})$ gauge group $G_{\TC}$, where $N_\TC$  is the number of 
technicolours, and couples to one or more doublets of massless Dirac 
\textit{technifermions}. In analogy to QCD, the running gauge coupling $\alpha_{\TC}$ 
triggers a spontaneous chiral symmetry breaking, which leads to a dynamical 
mass generation of the technifermions and, in addition, to a large number 
of massless Goldstone bosons. It is further postulated that the technifermions 
transform chirally under the electroweak gauge group $SU(2)\times U(1)$ 
so that three linear combinations of the Goldstone bosons couple to three 
electroweak gauge currents. It was shown in Ref.~\cite{Weinberg:1979bn} that 
these Goldstone bosons (the so-called technipions $\pi_T$) can give mass to the 
electroweak gauge bosons by the usual Higgs mechanism. The properties of any remaining 
technipions (their corresponding quantum numbers and masses) are model dependent. 
Similar to the vector mesons in QCD, further technicolour resonances with masses
in the TeV range are expected~\cite{Nakamura:2010zzi}. Direct searches for such 
resonances were performed at several collider experiments studying dilepton and 
dijet resonances~\cite{Nakamura:2010zzi,Abazov:2006iq}. Model dependent 95\% CL 
exclusion bounds on technipion and technirho masses of $80\:\gev<m_{\pi_{T}}<115\:\gev$ 
and $170\:\gev<m_{\rho_{T}}<215\:\gev$ were obtained. 

Such simple versions of technicolour models do not explain the explicit breaking 
of chiral symmetries of quarks and leptons. Extended technicolour models (ETC)
have been developed to address this issue by assuming that ordinary $SU(3)$ 
colour, $SU(N_{\TC})$ technicolour, and flavour symmetries are unified into one 
gauge group $G_{\ETC}$, which allows the technifermions to couple to quarks and 
leptons via gauge bosons of the enlarged group. In $G_{\ETC}$ technifermions, 
quarks, and leptons belong to the same representations. Hence flavour, colour 
and technicolour can be interpreted as a subset of the ETC quantum numbers.
It is assumed that the ETC gauge symmetry breaking into $SU(3)\times SU(N_{\TC})$ 
occurs at scales well above the TC scale of $0.1$--$1\:\tev$.
The broken gauge interactions give mass to the quarks and leptons by connecting 
them to technifermions. An introduction to technicolour models can 
be found, for instance, in Ref.~\cite{Lane:2002wv}. 

Because technicolour is a strongly interacting theory, the oblique corrections 
of technicolour models cannot be calculated by ordinary perturbation theory. Two 
approaches are followed to address these difficulties~\cite{Chivukula:1992nw}. 
The first approach assumes that $S$ and $T$ can be 
expressed as a spectral integral, which is evaluated with the use of QCD data and then 
extrapolated to technicolour energies~\cite{Peskin:1991sw}. The second approach
is based on the relations of $S$, $T$, and $U$ to the coefficients of four-derivative 
operators in the chiral Lagrangian~\cite{ Holdom:1990tc, Coleman:1969sm}. Both approaches give 
consistent results for QCD-like technicolour models.

The magnitude of the radiative corrections in technicolour models increases with 
the number of technicolours ($N_{\TC}$) and the number of techniflavours ($N_{\TF}$). It 
is therefore justified (conservative) to choose a minimal ETC model~\cite{Ellis:1994pq}, 
which has one technicolour generation with $N_{\TC}=2,3$, to study the compatibility 
of ETC models with the electroweak data. The model chosen here contains a colour triplet 
of techniquarks ($U,D$) with degenerate mass, and a doublet of technileptons ($N,E$)
with $m_N \le m_E$ to allow for isospin splitting. The technineutrino $N$ can be of 
either Dirac or Majorana type. Following 
Refs.~\cite{Peskin:1991sw,Ellis:1994pq,Golden:1990ig,Sundrum:1991rf}, 
the oblique corrections for Dirac technineutrinos are given by
\begin{eqnarray}
\label{eq:STTechnicolourDirS}
   S_D & \!\!=\!\! & 0.1\cdot(\NC+1)\cdot N_{\TC} - \frac{N_{\TC}}{6\pi}\cdot Y\cdot \ln r\,, \\[0.2cm]
\label{eq:STTechnicolourDirT}
   T_D & \!\!=\!\! & \frac{N_{\TC}}{16\pi s_0^2 c_0^2} \frac{m_E^2}{M_Z^2} 
        \left( 1 + r - 2 \frac{r}{r-1} \ln r \right) \underset{\lim r\rightarrow 1}{=} \frac{N_{\TC}}{12\pi s_0^2 c_0^2} \frac{\Delta m^2}{M_Z} \,,\\[0.2cm] 
\label{eq:STTechnicolourDirU}
   U_D & \!\!=\!\! & \frac{N_{\TC}}{6\pi} \left[ -\frac{5r^2-22r+5}{3(r-1)^2} + 
                               \frac{r^3-3r^2-3r+1}{(r-1)^3} \ln r \right]  \underset{\lim r\rightarrow 1}{=} \frac{2N_{\TC}}{15\pi} \frac{\Delta m^2}{m_E^2} \,, 
\end{eqnarray}
and for Majorana technineutrinos by
\begin{eqnarray}
\label{eq:STTechnicolourMajS}
S_M & \!\!=\!\! & (0.04+0.1\NC)\cdot N_{\TC} + \frac{N_{\TC}}{6\pi} 
        \bigg(-\frac{r}{(1+r)^2} \cdot \bigg[\frac{8}{3} + \frac{3r-4r^2+3r^3}{(1-r^2)^2} \\\nonumber
    & & +\; 2\frac{r^6 - 3r^4 + 6r^3 - 3r^2 + 1}{(1-r^2)^3} \ln r \bigg] 
        + \frac{1-r}{1+r} \ln r + \frac{3}{2} \bigg)\,, \\[0.2cm]
\label{eq:STTechnicolourMajT}
T_M & \!\!=\!\! & \frac{N_{\TC}}{16\pi s_0^2 c_0^2} \frac{m_E^2}{M_Z^2} 
        \left(2-\frac{4r}{r^2-1} \ln r 
              + \frac{4r}{(r+1)^2} \left[1-\frac{r^2+1}{4r} - \frac{r^2-r+1}{r^2-1} \ln r\right] 
        \right) \\\nonumber 
        & \!\!=\!\! & - \frac{N_{\TC}}{12\pi s_0^2 c_0^2} \frac{\Delta m^2}{M_Z^2}\bigg|_{\lim r\rightarrow 1}\,, \\[0.2cm]
\label{eq:STTechnicolourMajU}
U_M & \!\!=\!\! & \frac{N_{\TC}}{6\pi} \bigg( \frac{r}{(r+1)^2} \left[ \frac{8}{3} + \frac{3r^3-4r^2+3r}{(r^2-1)^2} -2\frac{r^6-3r^4+6r^3-3r^2+1}{(r^2-1)^3} \ln r \right] \\\nonumber
    & & +\; \frac{r^3-3r^2-3r+1}{(r-1)^3} \ln r - \frac{13}{6} + \frac{4r}{(r-1)^2} \bigg) \\\nonumber 
        & \!\!=\!\! & -  \frac{2N_{\TC}}{15\pi} \frac{\Delta m^2}{m_E^2}\bigg|_{\lim r\rightarrow 1}\,,
\end{eqnarray}
where $Y=-1$ denotes the weak hypercharge of the technilepton doublet, $\NC=3$ defines 
the number of QCD colours, and  $r=m_N^2/m_E^2$. Several aspects should be noted: 
in the limit of $\Delta m = m_E - m_N \ll m_E$, the contributions to the $T$
and $U$ parameters depend linearly on $\Delta m^2$ and are up to a sign-flip 
equivalent for the Dirac and Majorana cases. Similar formulas for $T$ and $U$ can be 
derived for the techniquark sector, but since $m_U=m_D$ is postulated their contributions
vanish. The situation is different for the $S$ parameter where a term proportional to 
$\NC$ arises from the techniquark sector due to nonperturbative contributions~\cite{Sundrum:1991rf}. 
The corresponding oblique corrections from the techniquark doublet ($U,D$) are obtained 
by Eqs.~(\ref{eq:STTechnicolourDirS}--\ref{eq:STTechnicolourDirU}) with an extra 
factor $\NC$ and the appropriate value of $Y$.  

\begin{figure}[p]
   \centerline{\epsfig{file=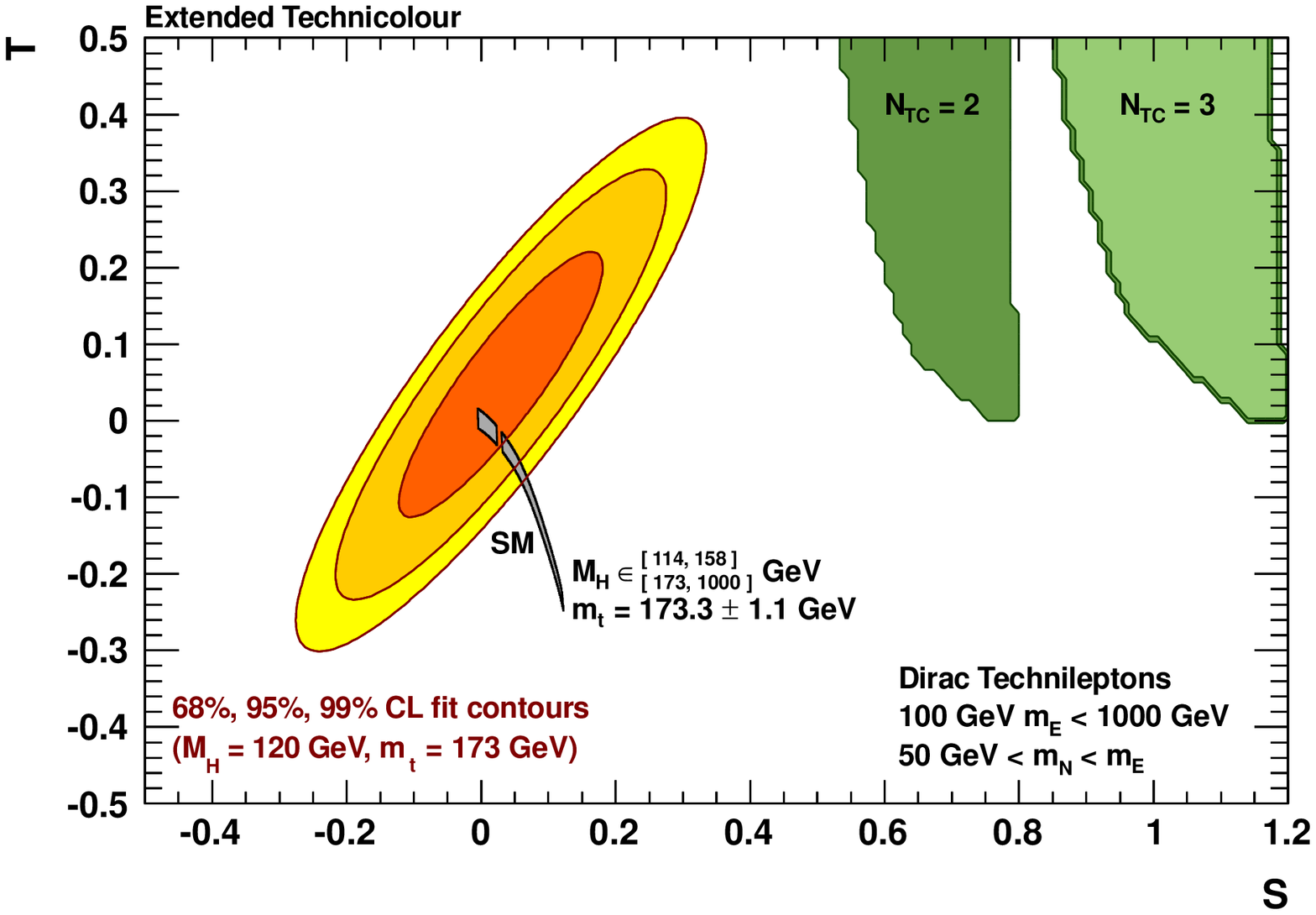, scale=\defaultFigureScale}}
  \vspace{0.4cm}
   \centerline{\epsfig{file=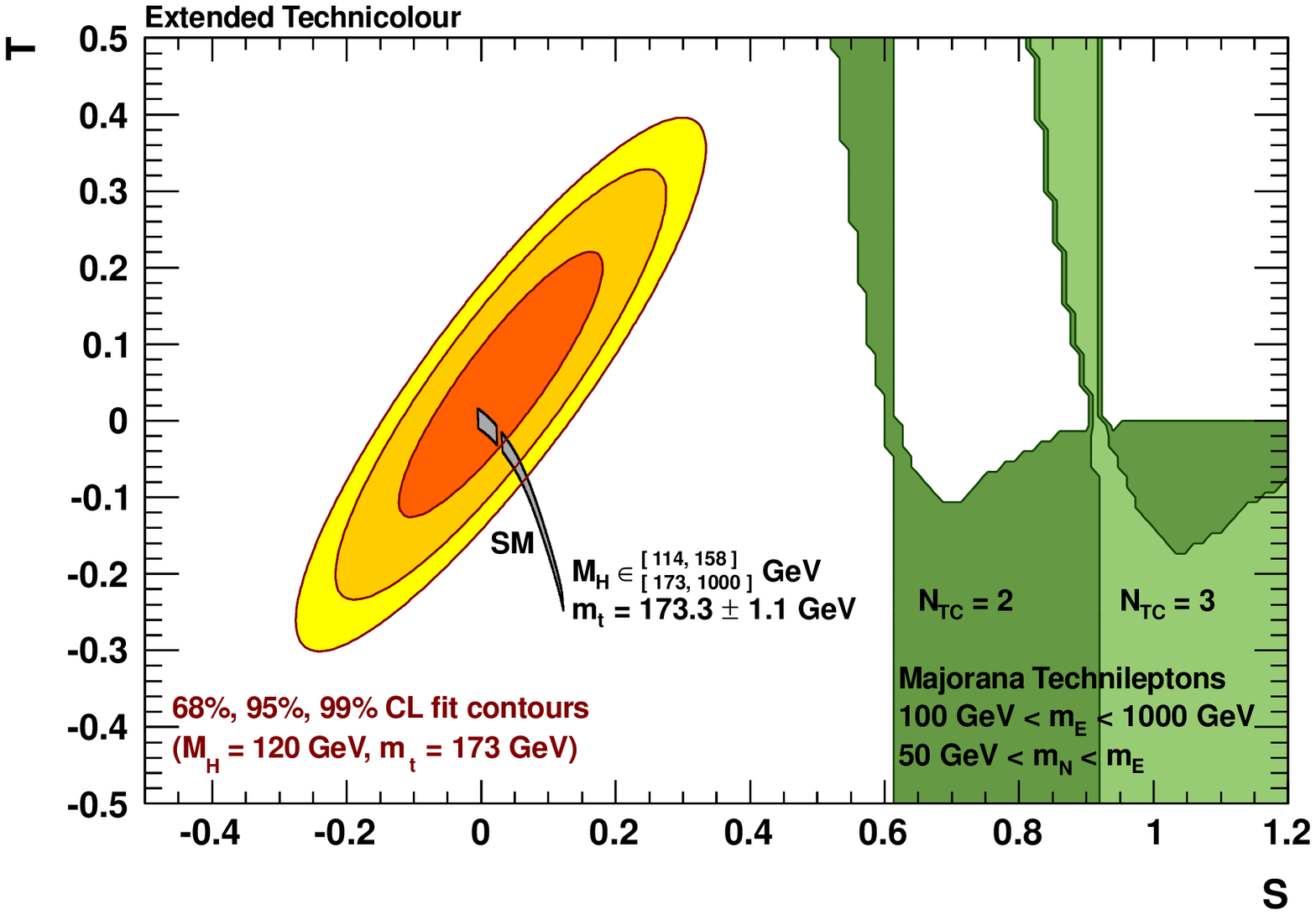, scale=\defaultFigureScale}}
  \vspace{ 0.3cm}
  \caption{\label{fig:FigTC_STU_Dirac_and_Majorana} 
           Oblique parameters for extended technicolour models. 
           Shown are the $S$, $T$ fit results (without $U$ constraint) compared
           with predictions from the SM (grey) and the ETC model 
           for 2 (dark green) and 3 (light green) technicolours, and assuming 
           Dirac (top) and Majorana (bottom) technineutrinos, respectively. 
           The green ETC areas correspond to the predicted parameter regions when 
           varying the technilepton masses in the ranges indicated on the plots. 
           The techniquark doublet is assumed to have degenerate mass. }
\end{figure}

Figure~\ref{fig:FigTC_STU_Dirac_and_Majorana} shows the predicted $S$ and $T$ 
values for the Dirac (top) and Majorana (bottom) technineutrino
cases together with the SM prediction and the electroweak data (ellipses). 
The shaded (green) areas correspond to the allowed parameter regions when varying 
the technilepton masses in the ranges\footnote
{
   The lower mass limits are determined by where the calculation of 
   the oblique parameters can be trusted. 
} 
$100\:\gev < m_E < 1\:\tev$ and $50\:\gev < m_N < m_E$. 
For both technineutrino hypotheses, the $S$ parameter is 
disfavoured by the electroweak precision data as was first discussed in 
Refs.~\cite{Peskin:1991sw,Ellis:1994pq}. The main difference in both figures for 
either model is the allowed region of $r=m_N^2/m_E^2$, which governs the amount of
isospin violation in the technilepton doublet. The sharp vertical edges for the 
different $N_{TC}$ regions correspond to the smallest allowed value of $r$. Small 
values of $r$, corresponding to large isospin violation in the technilepton sector, 
lower $S$ but increase $T$. 
Even though the $T$ parameter prediction differs in sign between the Dirac and 
Majorana technineutrino cases, both hypotheses remain compatible with the 
current data. Compatible results are also found for the $U$ parameter 
for which, assuming $100\:\gev < m_E < 500\:\gev$ and $50\:\gev < m_N < m_E$,
we find the predicted ranges
\begin{equation}
  U=\begin{cases}
   \; [0.04,\, 0.31]    & \text{Dirac technineutrinos, }N_{TC}=2\\
   \; [0.06,\, 0.47]    & \text{Dirac technineutrinos, }N_{TC}=3\\
   \; [-0.01,\, 0.25]  & \text{Majorana technineutrinos, }N_{TC}=2\\
   \; [-0.01,\, 0.38]  & \text{Majorana technineutrinos, }N_{TC}=3
\end{cases}
\end{equation}
where the large upper bounds in all cases arise from small values of $r$, that is,
large isospin violation.

It should be noted that the same mechanism that generates the $u,d$ and $c,s$ quark 
masses and their splitting is also responsible for the technifermion mass splitting. 
The splitting is therefore expected to be significantly smaller than the dynamically 
generated masses of the technifermions so that large isospin violation is 
disfavoured in extended technicolour models.

The $S$ incompatibility problem is present in all technicolour models that are built
upon scaling up ordinary QCD. It should be noted, that allowing isospin violation
in the techniquark sector leads to a further increase of the $S$ parameter and 
therefore an even larger incompatibility. The problem may be remedied by introducing 
non-QCD like technicolour gauge dynamics with a slowly evolving (or {\em walking}) 
gauge coupling $\alpha_{\TC}(\mu)$ over the large energy range from the TC to the ETC symmetry
breaking scales. A prediction of the oblique parameters for these so-called walking
technicolour models turns however out to be difficult as the QCD renormalisation group
equations cannot be applied anymore. In recent years, predictions for so-called 
holographic walking technicolour models have been made, which indicate a possible 
consistency with the electroweak 
data~\cite{Dietrich:2005jn, Mintakevich:2009wz, Hirn:2006nt, Hirn:2008tc}.

%% file: Conclusions.tex
\section{Conclusions and Perspectives}
\label{sec:conclusions}

We have updated in this paper the results of the  Standard Model 
fits to electroweak precision data with the Gfitter package, and 
revisited the electroweak constraints on several Standard Model extensions. The fit 
uses newest experimental results on the top quark and $W$ boson masses, and a new
evaluation of the hadronic contribution to the electromagnetic fine-structure constant
at $M_Z$. The update of the latter parameter reduces the tension between the electroweak 
fit and the LEP 
limit on the Higgs boson mass. The LEP and Tevatron data on the direct Higgs searches 
have been extended by results from the 2010 Higgs searches at the LHC, using data 
corresponding to approximately 35\invpb of integrated luminosity. 

From the complete fit, including the results from direct Higgs
searches, we find for the mass of the Higgs boson an upper limit of
143\:\gev at 95\% confidence level.  This bound is alleviated to
163\:\gev when not including the direct Higgs searches (standard fit).
Theoretical errors parametrising the uncertainties in the perturbative
predictions of $M_W$, $\sinfeff$, and the electroweak form factors,
contribute with approximately $8\:\gev$ to the total fit error found
for $M_H$ in the standard fit.  In a fit excluding the measurement of
the top quark mass (but including the direct Higgs searches) we obtain
the indirect determination $m_t= (177.2\pm3.4) \:\gev$, in fair
agreement with the experimental world average. This error being much
larger than that of the direct measurement, a reduction in the
experimental error will not significantly impact the electroweak fit
if the central value does not move by an unexpected amount. The
experimental and theoretical effort should therefore concentrate on
clarifying the relation between the measured top mass and the top pole
mass used in the electroweak formalism, and the uncertainty inherent
in identifying the latter mass with the former one. From the indirect
determination of the mass of the $W$ boson from the complete fit we
find $(80.360^{+0.014}_{-0.013})\:\gev$, which is more precise and $1.6\sigma$
below the experimental world average.  The indirect determination of
the effective weak mixing angle from the complete fit gives 
$\sinleff =0.23148\pm0.00011$, which is compatible with and more
precise than the direct experimental average from the asymmetry
measurements at LEP and SLD. The strong coupling constant to \NNNLO
order at the $Z$-mass scale is found to be $\asZ=0.1194\pm0.0028$,
with negligible theoretical uncertainty due to the good convergence of
the perturbative series at that scale.

Using the oblique parameter approach encoded in the \STU formalism, together 
with a fixed Standard Model reference of $M_{H,\rm ref}=120\:$GeV and
$m_{t,\rm ref}=173\:$GeV, we derive the experimental constraints
$S=0.03\pm 0.10$, $T=0.05\pm 0.12$ and $U=0.07\pm 0.11$, with large 
correlations between the parameters. These results are used to revisit the
oblique parameter constraints of the Standard Model and selected extensions, such 
as a fourth family, two Higgs doublet and inert Higgs models, littlest Higgs, 
models with large, universal or warped extra dimensions and technicolour. 
The constraints from the data are used to derive allowed regions in the 
parameter spaces of these models, where we confirm results from earlier studies. 
In most of these models a heavy Higgs boson can be made compatible with the 
electroweak precision data by adjusting the required amount of weak isospin 
breaking.

Given the strong performance of the LHC and its experiments, with already over 
1\invfb integrated
luminosity accumulated at the date of this paper, the present analysis might be 
among the last global electroweak fits working with Higgs limits only. In case 
of a Higgs discovery, the electroweak fit does not cease to be important. For
example, as a test of the Standard Model the indirect prediction of 
the $W$ mass will achieve an accuracy of $11\:\mev$ that can be confronted 
with experiment. The precision of the $M_W$ world average measurement will further 
improve with forthcoming Tevatron analyses and, eventually, by a measurement 
at the LHC with (expectantly) competitive error with the indirect determination 
or better. A discovery of the Higgs would also strongly impact the allowed 
parameter space of many new physics models via mainly the reduced flexibility of 
the $S$ oblique parameter and the then known amount of weak isospin violation in 
the electroweak Standard Model. 

\subsubsection*{Acknowledgements}
\addcontentsline{toc}{section}{Acknowledgements}
\label{sec:Acknowledgments}

\begin{details}
We are indebted to the LEP-Higgs and Tevatron-NPH working groups for providing the numerical ${\rm CL}_{\rm s+b}$ results of 
their direct Higgs-boson searches.
It is a pleasure to thank Alexander Lenz for instructive discussions 
on extended fermion generations. 
We are grateful to Hong-Jian He und Shu-fang Su for their help on
the implementation of the 2HDM oblique corrections. 
We thank Gian Giudice for helpful correspondance on large extra dimensions. 
We also thank Kenneth Lane for his helpful comments and suggestions 
on the technicolour studies presented here.
This work is funded by the German Research Foundation (DFG) in the
Collaborative Research Centre (SFB) 676 ``Particles, Strings and the
Early Universe'' located in Hamburg.
A.H. thanks the Aspen Center for Physics, which is supported by an NSF grant, 
for its hospitality during the finalisation of this paper.

\end{details}

%% file: Appendix.tex
\begin{appendix}

\section{Oblique Parameter Formalism}
\label{app:obliqueParams}

\subsubsection*{Absorption of radiative corrections} \label{sec:absorbtion}

Oblique corrections can generally be absorbed into the fundamental 
constants occurring at the tree-level of the SM. Kennedy and 
Lynn~\cite{Kennedy:1988sn} have shown that this statement is general to 
all vacuum polarisation orders. In this appendix we illustrate 
the absorption process with some explicit examples.

The effects of oblique corrections on fermion scattering can be 
determined by examining how the gauge boson vacuum polarisation functions
\begin{equation}
\label{eq:vacpol}   
   \Pi_{ab}^{\mu\nu}(q)=\Pi_{ab}(q^{2})g^{\mu\nu} + (q^{\mu}q^{\nu}\ \textrm{terms})\,,
\end{equation}
with $a,b = \gamma, W, Z$, appear in the electroweak observables of 
interest.\footnote
{
   Owing to $U(1)_{Q}$ gauge symmetry, for the photon propagator the 
   term $q^{\mu}q^{\nu}$ has no physical effect. 
   For the (massive) $W$ and $Z$ propagators the terms are also 
   negligible, since, in the interaction with light fermions, 
   they are suppressed by the fermion mass scale compared with the 
   $g^{\mu\nu}$ parts. From now on we shall ignore the $q^{\mu}q^{\nu}$ terms.
}
The functions~(\ref{eq:vacpol}) have an SM and an unknown new physics 
component: $\Pi_{ab}^{\mu\nu}(q^2)=\Pi_{ab}^{S\!M}(q^2) + \delta\Pi_{ab}^{NP}(q^2)$.

For the $W$ and $Z$ bosons one finds the following mass corrections to 
the tree-level quantities\footnote
{
   Throughout this appendix the superscript $(0)$ is used to label tree-level 
   quantities.
}
\beq 
\label{eq:bosonmasscorrections}
\begin{array}{rclcl}
   M_{W}^{2} & \!\!\equiv\!\! & M_{W}^{2}(M_{W}^{2}) &\!\!=\!\!& M_{W}^{(0)\,2} + \Pi_{WW}(M_{W}^{2})\,,\\
   M_{Z}^{2} & \!\!\equiv\!\! & M_{Z}^{2}(M_{Z}^{2}) &\!\!=\!\!& M_{Z}^{(0)\,2} + \Pi_{ZZ}(M_{Z}^{2})\,,
\end{array}
\eeq
where the vacuum polarisation functions are evaluated at the poles of 
the propagators.

For the massless photon one has
\begin{equation}
   \Pi_{\gamma\gamma}(0)=\Pi_{\gamma Z}(0)=0\,.
\end{equation}

The impact on the electromagnetic constant $\alpha$ is obtained by taking the 
leading-order photon propagator plus the first-order correction. Together 
these yield
\begin{equation}
   \frac{-ie^{2}}{q^{2}}\Big(1+i\Pi_{\gamma\gamma}(q^{2})\cdot\frac{-i}{q^{2}}\Big)\,.
\end{equation}
The observed value of the electric charge is then found by taking the limit 
$q^{2}\rightarrow 0$ of this expression
\begin{equation} \label{eq:alphacorrected}
   4\pi\alpha_{*}(0) \equiv e_{*}^{2}(0)=\frac{g^{2}g^{\prime 2}}{g^{2}+g^{\prime 2}}
   \left(1+\Pi_{\gamma\gamma}^{\prime}(0)\right)\,,
\end{equation}
where 
\begin{equation}
   \Pi_{\gamma\gamma}^{\prime}(0)=\frac{d\Pi_{\gamma\gamma}}{dq^{2}}\bigg|_{q^{2}=0}\,.
\end{equation}

The weak mixing angle $s_{W}$ appears in the interactions of $Z$ bosons to 
fermions, and is shifted by the vacuum polarisation amplitude $\Pi_{Z\gamma}$.
The corrections change a $Z$ into a photon that decays to two fermions, with 
coupling strength $Qe$, leading to the contribution
\begin{equation}
   i\Pi_{Z\gamma}(q^{2})\frac{-i}{q^{2}}\cdot(ieQ)\,.
\end{equation}
Including this correction, the $Z$-fermion interaction takes the form
\begin{equation}
   i\sqrt{g^{2}+g^{\prime 2}}\Big(T^{3}-s_{*}^{2}Q\Big)\,,
\end{equation}
with
\begin{equation}
   s_{*}^{2}(M_{Z}^{2}) = s_{W}^{(0)\,2} - \frac{e}{\sqrt{g^{2}+g^{\prime 2}}}
   \frac{\Pi_{Z\gamma}(M_{Z}^{2})}{M_{Z}^{2}}\,,
\end{equation}
as evaluated at $q^{2}=M_{Z}^{2}$.

The Fermi constant, obtained from muon decays, as mediated by $W$ propagator,  
receives the first-order correction from the $W$ vacuum polarisation function
\begin{equation}
   \frac{-ig^{2}}{q^{2}-M_{W}^{2}}\Big(1+i\Pi_{WW}(q^{2})\frac{-i}{q^{2}-M_{W}^{2}}\Big)\,.
\end{equation}
At $q^{2}=0$, the observed Fermi constant process shifts to
\begin{equation} \label{eq:fermicorrected}
   \frac{G_{F*}}{\sqrt{2}}=\frac{1}{2v^{2}}\Big(1-\frac{\Pi_{WW}(0)}{M_{W}^{2}}\Big)\,.
\end{equation}

These examples illustrate that oblique corrections can be 
absorbed into the fundamental constants occurring of the SM.
This conclusion is applied in the following Section.


\subsubsection*{Introduction of the {\em S, T, U} parameters}

In the SM with a single Higgs doublet the relationship between the neutral and 
charged weak couplings is fixed by the ratio of $W$ and $Z$ boson masses
\beq \label{eq:rhorelation}
   \rho\ =\ \frac{M_{W}^{2}}{M_{Z}^{2}\cos^{2}\!\tzw}\,,
\eeq
where $\rho_{0}=1$ at tree level. Generally one writes
\beq
   \rho\ =\ 1+\Delta\rho\,,
\eeq
where $\Delta\rho$ captures the radiative corrections to the gauge 
boson propagators and vertices. Inserting the first-order mass-corrections 
of Eqs.~(\ref{eq:bosonmasscorrections}) into Eq.~(\ref{eq:rhorelation}) gives
\begin{equation} 
\label{eq:rhofirstorder}
   \Delta\rho\ =\ \frac{\Pi_{WW}(0)}{M_{W}^{2}}-\frac{\Pi_{ZZ}(0)}{M_{Z}^{2}}\,.
\end{equation}

The tree-level vector and axial-vector couplings occurring in the $Z$ 
boson to fermion-antifermion vertex 
$i\fbar\gamma_\mu(\gvz{f}+\gvz{f}\gamma_5)f Z_\mu$ are given by
\beqn
   \gvz{f} & \!\!=\!\! & I^{f}_{3} - 2Q^{f} \sin^{2}\!\tzw\,, \\
   \gaz{f} & \!\!=\!\! & I^{f}_{3}\,, 
   \label{eq:treecouplings2}
\eeqn
where $Q^f$ and $I^{f}_{3}$ are respectively the charge and the third 
component of the weak isospin. In the (minimal) SM, containing only 
one Higgs doublet, the weak mixing angle is defined by 
\beq
   \sin^{2}\!\tzw\ =\  1-\frac{M_{W}^2}{M_{Z}^2}\,.
\label{eq:massrelation}
\eeq
Electroweak radiative corrections modify these relations, leading to 
the effective weak mixing angle and effective couplings
\begin{align}
\label{eq:Zcouplings_sinfeff}
   \sinfeff & \ =\   \kZ{f} \sin^{2}\!\tzw\,, \\
\label{eq:Zcouplings_rZ}
   \gv{f}   & \ =\   \sqrt{\rZ{f}} \left( I^{f}_{3} - 2Q^{f} \sinfeff \right)\,, \\
\label{eq:Zcouplings_kZ}
   \ga{f}   & \ =\   \sqrt{\rZ{f}}  I^{f}_{3}\,,
\end{align} 
where the radiative corrections are absorbed in the form factors 
$\kZ{f}=1+\Delta\kZ{f}$ and $\rZ{f}=1+\Delta\rZ{f}$.

Electroweak unification leads to a relation between weak and electromagnetic 
couplings, which at tree level reads
\beq
\label{eq:GF}
     \GF \ =\  \frac{\pi \alpha}{\sqrt{2} \left(M_{W}^{(0)}\right)^2 
               \left(1-\frac{(M^{(0)}_{W})^2}{M_{Z}^2}\right)}\,.
\eeq
The radiative corrections are parametrised by multiplying the r.h.s. of 
Eq.~(\ref{eq:GF}) with the form factor $(1-\Delta r)^{-1}$. 
Using Eq.~(\ref{eq:massrelation}) and resolving for $M_W$ gives
\beq
\label{eq:MWdeltar}
    M_{W}^2 \ =\  \frac{M_{Z}^2}{2}
              \left(1+\sqrt{1-\frac{\sqrt{8}\,\pi\alpha(1+\Delta r)}{\GF{M_{Z}^2}}}\right)\,.
\eeq

An extra correction is required for the $Z\rightarrow b\overline{b}$ 
decay vertex. The bottom quark is the only fermion that receives 
unsuppressed vertex corrections from the top quark.
These corrections turn out to be significant -- at the level of 
$G_{F}m_{t}^{2}$ -- and must be accounted for.
The vector and axial couplings receive an extra contribution $\varepsilon_{b}$
\begin{equation}
   \gv{b}= -\frac{1}{2}\sqrt{\rZ{b}}\left( 1-\frac{4}{3}\sin^2\!\theta_{\rm eff} + \varepsilon_{b} \right) 
           \ \ \textrm{and}\quad
   \ga{b}= -\frac{1}{2}\sqrt{\rZ{b}}\left( 1+\varepsilon_{b} \right) \,,
\end{equation}
where $\varepsilon_{b}$ contains all top-quark induced vertex corrections.

The entire dependence of the electroweak theory on $m_{t}$ and $M_{H}$, arising 
from one-loop diagrams and higher, only enters through the four parameters 
$\Delta \kappa$, $\Delta \rho$, $\Delta r$, and $\varepsilon_{b}$. 
The quantities $\Delta \kappa$, $\Delta \rho$, and $\Delta r_{W}$ are 
mostly sensitive to the absolute mass splittings between different 
weak-isospin partners. In practise this means the mass differences 
between the top and bottom quarks, and the $Z$ and $W$ bosons. 
For example, the dominant contributions to $\Delta \rho$ are~\cite{Kennedy:1988sn}
\begin{eqnarray} \label{eq:masssplitting}
   \Delta \rho_{t}& \!\!=\!\! &\frac{3G_{F}}{8\sqrt{2}\pi^{2}}\left[
   m_{t}^{2}+m_{b}^{2}-\frac{2m_{t}^{2}m_{b}^{2}}{m_{t}^{2}-m_{b}^{2}}
   \ln\left(\frac{m_{t}^{2}}{m_{b}^{2}}\right)\right]\,\ \geq\,\ \frac{3G_{F}}{8\sqrt{2}\pi^{2}}(m_{t}-m_{b})^{2} \nonumber \\
   &\!\!\longrightarrow\!\!& \frac{3G_{F}m_{t}^{2}}{8\sqrt{2}\pi^{2}}\,,\ \ \textrm{as}\quad
   m_{t}^{2}\gg m_{b}^{2}\,, \nonumber \\
   & & \nonumber\\
   \Delta \rho_{H} & \!\!=\!\! & \frac{3G_{F}}{8\sqrt{2}\pi^{2}}\left[
   M_{W}^{2}\ln\left(\frac{M_{H}^{2}}{M_{W}^{2}}\right)
   -M_{Z}^{2}\ln\left(\frac{M_{H}^{2}}{M_{Z}^{2}}\right)\right]\,,
\end{eqnarray}
exhibiting a quadratic dependence on the top mass, and a logarithmic 
dependence on the Higgs mass. Since $M_{H}>M_{Z}>M_{W}$, $\rho_H$ is negative.

Ignoring terms proportional to $\ln {m_t/M_Z}$ and vertex corrections, 
which do not contain sizable terms containing $M_H$ and $m_t$, the 
parameters on one-loop level can can be written as~\cite{burgersjegerlehner}:
\begin{eqnarray}
  \label{eq:rhokapr}\nonumber
  \Delta \rho & \!\!=\!\! & \frac{3 G_F M_W^2}{8 \sqrt{2} \pi^2}
                    \left[\frac{m_t^2}{M_W^2} - \frac{\sin^2\!
                        \theta_W}{\cos^2\! \theta_W} \left(
                        \ln{\frac{M_H^2}{M_W^2}} - \frac{5}{6} \right) +
                        ... \right] \\
  \Delta \kappa & \!\!=\!\! & \frac{3 G_F M_W^2}{8 \sqrt{2} \pi^2}
                    \left[\frac{m_t^2}{M_W^2} \frac{\cos^2\!
                        \theta_W}{\sin^2\! \theta_W} - \frac{10}{9} \left(
                        \ln{\frac{M_H^2}{M_W^2}} - \frac{5}{6} \right) +
                        ... \right] \\ \nonumber
  \Delta r_W    & \!\!=\!\! & \frac{3 G_F M_W^2}{8 \sqrt{2} \pi^2}
                    \left[- \frac{m_t^2}{M_W^2}\frac{\cos^2\! \theta_W}
                    {\sin^2\! \theta_W}+ \frac{11}{3} \left(
                        \ln{\frac{M_H^2}{M_W^2}} - \frac{5}{6} \right) +
                        ... \right] \\ \nonumber
  \varepsilon_b & \!\!=\!\! & -\frac{G_{F}m_{t}^{2}}{4\sqrt{2}\pi^{2}}+\ldots\,.
\end{eqnarray}

All quantities are dominated by terms of $G_{F}m_{t}^{2}$. Considering this 
term only, $\Delta k$, $\Delta \rho$, $\Delta r_{W}$ are related as follows
\begin{equation}
   \Delta r_{W} = \frac{c^{2}-s^{2}}{s^{2}} \Delta k=-\frac{c^{2}}{s^{2}} \Delta \rho\,.
\end{equation}

Restoring the $\ln\!\frac{m_t}{m_z}$ terms, the $\varepsilon_{1,2,3}$ parameters 
defined in Eqs.~(\ref{eq:epsilon1}--\ref{eq:epsilon3}) on page~\pageref{eq:epsilon1}
are given by
\begin{eqnarray}
  \label{eq:epssm}\nonumber
  \varepsilon_1 &\!\! =\!\! & \frac{3 G_F M_W^2}{8 \sqrt{2} \pi^2}
                    \left[\frac{m_t^2}{M_W^2} - \frac{\sin^2\!
                        \theta_W}{\cos^2\! \theta_W} \left(
                        \ln{\frac{M_H^2}{M_W^2}} - \frac{5}{6} \right) +
                        ... \right] \\ 
  \varepsilon_2 &\!\! = \!\!& \frac{3 G_F M_W^2}{2 \sqrt{2} \pi^2}\ln \frac{m_t}{M_Z}
                      + ...\\ \nonumber
  \varepsilon_3 &\!\! = \!\!& \frac{3 G_F M_W^2}{8 \sqrt{2} \pi^2}
                      \left[ \frac{2}{9} \left(\ln{\frac{M_H^2}{M_W^2}}
                      - \frac{5}{6} \right) - \frac{4}{9} \ln{\frac{m_t}{M_Z}}
                      +... \right]\,.
\end{eqnarray}

The SM subtraction results in the parameter set $\hat{\varepsilon}$.
In terms of propagator functions one has~\cite{Nakamura:2010zzi} 
\begin{eqnarray} \label{eq:epshats}
  \hat{\varepsilon_{1}}& \!\!=\!\! &\frac{\Pi_{WW}^{NP}(0)}{M_{W}^{2}}-\frac{\Pi_{ZZ}^{NP}(0)}{M_{Z}^{2}}\,, \nonumber \\
  \frac{\hat{\varepsilon_{3}}}{c^{2}}& \!\!=\!\! &
  \frac{\Pi_{ZZ}^{NP}(M_{Z}^{2})-\Pi_{ZZ}^{NP}(0)}{M_{Z}^{2}}
  -\Pi_{\gamma\gamma}^{\prime\,NP}(0)
  -\Big(\frac{c^{2}-s^{2}}{cs}\Big)\frac{\Pi_{Z\gamma}^{NP}(M_{Z}^{2})}{M_{Z}^{2}}\,, \nonumber \\
  \hat{\varepsilon_{3}}\!-\!\hat{\varepsilon_{2}}& \!\!=\!\! &
  \frac{\Pi_{WW}^{NP}(M_{W}^{2})-\Pi_{WW}^{NP}(0)}{M_{W}^{2}}
  -\Pi_{\gamma\gamma}^{\prime\,NP}(0)
  -\Big(\frac{c}{s}\Big)\frac{\Pi_{Z\gamma}^{NP}(M_{Z}^{2})}{M_{Z}^{2}}\,.
\end{eqnarray}
Equivalently, contributions to $\hat{\varepsilon}_{b}$ are the NP vertex correction
to $Z\rightarrow b\overline{b}$. The \STU parameters expressed in terms of the 
$\hat{\varepsilon}$ parameters read
\begin{equation}
   S = \frac{4s^{2}\hat{\varepsilon}_{3}}{\alpha(M_{Z}^{2})}\,,\hspace{0.5cm}
   \label{eq:stu_eps}
   T = \frac{\hat{\varepsilon}_{1}}{\alpha(M_{Z}^{2})}\,,\hspace{0.5cm}
   U = \frac{-4s^{2}\hat{\varepsilon}_{2}}{\alpha(M_{Z}^{2})}\,. 
\end{equation}

\end{appendix}